\documentclass[twoside,11pt]{article}
\pdfoutput=1
\usepackage{jair, theapa, rawfonts}

\usepackage[hyphens]{url}
\usepackage{graphicx}

\usepackage{algorithm}
\usepackage{algorithmic}

\usepackage{booktabs}       
\usepackage{amsfonts}       
\usepackage{nicefrac}       
\usepackage[usenames]{color}
\usepackage[dvipsnames]{xcolor}         

\usepackage{amsmath,amssymb}       
\usepackage{mathbbol}
\usepackage{bm}
\usepackage{bbm}
\DeclareSymbolFontAlphabet{\amsmathbb}{AMSb}
\usepackage{mathtools}
\usepackage{accents}

\newcommand\munderbar[1]{%
  \underaccent{\bar}{#1}}

\usepackage{siunitx}

\DeclareMathOperator*{\minimize}{minimize}
\DeclareMathOperator*{\argmax}{arg\,max}

\usepackage[acronym]{glossaries}

\usepackage{fontawesome5}

\usepackage{xspace}
\usepackage{todonotes}
\usepackage{colonequals}
\usetikzlibrary{patterns,arrows,arrows.meta,calc,shapes,shadows,decorations.pathmorphing,decorations.pathreplacing,automata,shapes.multipart,positioning,shapes.geometric,fit,circuits,trees,shapes.gates.logic.US,fit, shadows.blur,shapes.symbols}

\usetikzlibrary{backgrounds,scopes}
\usepackage{marvosym}
\usepackage{multirow}

\usepackage{float}

\usepackage{tabularx}
\usepackage{booktabs}

\usepackage{caption}
\usepackage{subcaption}
\usepackage{appendix}
\usepackage{cleveref}

\crefname{equation}{Eq.}{Eqs.}
\crefname{algorithm}{Algorithm}{Algorithms}
\crefname{figure}{Fig.}{Figs.}
\crefname{section}{Sect.}{Sects.}
\crefname{table}{Table}{Tables}
\crefname{definition}{Def.}{Defs.}
\crefname{theorem}{Theorem}{Theorems}
\crefname{lemma}{Lemma}{Lemmas}
\crefname{example}{Example}{Examples}
\crefname{assumption}{Assumption}{Assumptions}
\crefname{remark}{Remark}{Remarks}
\crefname{appendix}{Appendix}{Appendices}
\crefname{corollary}{Corollary}{Corollaries}

\usepackage{algorithmic}
\usepackage{tikz} 
\usepackage{pgfplots}
\usetikzlibrary{pgfplots.groupplots}
\usepgfplotslibrary{external,fillbetween}
\pgfplotsset{compat=1.8}

\definecolor{red}{rgb}{0.745,0.192,0.102}
\definecolor{darkgreen}{RGB}{34,161,55}
\definecolor{ruhuisstijlrood}{rgb}{0.745,0.192,0.102}
\definecolor{ruhuisstijlzwart}{rgb}{0,0,0}
\definecolor{ruhuisstijlwit}{rgb}{0.98,0.98,0.98}
\newcommand{\rured}[1]{\textcolor{ruhuisstijlrood}{#1}}
\newcommand{\darkgreen}[1]{\textcolor{darkgreen}{#1}}

\usepackage{mdframed}
\usepackage{enumitem}

\usetikzlibrary{arrows,decorations.pathmorphing,positioning,fit,trees,shapes,shadows,automata,calc}

\newcommand{\prism}{\textrm{PRISM}\xspace}

\newcommand{\stochy}{\textrm{StocHy}\xspace}
\newcommand{\sreachtools}{\textrm{SReachTools}\xspace}
\newcommand{\probreach}{\textrm{ProbReach}\xspace}
\newcommand{\matlab}{\textrm{MATLAB}\xspace}

\newcommand{\R}{\amsmathbb{R}}

\newcommand{\N}{\amsmathbb{N}}
\newcommand{\distr}[1]{\ensuremath{\mathit{Dist(#1)}}}

\newcommand{\card}[1]{\ensuremath{\vert #1 \vert}}

\newcommand{\States}{S}
\newcommand{\goalStates}{\ensuremath{\States_{G}}}
\newcommand{\criticalStates}{\ensuremath{\States_{C}}}
\newcommand{\Actions}{Act}
\newcommand{\initState}{\ensuremath{s_I}}
\newcommand{\transfunc}{P}
\newcommand{\transfuncImdp}{\ensuremath{\mathcal{P}}}

\newcommand{\Interval}{\ensuremath{\amsmathbb{I}}}
\newcommand{\MDP}{\ensuremath{(\States,\Actions,\initState,\transfunc)}}
\newcommand{\iMDP}{\ensuremath{(\States,\Actions,\initState,\transfuncImdp)}}
\newcommand{\mdp}{\ensuremath{\mathcal{M}}}
\newcommand{\imdp}{\ensuremath{\mathcal{M}_\Interval}}

\newcommand{\policy}{\ensuremath{\pi}}
\newcommand{\policySpace}{\ensuremath{\Pi}}

\newcommand{\controller}{\ensuremath{\phi}}

\newcommand{\cStateSpace}{\ensuremath{\mathcal{X}}}
\newcommand{\cControlSpace}{\ensuremath{\mathcal{U}}}

\newcommand{\Partition}{\ensuremath{\mathcal{R}}}
\newcommand{\goalRegion}{\ensuremath{\mathcal{X}_G}}
\newcommand{\criticalRegion}{\ensuremath{\mathcal{X}_C}}
\newcommand{\horizon}{\ensuremath{K}}

\newcommand{\property}{\ensuremath{\varphi_{\bm{x}_0}^\horizon}}
\newcommand{\propertyMDP}{\ensuremath{\varphi_{\initState}^\horizon}}

\newcommand{\reachProbCont}{\ensuremath{\mathit{Pr}^{\controller}( \property )}}

\newcommand{\reachProbDiscr}{\ensuremath{\mathit{Pr}^\policy(\propertyMDP)}}
\newcommand{\reachProbDiscrStar}{\ensuremath{\mathit{Pr}^{\policy^\star}(\propertyMDP)}}
\newcommand{\reachProbDiscrMax}{\ensuremath{\mathit{\overline{Pr}}^\policy(\propertyMDP)}}
\newcommand{\reachProbDiscrMin}{\ensuremath{\mathit{\underline{Pr}}^\policy(\propertyMDP)}}
\newcommand{\reachProbDiscrMinStar}{\ensuremath{\mathit{\underline{Pr}}^{\munderbar{\policy}^\star}(\propertyMDP)}}
\newcommand{\reachProbDiscrMaxStar}{\ensuremath{\mathit{\overline{Pr}}^{\bar{\policy}^\star}(\propertyMDP)}}

\newcommand{\Gauss[2]}{\ensuremath{\mathcal{N}(#1 , #2)}}
\newcommand{\mean}{\ensuremath{\mu}}
\newcommand{\cov}{\ensuremath{\Sigma}}
\usepackage{amsthm}
\newtheorem{theorem}{Theorem}
\newtheorem{assumption}{Assumption}
\newtheorem{definition}{Definition}
\newtheorem{example}{Example}
\newtheorem{lemma}{Lemma}
\newtheorem{remark}{Remark}
\newtheorem{proposition}{Proposition}
\newtheorem{corollary}{Corollary}

\newacronym[]{LTL}{LTL}{linear temporal logic}
\newacronym[]{iid}{i.i.d.}{independent and identically distributed}
\newacronym[]{UAV}{UAV}{unmanned aerial vehicle}

\newacronym[]{PAC}{PAC}{probably approximately correct}
\newacronym[]{DRO}{DRO}{distributionally robust optimization}
\newacronym[plural=MDPs,firstplural=Markov decision processes (MDPs)]{MDP}{MDP}{Markov decision process}
\newacronym[plural=iMDPs,firstplural=interval Markov decision processes (iMDPs)]{iMDP}{iMDP}{interval Markov decision process}
\newacronym[plural=aMDPs,firstplural=augmented MDPs (aMDPs)]{aMDP}{aMDP}{augmented MDP}
\newacronym[plural=POMDPs,firstplural=partially observable Markov decision processes (POMDPs)]{POMDP}{POMDP}{partially observable Markov decision process}

\newcommand{\state}{\ensuremath{\bm{x}}}
\newcommand{\control}{\ensuremath{\bm{u}}}
\newcommand{\noise}{\ensuremath{\bm{w}}}
\newcommand{\drift}{\ensuremath{\bm{q}}}

\newcommand{\region}{\ensuremath{R}}
\newcommand{\inv}{\ensuremath{{-1}}}

\glsdisablehyper

\ShortHeadings{Robust Control for Dynamical Systems with Non-Gaussian Noise}
{Badings, Romao, Abate, Parker, Poonawala, Stoelinga \& Jansen}
\firstpageno{1}

\begin{document}

\title{Robust Control for Dynamical Systems \\ With Non-Gaussian Noise via Formal Abstractions}

\author{\name Thom Badings \email thom.badings@ru.nl \\
        \addr Radboud University, Nijmegen, the Netherlands
        \AND
        \name Licio Romao \email licio.romao@cs.ox.ac.uk \\
        \name Alessandro Abate \email alessandro.abate@cs.ox.ac.uk \\
        \name David Parker \email david.parker@cs.ox.ac.uk \\
        \addr University of Oxford, Oxford, United Kingdom
        \AND
        \name Hasan A. Poonawala \email hasan.poonawala@uky.edu \\
        \addr University of Kentucky, Kentucky, USA
        \AND
        \name Marielle Stoelinga \email m.stoelinga@cs.ru.nl \\
        \addr Radboud University, Nijmegen, the Netherlands; University of Twente, Enschede, the Netherlands
        \AND
        \name Nils Jansen \email n.jansen@science.ru.nl \\
        \addr Radboud University, Nijmegen, the Netherlands}

\maketitle

\begin{abstract}
Controllers for dynamical systems that operate in safety-critical settings must account for stochastic disturbances.
Such disturbances are often modeled as process noise in a dynamical system, and common assumptions are that the underlying distributions are known and/or Gaussian.
In practice, however, these assumptions may be unrealistic and can lead to poor approximations of the true noise distribution.
We present a novel controller synthesis method that does not rely on any explicit representation of the noise distributions.
In particular, we address the problem of computing a controller that provides probabilistic guarantees on safely reaching a target, while also avoiding unsafe regions of the state space.
First, we abstract the continuous control system into a finite-state model that captures noise by probabilistic transitions between discrete states.
As a key contribution, we adapt tools from the scenario approach to compute probably approximately correct (PAC) bounds on these transition probabilities, based on a finite number of samples of the noise.
We capture these bounds in the transition probability intervals of a so-called interval Markov decision process (iMDP).
This iMDP is, with a user-specified confidence probability, robust against uncertainty in the transition probabilities, and the tightness of the probability intervals can be controlled through the number of samples.
We use state-of-the-art verification techniques to provide guarantees on the iMDP and compute a controller for which these guarantees carry over to the original control system.
In addition, we develop a tailored computational scheme that reduces the complexity of the synthesis of these guarantees on the iMDP. 
Benchmarks on realistic control systems show the practical applicability of our method, even when the iMDP has hundreds of millions of transitions.
\end{abstract}

\section{Introduction}

\label{sec:Introduction}

\paragraph{Dynamical systems.}
Dynamical systems are widely used to model the state dynamics of complex systems~\cite{Kulakowski2007dynamic}.
For example, an \gls{UAV} can be modeled as a dynamical system, in which the (possibly continuous) \emph{state} reflects its current position and velocity, and the (possibly continuous) \emph{control inputs} reflect choices that may change the state over time~\cite{Kulakowski2007dynamic}.
The dynamical system is \emph{linear} if the state transition is linear in the current state and control input.

\paragraph{Controller synthesis.}
Verifying that a controlled dynamical system satisfies a desired property is of paramount importance, especially in safety-critical applications.
Traditional methods from control theory are powerful tools for addressing properties about stability and (asymptotic) convergence, but these methods by and large do not provide formal guarantees about richer, temporal properties~\cite{DBLP:books/daglib/BaierKatoen2008,DBLP:journals/tac/FanQMNMV22}, namely requirements on state trajectories of the system in time.
A common example is the \emph{reach-avoid property}, where the task is to reach a desirable region within a given time horizon, while always remaining in a (possibly non-convex) safe region.
The general \emph{controller synthesis problem} is to compute (in an automated fashion) a \emph{feedback controller} for the dynamical system, such that any state trajectory that the system generates satisfies the given reach-avoid property~\cite{DBLP:conf/atva/KwiatkowskaP13}. 

\paragraph{Modeling process noise.}
However, for a system such as a \gls{UAV}, factors including turbulence and wind gusts cause \emph{uncertainty} affecting the state dynamics of the system~\cite{DBLP:journals/trob/BlackmoreOBW10}.
We model such uncertainty as \emph{process noise}, which is an additive random variable (with possibly infinitely many outcomes) entering the dynamical system and thus affecting state transitions.
Due to this additional noise term, it is generally impossible to guarantee that \emph{every} state trajectory satisfies a reach-avoid property.
Instead, we reason over the \emph{probability} that a reach-avoid property is satisfied, and the synthesis problem is then to compute a controller that maximizes this probability.

\paragraph{Distribution of the noise.}
A common assumption to achieve computational tractability for the controller synthesis problem is that the process noise follows a Gaussian distribution~\cite{DBLP:journals/spm/ParkSQ13}, e.g., as is classically assumed in linear-quadratic-Gaussian control~\cite{anderson2007optimal}.
However, in realistic problems, such as a \gls{UAV} operating under turbulence, this assumption yields a poor approximation of the uncertainty~\cite{DBLP:journals/trob/BlackmoreOBW10}.
Distributions may even be \textit{unknown}, meaning that one cannot derive a set-bounded or a precise probabilistic representation of the noise. 
In this case, it is generally hard or even impossible to derive \emph{hard guarantees} on the probability that a given controller ensures the satisfaction of the considered reach-avoid property.

\paragraph{Problem statement.}
In this paper, we consider the controller synthesis problem for dynamical systems with additive process noise of an unknown distribution.
For the synthesized controller, we provide a \emph{\gls{PAC}} guarantee on the probability of satisfying a given reach-avoid problem.
As such, we solve the following problem:

\begin{mdframed}[backgroundcolor=gray!20, nobreak=true, innerleftmargin=8pt, innerrightmargin=8pt]
Given a linear dynamical system perturbed by additive noise of unknown distribution, compute a controller under which, with a user-specified confidence level, the probability to satisfy a reach-avoid problem is above a given threshold value.
\end{mdframed}

\paragraph{Finite-state abstraction.}
We solve this problem by computing a finite-state abstraction of the original dynamical system \cite{DBLP:journals/siamads/SoudjaniA13}, which we obtain from a \emph{partition} of its continuous state space into a set of disjoint convex \emph{regions}.
Actions in this abstraction correspond to continuous control inputs that yield transitions between these regions.
Due to the process noise, the outcome of an action is stochastic, rendering transitions probabilistic. 
We capture these probabilities in a \gls{MDP}~\cite{DBLP:books/wi/Puterman94}.
A defining characteristic of our approach is that we leverage \emph{backward reachability computations} on the dynamical system to determine which actions are enabled at each discrete region.
By contrast, most other abstraction methods \cite<see the related work in \cref{sec:Related} and the survey article>{LSAZ21} rely on \emph{forward} reachability computations, which are associated with errors that grow with the time horizon of the considered property.
Our backward scheme avoids such abstraction errors, at the cost of requiring slightly more restrictive assumptions on the system dynamics.

\paragraph{Probability intervals.}
Since the distribution of the noise is unknown, it is not possible to compute the transition probabilities of the abstract \gls{MDP} exactly, e.g., as in~\citeA{DBLP:journals/siamads/SoudjaniA13}.
Instead, we estimate the probabilities based on a finite number of \emph{samples} of the noise, which may be obtained from a high fidelity (black box) simulator, from historical data, or from (physical or numerical) experiments.
To be \emph{robust} against estimation errors in these probabilities, we
formulate the error estimation process as a so-called scenario optimization problem and leverage state-of-the-art tools~\cite{romao2022tac} from the \emph{scenario approach}, which is a methodology to deal with stochastic optimization in a data-driven fashion~\cite{DBLP:journals/siamjo/CampiG08,DBLP:journals/arc/CampiCG21}.
We compute \emph{upper and lower bounds} on the transition probabilities with a desired \emph{confidence level}, which we choose up front.
These bounds are \emph{PAC}, as they contain the true probabilities with at least this confidence level.

\paragraph{Interval MDPs.}
We formalize our abstractions with the \gls{PAC} probability intervals using so-called \glspl{iMDP}, which are an extension of \glspl{MDP} with intervals of probabilities~\cite{DBLP:journals/ai/GivanLD00}. 
More generally, \glspl{iMDP} are a particular case of uncertain or non-deterministic \glspl{MDP}; see the references in \cref{sec:Related} for more details.
\glspl{iMDP} have recently been proposed as an alternative to standard \glspl{MDP} for abstracting stochastic dynamical systems~\cite{DBLP:conf/hybrid/CauchiLLAKC19,LSAZ21}. 
We show explicitly how to lift the \gls{PAC} guarantees on individual transition probabilities to a correctness guarantee on the whole \gls{iMDP}.
Policies for \glspl{iMDP} have to robustly account for all possible probabilities within the intervals, and one usually provides upper and lower bounds on maximal or minimal reachability probabilities or expected rewards~\cite{DBLP:conf/qest/HahnHHLT17,DBLP:conf/cav/PuggelliLSS13,DBLP:conf/cdc/WolffTM12}.
Note that given the \gls{PAC}-correct \gls{iMDP}, these bounds on the reachability probabilities are exact and thus do not introduce another error bound.
For \glspl{MDP}, mature tool support exists, e.g., via \prism~\cite{DBLP:conf/cav/KwiatkowskaNP11}, and this support was previously extended to \glspl{iMDP} in \citeA{Badings2022AAAI}.

\paragraph{Infinite-horizon properties.}
One notable feature of our abstraction scheme is that we can handle reach-avoid properties over \emph{infinite-time horizons}, also known as unbounded properties~\cite{TA14}. 
Most existing abstraction-based controller synthesis methods cause abstraction errors that grow with the time horizon and are thus limited to finite-horizon properties~\cite{AKLP10,DBLP:journals/siamads/SoudjaniA13,LSAZ21,Zikelic2022}.
By contrast, we provide a \gls{PAC} guarantee on each transition probability interval of the \gls{iMDP} being correct, without errors that grow with the horizon of the considered properties.

\paragraph{Iterative abstraction improvement.}
The tightness of the probability intervals in the iMDP depends on the number of noise samples. 
Hence, we propose an \emph{iterative abstraction scheme} to improve these intervals by sequentially increasing the number of noise samples used to compute probability intervals.
Our scheme can be summarized as follows.
We first compute a robust policy that maximizes the probability of safely reaching the goal states.
Then, we check whether this optimal reach-avoid probability is satisfactory or not with respect to the pre-defined threshold value on the given property.
In case it is not, we collect additional samples to reduce the uncertainty in the probability intervals; otherwise, we extract and use the optimal policy to compute ``on the fly'' a feedback controller for the original dynamical system.
The specified confidence level reflects the likelihood that the optimal reach-avoid probability on the \gls{iMDP} is a \emph{lower bound} for the probability that the dynamical system satisfies the reach-avoid problem under this derived controller.

\paragraph{Improved policy synthesis scheme.}
The generated \gls{iMDP} abstractions can potentially have hundreds of millions of transitions, especially if the state dimension is high.
To reduce the computational complexity related to the policy synthesis algorithm on the iMDP, we propose an algorithmic optimization.
Instead of employing the large iMDP abstraction, we soundly merge states that show similar reach-avoid probabilities.
In particular, we associate a merged state with the minimum reach-avoid probability of the set of states it is comprised of, resulting in a conservative but sound approximation of the original \gls{iMDP}.
This strategy reduces the overall size of the \gls{iMDP} significantly and additionally enables us to solve more complex reach-avoid problems, as shown in our experiments.

\paragraph{Contributions.}
Our contributions are threefold:
(1)~We propose a novel method to compute controllers with \gls{PAC} guarantees for dynamical systems with unknown noise distributions.
Specifically, the probability of satisfying an (in)finite-horizon reach-avoid problem is guaranteed, with a user-specified confidence probability, to exceed a pre-defined threshold.
(2)~We propose a sound and scalable policy synthesis algorithm that allows us to verify abstract models with hundreds of millions of transitions.
(3)~We apply our method to multiple realistic control problems and benchmark against two other tools: \stochy and \sreachtools.
We demonstrate that the guarantees obtained for the \gls{iMDP} abstraction carry over to the dynamical system of interest.
Moreover, we show that using probability intervals instead of point estimates of probabilities yields significantly more robust results.

This paper extends~\citeA{Badings2022AAAI} in several ways.
First, we expand on the intuition and presentation of our abstraction method, comparing the scenario optimization with other methods for computing probability intervals.
Second, we provide deeper theoretical results on the correctness of our method by formalizing the relationship between the dynamical system and the \gls{iMDP}.
Moreover, we lift the \gls{PAC} guarantees on individual transitions, as done in~\citeA{Badings2022AAAI}, to a correctness guarantee on \emph{the whole} \gls{iMDP} abstraction.
Finally, we present a new algorithmic extension and an additional experiment on an autonomous satellite rendezvous problem from~\citeA{DBLP:conf/cdc/JewisonE16}.
\section{Foundations and Outline}

\label{sec:Preliminaries}

A \emph{discrete probability distribution} over a finite set $X$ with cardinality $|X|$ is a function $\mathit{prob} \colon X \to [0,1]$ with $\sum_{x \in X} \mathit{prob}(x) = 1$.
The set of all distributions over $X$ is $\distr{X}$.
A \emph{probability density function} over a random variable $x$ conditioned on $y$ is written as $p(x \vert y)$.
All vectors $\state \in \R^n$, $n \in \N$, are column vectors and are denoted by bold letters.
We use the term \emph{controller} when referring to a map selecting inputs for dynamical systems, while we use the term \emph{policy} for (i)\glspl{MDP}.

\subsection{Linear Dynamical Systems}
\label{subsec:LinDynSystems}

We consider discrete-time, continuous-state systems, where the progression of the state $\state \in \R^n$ depends \emph{linearly} on the current state, on a control input, and on a process noise term.
Given a state $\state_k$ at discrete time $k \in \N$, the successor state at time $k+1$ is given as
\begin{equation}
    \state_{k+1} = A \state_{k} + B \control_{k} + \drift_k + \noise_k,
    \label{eq:continuousSystem}
\end{equation}
where $\control_k \in \cControlSpace \subset \R^p$ is the (continuous) control input at time $k$, $A \in \R^{n \times n}$ and $B \in \R^{n \times p}$ are appropriate matrices, and $\drift_k \in \R^n$ is a deterministic disturbance.
The term $\noise_k \in \Delta \subset \R^n$ is an arbitrary additive process noise term, which is a random variable defined on a probability space $(\Delta, \mathcal{D}, \amsmathbb{P})$, with $\sigma$-algebra $\mathcal{D}$ and probability measure $\amsmathbb{P}$ defined over $\mathcal{D}$. 
In our formulation, the sample space $\Delta$ is an abstract set of possible noise values at any time $k$, and the probability measure $\amsmathbb{P}$ is unknown but time-invariant. 
To deal with this lack of knowledge, we employ a sampling-based approach, for which it suffices to obtain a finite number of samples of the process noise, whose distribution is independent of time. 
This underlying source of uncertainty induces a probabilistic model over the successor state $\state_{k+1}$. We denote the probability density function over successor states for a given state $\state_k$ and control input $\control_k$ as $p_{\noise_k}(\state_{k+1} \ | \ \state_k, \control_k)$.

The linear dynamical system defined by \cref{eq:continuousSystem} is controlled by a \emph{time-varying linear feedback controller} of the following form:
\begin{definition}
    \label{def:controller}
    A piece-wise linear feedback controller is a function $\controller \colon \R^n \times \N \to \cControlSpace$, which maps a state $\state_k \in \R^n$ and a time step $k \in \N$ to a control input $\control_k \in \cControlSpace$.
\end{definition}
\noindent
We use time-varying controllers because, for finite-horizon control objectives, the optimal control generally depends on the time index.
For infinite-horizon properties instead, the optimal control is independent of the time index \cite{DBLP:books/daglib/BaierKatoen2008}.

\begin{example}
    The longitudinal dynamics of a \gls{UAV} are modeled as
    \begin{equation}
        \state_{k+1} = \begin{bmatrix}
            p_{k+1} \\ v_{k+1}
        \end{bmatrix} = \begin{bmatrix}
            1 & 1 \\ 0 & 1
        \end{bmatrix} \state_k 
        +
        \begin{bmatrix}
            0.5 \\ 1
        \end{bmatrix} \control_k
        + \noise_k,\label{example:UAV}
    \end{equation}
    where $p_k$ and $v_k$ are the position and velocity at time $k$, and the control is bounded by $\bm{u}_k \in \cControlSpace = [-4, 4]$.
    The distribution of the noise $\noise_k$ is unknown, thus possibly not Gaussian. 
\end{example}

\begin{remark}[Restriction to linear systems]
    \label{remark:LinearSystems}
    Our methods are theoretically amenable to work with nonlinear drift dynamics rather than the linear drift terms we see in \cref{eq:continuousSystem,example:UAV}. However, this requires more advanced $1$-step reachability computations, which are not core to our main contributions.
    Hence, we restrict ourselves to the linear drift of the model in \cref{eq:continuousSystem} and discuss extensions to nonlinear models in \cref{sec:Conclusion}.
\end{remark}

\subsection{Problem Statement}
\label{sec:prelim:problem}
We consider control objectives expressed as \emph{reach-avoid properties} over an infinite or a finite horizon.
A reach-avoid property $\property$ is satisfied if, starting from an initial state $\state_0 \in \R^n$ at time $k=0$, the system reaches a desired \emph{goal region} $\goalRegion \subset \R^n$ within a horizon of $\horizon \in \N \cup \infty$ steps, while avoiding a \emph{critical region} $\criticalRegion \subset \R^n$.
We write the probability of satisfying a reach-avoid property $\property$ under a controller $\controller$ as $\reachProbCont$.
We formally state the problem that we solve in this paper as follows.
\begin{mdframed}[backgroundcolor=gray!20, nobreak=true, innerleftmargin=8pt, innerrightmargin=8pt]
Compute a controller $\controller$ for the dynamical system in \cref{eq:continuousSystem} that, with a confidence probability of at least $\alpha \in [0,1]$, guarantees that $\reachProbCont \geq \eta$, where $\eta \in [0,1]$ is a pre-defined probability threshold.
\end{mdframed}

\subsection{Markov Decision Processes}
We solve the problem above via a finite-state abstraction of the dynamical system, which we formalize as a variant of \glspl{MDP}~\cite{DBLP:books/wi/Puterman94}:
\begin{definition}[MDP]
    \label{def:MDP}
    A \emph{\glsfirst{MDP}} is a tuple $\mdp=\MDP$ where $\States$ is a (finite) set of states, $\Actions$ is a (finite) set of actions, $\initState$ is the initial state, and $\transfunc \colon \States \times \Actions \rightharpoonup \distr{\States}$ is the (partial)\footnote{The transition function is partial in general, meaning that not all state-action pairs $(s,a) \in \States \times \Actions$ are in the domain of $\transfunc$. We need to define this partial transition function because not all actions may be enabled in each state.} probabilistic transition function.   
\end{definition}
\noindent
The set of actions enabled in state $s \in \States$ is $\Actions(s) \subseteq \Actions$.
We call $(s,a,s')$ with probability $\transfunc(s,a)(s')>0$ a \emph{transition}.
A time-varying deterministic policy for an \gls{MDP} $\mdp$ is a function $\policy \colon \States \times \N \to \Actions$ mapping states and time indices to actions~\cite{DBLP:books/wi/Puterman94}.
The set of all possible policies for $\mdp$ is denoted by $\policySpace_\mdp$.

A \emph{probabilistic reach-avoid property} $\reachProbDiscr$ for an \gls{MDP} describes the probability of reaching a set of goal states $\goalStates \subset \States$ within $\horizon \in \N \cup \infty$ steps under policy $\policy \in \policySpace_\mdp$, while avoiding a set of critical states $\criticalStates \subset \States$, where $\goalStates \cap \criticalStates = \emptyset$.
An optimal policy $\policy^\star \in \policySpace_\mdp$ for \gls{MDP} $\mdp$ maximizes the \emph{reach-avoid probability}:
\begin{equation}
    \policy^\star =  \argmax_{\policy \in \policySpace_\mdp} \reachProbDiscr.
	\label{eq:optimalPolicy}
\end{equation}
\begin{remark}[Dynamical systems as continuous MDPs]
    \label{remark:infinite_MDP}
    The dynamical system in \cref{eq:continuousSystem} can equivalently be seen as an \gls{MDP} with an uncountably infinite number of states $\States$ (representing all $\state_k \in \R^n$), and an infinite number of actions $\Actions$ (representing all $\control_k \in \cControlSpace$). Note that every action is enabled at any state, i.e. $\Actions(s) = \Actions, \, \forall s \in \States$.
    Moreover, each state-control pair $(\state_k, \control_k)$, i.e., applying control input $\control_k$ in state $\state_k$ induces a probability distribution (or in fact, a density function) over successor states $\state_{k+1} \in \R^n$.%
\end{remark}%
\noindent
We now relax the assumption that the transition probabilities of \glspl{MDP} are precisely given.

\begin{definition}[iMDP]
    \label{def:iMDP}
    An \emph{\glsfirst{iMDP}} is a tuple $\imdp=\iMDP$ where $\States$, $\Actions$, and $\initState$ are defined by \cref{def:MDP}, and where the uncertain (partial) probabilistic transition function $\transfuncImdp \colon \States \times \Actions \times \States \rightharpoonup \Interval \cup \{ [0,0] \}$ is defined over intervals $\Interval = \{ [a,b] \ | \ a,b \in (0,1] \text{ and } a \leq b \}$. 
\end{definition}
\noindent
Note that an interval cannot have a lower bound of $0$ except for the $[0,0]$ interval.
Since the transition function $\transfuncImdp$ is partial, we do not require each action to be enabled at any state.
An \gls{iMDP} defines a (possibly empty) set of \glspl{MDP} that vary only in their transition function.
In particular, for an \gls{MDP} with transition function $\transfunc$, we write $\transfunc \in \transfuncImdp$ if for all $s,s' \in \States$ and $a \in \Actions$ we have $\transfunc(s,a)(s') \in \transfuncImdp (s,a,s')$ and $P(s, a)\in\distr{\States}$.

\begin{remark}[Geometry of uncertainty sets]
    \label{remark:shape_uncertainty_sets}
    For each state-action pair $(s,a)$, the set $\{ \transfunc(s,a) \mid \transfunc \in \transfuncImdp \}$ of feasible probability distributions is a probability simplex with interval (box) constraints defined by the intervals in $\transfuncImdp$.
    The resulting uncertainty set $\{ \transfunc(s,a) \mid \transfunc \in \transfuncImdp \}$ is a convex polytope by construction.
    Moreover, the transition probabilities $P(s,a)$ for different state-action pairs $(s,a)$ are independent and thus respect the so-called \emph{rectangularity assumption}, which is common in the literature on robust MDPs~\cite{DBLP:journals/ior/WiesemannKS14} and robust optimization~\cite{DBLP:journals/siamrev/BertsimasBC11}, and is necessary to guarantee computational tractability.
\end{remark}

For \glspl{iMDP}, a policy needs to be \emph{robust} against all possible transition functions $\transfunc \in \transfuncImdp$.
We employ a robust variant of value iteration from~\citeA{DBLP:conf/cdc/WolffTM12} to compute a policy $\munderbar{\policy}^\star \in \policySpace_{\imdp}$ for \gls{iMDP} $\imdp$ that maximizes the lower bound  $\reachProbDiscrMin$ of on the reach-avoid probability within horizon $\horizon$:
\begin{equation}
	\munderbar{\policy}^\star 
	= \argmax_{\policy \in \policySpace_{\imdp}} \reachProbDiscrMin 
	= \argmax_{\policy \in \policySpace_{\imdp}} \, \min_{\transfunc \in \transfuncImdp} \reachProbDiscr.
	\label{eq:optimalPolicy_lb}
\end{equation}
Similarly, we can also maximize an upper bound $\reachProbDiscrMax$ on the reach-avoid probability:
\begin{equation}
	\bar{\policy}^\star = \argmax_{\policy \in \policySpace_{\imdp}} \reachProbDiscrMax 
	= \argmax_{\policy \in \policySpace_{\imdp}} \, \max_{\transfunc \in \transfuncImdp} \reachProbDiscr.
	\label{eq:optimalPolicy_ub}
\end{equation}
We will use the lower bound in \cref{eq:optimalPolicy_lb} to compute robust optimal policies, while we use \cref{eq:optimalPolicy_ub} to determine if the threshold $\eta$ specified in the formal problem statement is satisfiable at all.
Note that deterministic policies suffice to obtain optimal values for (i)\glspl{MDP}~\cite{DBLP:books/wi/Puterman94,DBLP:conf/cav/PuggelliLSS13}, and particularly also for reach-avoid properties \cite{Abate2008probabilisticSystems}.

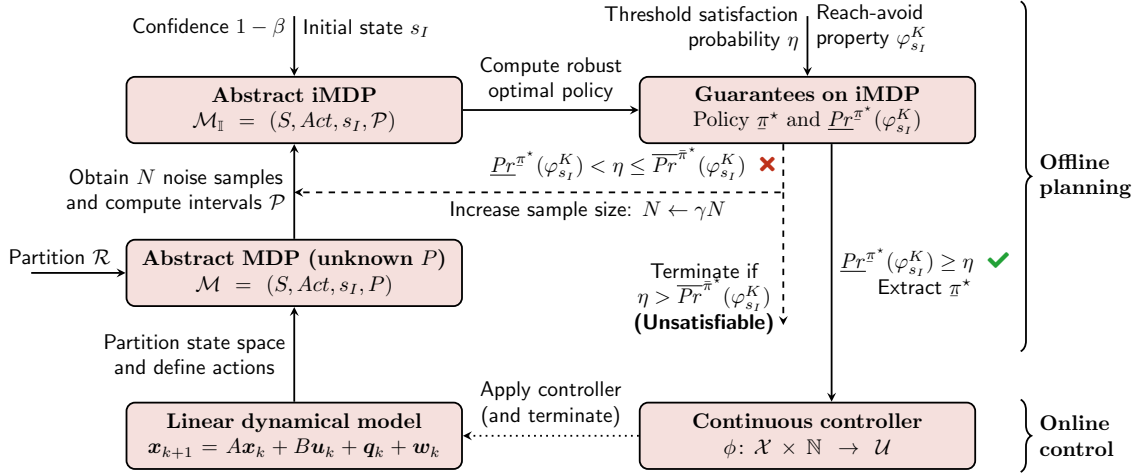
\begin{figure}[t!]
	\centering
	\usetikzlibrary{calc}
\usetikzlibrary{arrows.meta}
\usetikzlibrary{positioning}
\usetikzlibrary{fit}
\usetikzlibrary{shapes}

\tikzstyle{flowNodeRed} = [rectangle, rounded corners, minimum width=5.4cm, text width=5.3cm, minimum height=1.1cm, text centered, draw=black, fill=red!15]

\resizebox{\linewidth}{!}{%
	\begin{tikzpicture}[node distance=4.5cm,->,>=stealth,line width=0.3mm,font=\small]
		
		\newcommand\xshift{3.8cm}
		\newcommand\xshiftB{4.0cm}
		\newcommand\ydist{-1.8cm}
		\newcommand\ydistInput{-2.8cm}
		
		\node (dynSys) [flowNodeRed] {\textbf{Linear dynamical model} \\ $\bm{x}_{k+1} = A\bm{x}_k + B\bm{u}_k + \bm{q}_k + \bm{w}_k$};
		
		\node (partition) [flowNodeRed, above of=dynSys, yshift=\ydist] {\textbf{Abstract \gls{MDP} (unknown $P$)} \\ $\mdp = \MDP$}; 
		
		\node (iMDP) [flowNodeRed, above of=partition, yshift=\ydist] {\textbf{Abstract \gls{iMDP}} \\ $\imdp = \iMDP$};
		
		\node (verify) [flowNodeRed, right of=iMDP, xshift=\xshiftB] {\textbf{Guarantees on \gls{iMDP}} \\ Policy $\munderbar{\policy}^\star$ and $\reachProbDiscrMinStar$};
		
		\node (policy) [flowNodeRed, right of=dynSys, xshift=\xshiftB] {\textbf{Continuous controller} \\ $\controller \colon \cStateSpace \times \N \to \cControlSpace$};
		
		\node (in) [above of=iMDP, yshift=\ydistInput] {};
		\node (in_mid) [left of=partition] {};
		\node (in2) [above of=verify, yshift=\ydistInput] {};
		
		\draw [->] (in) -- (iMDP) 
		node [pos=0.2, right, align=left, font=\sffamily\small] {Initial state $\initState$} 
		node [pos=0.2, left, align=right, font=\sffamily\small] {Confidence $1-\beta$};
		
		\draw [->] (in_mid) -- (partition) node [pos=0.3, above, align=left, font=\sffamily\small] {Partition $\Partition$};
		
		\draw [->] (in2) -- (verify) node [pos=0.2, right, align=left, font=\sffamily\small] {Reach-avoid\\property $\propertyMDP$} node [pos=0.2, left, align=right, font=\sffamily\small] {Threshold satisfaction\\probability $\eta$};
		
		\draw [->] (dynSys) -- (partition) node [midway, left, align=center, font=\sffamily\small] {Partition state space \\ and define actions};
		
		\draw [->] (partition) -- (iMDP) node [midway, left, align=center, font=\sffamily\small] {Obtain $N$ noise samples \\ and compute intervals $\transfuncImdp$};
		
		\draw [->] (iMDP) -- (verify) node [midway, above, align=center, font=\sffamily\small] 
		{Compute robust\\optimal policy};
		
		\draw [->, dashed] ($(verify.south) + (-0.4, 0)$) |- ($(iMDP.south) + (0, -0.8)$) node [pos=0.2, left, align=center, font=\sffamily\small] {$\reachProbDiscrMinStar < \eta \leq \reachProbDiscrMaxStar$ \ \rured{\faIcon{times}}} node [pos=0.70, below, align=center, font=\sffamily\small] {Increase sample size: $N \gets \gamma N$};
		
		\draw [->, dashed] ($(verify.south) + (-0.4, -0.8)$) -- ($(verify.south) + (-0.4, -3.0)$) node [pos=0.80, left, align=center, font=\sffamily\small] {Terminate if \\ $\eta > \reachProbDiscrMaxStar$ \\ \textbf{(Unsatisfiable)}};
		
		\draw [->] ($(verify.south) + (0.4, 0)$) -- ($(policy.north) + (0.4, 0)$) node [pos=0.5, right, align=center, font=\sffamily\small] {$\reachProbDiscrMinStar \geq \eta$ \ \darkgreen{\faIcon{check}} \\ Extract $\munderbar{\policy}^\star$};
		
		\draw [dotted, ->] (policy) -- (dynSys) node [midway, above, align=center, font=\sffamily\small] {Apply controller\\(and terminate)};
		
		\draw [-,decorate,decoration={brace,amplitude=5pt,mirror,raise=2ex}] (11.7cm, -0.6cm) -- (11.7cm, 0.6cm) node[midway, right, xshift=0.5cm, align=left] {\textbf{Online} \\ \textbf{control}};
    	\draw [-,decorate,decoration={brace,amplitude=5pt,mirror,raise=2ex}] (11.7cm, 1.4cm) -- (11.7cm, 7.2cm) node[midway, right, xshift=0.5cm, align=left] {\textbf{Offline} \\ \textbf{planning}};
		
	\end{tikzpicture}
}
	
	\caption{
		Our iterative approach between abstraction and verification, where $N$ is the number of samples used for the abstraction, and $\eta$ is the threshold reach-avoid probability.
	}
	\label{fig:Approach}
\end{figure}

\subsection{An Iterative Abstraction Scheme}

\label{sec:IterativeScheme}

Our approach is summarized in \cref{fig:Approach} and consists of an \emph{offline planning phase} in which we generate the abstraction and obtain guarantees on the abstract model, and an \emph{online control phase} in which we derive a controller for the dynamical system on the fly.
We describe in \cref{sec:Abstraction} how, for a given linear dynamical system, we generate a finite-state abstraction as an \gls{MDP} by partitioning the continuous state space.
We then describe in \cref{sec:ScenarioApproach} how we obtain \gls{PAC} bounds on the \gls{MDP}'s transition probabilities based on a finite set of $N$ samples of the noise.
The resulting abstraction is an \gls{iMDP}, for which we use \cref{eq:optimalPolicy_lb,eq:optimalPolicy_ub} to compute optimal policies $\munderbar{\policy}^\star$ and $\bar{\policy}^\star$ that maximize the lower/upper bounds on the probability of satisfying the given property.
We then proceed in one of the following three ways:
\begin{enumerate}
    \item If $\reachProbDiscrMinStar \geq \eta$, the formal problem is satisfied.
    Thus, we extract the optimal policy $\munderbar{\policy}^\star$, which we use to determine a controller $\controller$ for the dynamical system on the fly.
    \item If $\reachProbDiscrMaxStar < \eta$, the reach-avoid problem is (with high confidence) unsatisfiable.
    In this case, increasing the number of samples does not help, so we terminate the scheme.
    \item If $\reachProbDiscrMinStar < \eta$ but $\reachProbDiscrMaxStar \geq \eta$, we obtain additional samples by increasing $N$ by a fixed factor $\gamma > 1$.
    The updated \gls{iMDP} has tighter probability intervals but may also have more transitions.
    However, since the states and actions of the abstraction are independent of $N$ and defined at the first iteration, the \gls{MDP} abstraction is only performed once. 
\end{enumerate}
In \cref{sec:Algorithm} we describe each step of our approach in more detail, as well as a complete algorithm for solving the formal problem. 
We perform numerical experiments on benchmarks from multiple application domains in \cref{sec:Studies}.
We discuss the related work in \cref{sec:Related}.
\section{Finite-State MDP Abstraction}

\label{sec:Abstraction}
In this section, we describe how we generate an \gls{MDP} abstraction of the linear dynamical system in \cref{eq:continuousSystem}.
First, we partition the state space into a set of discrete convex regions.
We then use this partition to build a finite-state \gls{MDP} abstraction of the dynamical system.

\subsection{State Space Discretization}
We choose a \emph{partition} $\Partition = \{ \region_1, \ldots, \region_v \}$ of a bounded portion $\cStateSpace \subset \R^n$ of the continuous state space $\R^n$ into a set of $v$ disjoint \emph{regions}, such that $\bigcup_{i=1}^v \region_i = \cStateSpace$.
In addition, we define a single \emph{absorbing region} $\region_\ast$, representing $\R^n \backslash \cStateSpace$.
The absorbing region $\region_\ast$ captures the event that the continuous state leaves the bounded portion of the state space over which we plan.
We consider the regions in $\Partition$ to be $n$-dimensional bounded, convex polytopes.
Thus, each region $\region_i \in \Partition$ is the solution set of $m$ linear inequalities parameterized by $M_i \in \R^{m \times n}$ and $\bm{b}_i \in \R^m$, yielding $\region_i = \big\{ \state \in \R^n \, \vert \, M_i \state \leq \bm{b}_i \big\}.$
In addition, the following assumption allows us to translate properties for the dynamical system to properties on the \gls{iMDP} abstraction:

\begin{assumption}
    \label{assumption:RegionAlignment}
    The continuous goal region $\goalRegion$ and critical region $\criticalRegion$ are aligned with the union of a subset of regions in $\Partition$, i.e., $\goalRegion = \cup_{i \in I} R_i$ and $\criticalRegion = \cup_{j \in J} R_j$ for index sets $I,J \subset \{1,2,\ldots,\card{\Partition}\}$.
\end{assumption}

\subsection{MDP Abstractions}

We formalize the dynamical system discretized under $\Partition$ as an \gls{MDP} $\mdp = \MDP$, by defining its states, actions, and transition probabilities.
We assume that the initial state $\initState \in \States$ is known, and we capture time constraints by the reach-avoid property.

\paragraph{States.}
We define an \gls{MDP} state for every region of partition $\Partition$, as well as for the absorbing region $\region_\ast$.
Thus, we obtain the set of \gls{MDP} states $\States = \{ s_i \ \vert \ i = 1,\ldots,\vert\Partition\vert \} \cup \{ s_\ast \}$, which consists of $\vert \States \vert = \vert \Partition \vert + 1$ states: one for every region in $\Partition$, plus one state $s_\ast$ corresponding to the absorbing region $\region_\ast$ where the only outgoing transition leads back to $s_\ast$.
We denote by $\region_s \in \Partition$ the region of the partition associated with \gls{MDP} state $s \in \States$.
Formally, we define a function $T \colon \R^n \to \States$ that maps continuous states $\state \in \R^n$ to \gls{MDP} states $s \in \States$:\footnote{
Alternatively, we may define $T \subseteq \R^n \times \States$ as a binary relation between each continuous state $\state \in \R^n$ and \gls{MDP} state $s \in \States$.
In this paper, we use the function notation since we apply $T$ as a function only.
}
\begin{definition}
    \label{def:relation}
    A continuous state $\state \in \R^n$ and \gls{MDP} state $s \in \States$ are related, i.e., $T(\state) = s$, if $\state \in \region_s$.
    The inverse mapping $T^\inv(s) = \{ \state \in \R^n \ \vert \ (\state,s) \in T \}$ is equivalent to region $\region_s$.
\end{definition}

\noindent
The function $T$ expresses a relation from any state $\state \in \R^n$ to a (unique) discrete state $s \in \States$, and is, therefore, also called an \emph{abstraction function}.
Moreover, its inverse $T^{-1}(s) = R_s$ maps any state $s \in \States$ to a set of continuous states and is equal to the region $R_s$ of $s$ itself.

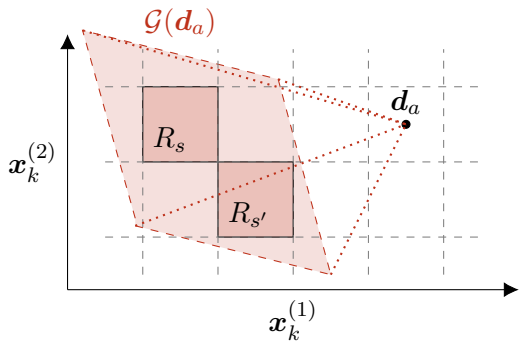
\begin{figure}[t!]
	\centering
	\begin{tikzpicture}[scale=1]

    \newcommand\orX{0}
    \newcommand\orY{0.3}
    \newcommand\boxW{1}

    \newcommand\uppX{5}
    \newcommand\uppY{3}
    
    \newcommand\originX{1}
    \newcommand\originY{1}
    \newcommand\targetX{4}
    \newcommand\targetY{2}
    
    \newcommand\dX{\targetX+0.5*\boxW}
    \newcommand\dY{\targetY+0.5*\boxW}
    
    	\draw[red, fill=red!15, dashed] (\originX-\boxW*0.1,\originY+\boxW*0.15) -- (\originX+\boxW*2.5,\originY-\boxW*0.5) -- (\originX+\boxW*1.8,\originY+\boxW*2.1) -- node[anchor=south, pos=0.5, yshift=0.1cm] {$\mathcal{G}(\bm{d}_a)$} (\originX-\boxW*0.8,\originY+\boxW*2.75) -- (\originX-\boxW*0.1,\originY+\boxW*0.25);
    	
    	\draw[fill=red!30] (\originX+\boxW,\originY) -- (\originX+\boxW*2,\originY) -- (\originX+\boxW*2,\originX+\boxW) -- (\originX+\boxW,\originX+\boxW) -- (\originX+\boxW,\originY) node[anchor=south west] {$R_{s'}$};
    	
    	\draw[fill=red!30] (\originX,\originY+\boxW) -- (\originX+\boxW,\originY+\boxW) -- (\originX+\boxW,\originX+\boxW*2) -- (\originX,\originY+\boxW*2) -- (\originX,\originY+\boxW) node[anchor=south west] {$R_{s}$};
    
        \foreach \x in {1,...,\uppX} 
        {
            \draw[gray, dashed] (\x,\boxW/2) -- (\x,\uppY+\boxW/2);
        }
        \foreach \y in {1,...,\uppY} 
        {
            \draw[gray, dashed] (\boxW/2,\y) -- (\uppX+\boxW/2,\y);
        }
        
        \draw[-{Latex[length=2mm,width=2mm]}] (\orX,\orY) -- (\uppX+\boxW-\orX,\orY);
        \draw[-{Latex[length=2mm,width=2mm]}] (\orX,\orY) -- (\orX,\uppY+\boxW-\orY);
        
        \draw[] (\uppX/2+\boxW/2,\orY) node[anchor=north] {$\state_{k}^{(1)}$};
        \draw[] (\orX,\uppY/2+\boxW/2) node[anchor=east] {$\state_{k}^{(2)}$};
        
        \filldraw[black] (\dX,\dY) circle (1.5pt) node[anchor=south] {$\bm{d}_a$};
        
        \draw[red, thick, dotted] (\originX-\boxW*0.1,\originY+\boxW*0.15) -- (\dX,\dY);
        \draw[red, thick, dotted] (\originX+\boxW*2.5,\originY-\boxW*0.5) -- (\dX,\dY);
        \draw[red, thick, dotted] (\originX+\boxW*1.8,\originY+\boxW*2.1) -- (\dX,\dY);
        \draw[red, thick, dotted] (\originX-\boxW*0.8,\originY+\boxW*2.75) -- (\dX,\dY);

\end{tikzpicture}
	\caption{
	A fragment of a rectangular partitioning of $\cStateSpace \subset \R^2$, showing the backward reachable set $\mathcal{G}(\bm{d}_a)$ of one particular action $a \in \Actions$ with target point $\bm{d}_a$. Action $a$ is enabled in states $s$ and $s'$, since $R_s \subseteq \mathcal{G}(d_a)$ and $R_{s'} \subseteq \mathcal{G}(d_a)$, respectively.}
	\label{fig:partition}
\end{figure}

\paragraph{Actions.}
Discrete actions correspond to the execution of a control input $\control_k \in \cControlSpace$ in the dynamical system in \cref{eq:continuousSystem}.
We define $q \in \N$ \gls{MDP} actions, so $\Actions = \{ a_1, \ldots, a_q \}$.
Let the noiseless successor state of $\state_k$ be defined as $\hat{\state}_{k+1} = A\state_k + B\control_k + \drift_k$ (i.e., $\state_{k+1}$ under the assumption that $\noise_k = 0$).
Every action $a$ is associated with a fixed continuous \emph{target point} $\bm{d}_a \in \R^n$, and is defined such that its noiseless successor state $\hat{\state}_{k+1} = \bm{d}_a$.
While not a restriction of our approach, we define one action for every \gls{MDP} state $s \in \States$ (except for the absorbing state $s_\ast$), and choose the target point to be the center of its region $\region_s$.\footnote{If, on the one hand, this choice for defining actions results in an \gls{MDP} that is too large, we may reduce the number of actions; if on the other the results are unsatisfactory, we may define additional actions.}

The \gls{MDP} must form a \emph{sound abstraction} of the dynamical system.
Thus, action $a \in \Actions$ only exists in an \gls{MDP} state $s \in \States$ if, for every continuous state $\state_k \in \region_s$, there exists a control $\control_k \in \cControlSpace$, such that $\hat{\state}_{k+1} = \bm{d}_a$.
To impose this constraint, we define the \emph{one-step backward reachable set} $\mathcal{G}(\bm{d}_a)$ of action $a \in \Actions$:
\begin{equation}
    \mathcal{G}(\bm{d}_a)
	= \{ \state \in \R^n \,\, | 
	\,\, \bm{d}_a = A \state + B \control_k + \drift_k, 
	\,\, \control_k \in \cControlSpace \}.
	\label{eq:BackwardReach}
\end{equation}
Then, action $a \in \Actions$ exists in state $s \in \States$ if and only if $\region_s \subseteq \mathcal{G}(\bm{d}_a)$.
As an example, \cref{fig:partition} shows the backward reachable set of a particular action $a \in \Actions$, which contains regions $R_s$ and $R_{s'}$, and hence this action is enabled in states $s$ and $s'$.
We write the set of actions that exist in state $s \in \States$ as $\Actions(s) = \{ a \in \Actions \ \vert \ \region_s \subseteq \mathcal{G}(\bm{d}_a) \}$.

Note that the existence of an action in an \gls{MDP} state merely implies that \emph{for every continuous state} in the associated region, \emph{there exists} a feasible control input that induces this transition.
Intuitively, this means that any possible transition in the \gls{MDP} must also be possible (with the same probability) in the dynamical system.
We make the following assumption on the controllability of the dynamical system~\cite{ogata2010modern}.
\begin{assumption}
    \label{assump:BackwardReachRank}
    The dynamical system in \cref{eq:continuousSystem} is controllable, meaning that the matrix $\mathcal{C} = \begin{bmatrix}B & A B & A^2 B & \cdots & A^{n-1} B\end{bmatrix}$ has full row rank.
\end{assumption}
\noindent
Intuitively, the dynamics of a controllable system can be `excited' (that is, we can make state transitions) to a subset of the $n$-dimensional state space that has a nonempty interior over a finite number of time steps.
As a result, under \cref{assump:BackwardReachRank} we can, by construction, derive a dynamical system for which the backward reachable set $\mathcal{G}(\bm{d}_a)$ has a non-empty interior.
For instance, in \cref{example:UAV}, the dimension of the control space ($p=1$) is lower than the dimension of the state space ($n=2$).
In this case, the backward reachable set $\mathcal{G}(\bm{d}_a)$ is a line segment in $\R^2$ and thus has an empty interior.
Hence, no region $\region_s \in \Partition$ of the partition can be contained in $\mathcal{G}(\bm{d}_a)$, so no action can ever exist in the \gls{MDP}.
However, since the system is assumed to be controllable (note that \cref{assump:BackwardReachRank} is indeed satisfied for \cref{example:UAV}), we can group together two discrete time steps, such that the dimension of the control space becomes equal to that of the state space, i.e., $p=n=2$.
Concretely, we redefine the dynamical system in \cref{example:UAV} as
\begin{equation}
\begin{split}
    \state_{k+2} 
    &= A^2 \state_k + \begin{bmatrix} AB & B \end{bmatrix} \begin{bmatrix} \control_k \\ \control_{k+1} \end{bmatrix} + A \noise_k + \noise_{k+1}
    \\
    &= \bar{A} \state_k + \bar{B} \control_{k,k+1} + \noise_{k,k+1},
\end{split}
\end{equation}
where $\state_k$ and $\control_{k,k+1}$ now have equal dimension, making $\mathcal{G}(\bm{d}_a)$ have a non-empty~interior.

\paragraph{Transition probabilities.}
To compute the control input $\control_k$ associated with action $a$ in a continuous state $\state_k$ at time $k$, we replace the successor state $\state_{k+1}$ by target point $\bm{d}_a$ in \cref{eq:continuousSystem} and solve for $\control_k$, yielding a control parameterized by state $\state_k$ and action $a$:
\begin{equation}
    \control_k(\state_k, a) = B^+ (\bm{d}_a - \drift_k - A\state_k),
    \label{eq:controlLaw}
\end{equation}
where $B^+$ is the pseudoinverse of $B$.\footnote{Even though we assume $B$ to be full row rank, it may have more columns than rows, so we use the pseudoinverse in \cref{eq:controlLaw}.}
It is easily verified that for every state $\state_k \in \region_s$ where action $a \in \Actions(s)$ exists, there exists a $\control_k$ such that \cref{eq:controlLaw} holds (depending on $B$, it may not be unique).
Due to the process noise $\noise_k$, the continuous successor state $\state_{k+1}$ upon choosing an action $a \in \Actions(s)$ in state $s \in \States$ is a random variable, which is written as
\begin{equation}
\begin{split}
    \state_{k+1} 
    &= A\state_k + B\control_k(\state_k, a) + \drift_k + \noise_k
    \\
    &= \bm{d}_a + \noise_k.
    \label{eq:continuous_successor}
\end{split}
\end{equation}
\begin{remark}[Independence of density functions]
    \label{remark:equivalence_probabilities}
    Recall that the probability density function of $\state_{k+1}$ under action $a$ is denoted by $p_{\noise_k}\left(\state_{k+1} \ | \ \state_k, \control_k(\state_k, a)\right)$.
    Importantly, we observe that by construction of \cref{eq:continuous_successor}, the continuous successor state $\bm{x}_{k+1}$ (and thus also the probability density function) upon choosing any action $a \in \Actions(s)$, with $s = T(\bm{x}_k$), is \emph{independent of the current state} $\state_k$ (as long as $a$ is enabled in $s$).
    We shall see how this simplifies the complexity of the abstract MDP. 
\end{remark}
\noindent
To define the transition probability function $\transfunc$ of the \gls{MDP}, we want to determine, for every pair of states $s, s' \in \States$ and action $a \in \Actions(s)$, the probability that the state-action pair $(s, a)$ leads to a continuous successor state $\state_{k+1} \in \region_{s'}$.
This transition probability is equal to the \emph{cumulative density function} $P_{\noise_k}\left(\state_{k+1} \in \region_{s'} \,\vert\, \state_k, \control_k(\state_k, a)\right)$ that the successor state $\state_{k+1}$ takes on a value in region $\region_{s'}$ upon choosing action $a$.
\begin{definition}
    \label{def:transition_probabilities}
    The transition probability $P(s,a)(s')$ for $s,s' \in \States$, $a \in \Actions(s)$ is defined as
    \begin{equation}
    \begin{split}
        P(s,a)(s') &= P_{\noise_k}\left(\state_{k+1} \in \region_{s'} \,\vert\, \state_k, \control_k(\state_k, a)\right)
        \\
        &= \int_{\region_{s'}} p_{\noise_k}\left(\state_{k+1} \ | \ \state_k, \control_k(\state_k, a)\right)d\state_{k+1}.
    	\label{eq:multivariateIntegral_mainProblem}
    \end{split}
    \end{equation}
    where $\control_k(\state_k, a)$ is defined as per \cref{eq:controlLaw}.
\end{definition}

\noindent
Due to its importance in the arguments of this paper, note that the \emph{cumulative} density function, denoted by capital $P_{\noise_k}(\state_{k+1} \in R_{s'}) \in [0,1]$ and being a probability, is the integral of the \emph{probability} density function over region $R_{s'}$, denoted by small $p_{\noise_k}(\state_{k+1})$ and being a function that assigns a probability to every possible value of $\state_{k+1}$.

\begin{remark}[Number of transition probabilities of the MDP]
    \label{remark:number_probabilities}
    For a given action $a \in \Actions$ and successor state $s' \in \States$, the transition probability $P(s, a)(s')$ is equal for any \gls{MDP} state $s$ for which $a \in \Actions(s)$.
    In other words, for any two states $s, \tilde{s} \in \States$ and an action $a$ for which $a \in \Actions(s)$ and $a \in \Actions(\tilde{s})$, we have that $P(s,a)(s') = P(\tilde{s},a)(s')$ for any $s' \in \States$.
    Thus, the \gls{MDP} has at most $\vert\Actions\vert \cdot \vert\States\vert$ unique transition probabilities (rather than at most $\vert\States\vert \cdot \vert\Actions\vert \cdot \vert\States\vert$ probabilities), which reduces the complexity of generating abstractions.
\end{remark}

\subsection{Soundness of the \gls{MDP} Abstractions}
The generated abstract \gls{MDP} $\mdp$ is an \emph{underapproximation} of the dynamical system, in the sense that every transition $(s,a,s')$ of $\mdp$ is contained in the dynamical system (note that the opposite is not true: the dynamical system contains transitions that are not in the \gls{MDP}).
We formalize this observation using the concept of probabilistic (bi)simulation, originally proposed for probabilistic transition systems by~\citeA{DBLP:journals/iandc/LarsenS91}.

\paragraph{Bisimilar states.}
We claim that any two continuous states $\state_k, \state_k' \in T^\inv(s) = R_s$ are \emph{probabilistically bisimilar} under the set of discrete actions $\Actions(s)$.
Intuitively, a sufficient condition for states $\state_k, \state_k'$ to be probabilistically bisimilar, is that the probability of transitioning to any $\bm{x}_{k+1} \in \region_{s'}$, $s' \in \States$ is the same for all actions $a \in \Actions(s)$~\cite{DBLP:journals/iandc/DesharnaisEP02}.
Mathematically, this is written as follows.
\begin{corollary}
    \label{lemma:bisimilar_states}
    Any two states $\state_k, \state_k' \in T^\inv(s) = R_s$ are probabilistically bisimilar, because for all actions $a \in \Actions(s)$ and for all successor states $s' \in \States$, it holds that
    \begin{equation}
        \label{eq:bisimilar_states}
        P_{\noise_k}\left(\state_{k+1} \in \region_{s'} \mid \state_k, \control_k(\state_k, a)\right)
        = 
        P_{\noise_k}\left(\state_{k+1} \in \region_{s'} \,\vert\, \state'_k, \control_k(\state'_k, a)\right).
    \end{equation}
\end{corollary}
\noindent
The proof of \cref{lemma:bisimilar_states} follows directly from \cref{remark:equivalence_probabilities}, which states that the density function $p_{\noise_k}\left(\state_{k+1} \ | \ \state_k, \control_k(\state_k, a)\right)d\state_{k+1}$, and thus also its integral over any discrete region of partition $\Partition$, is independent of the current state.
\cref{lemma:bisimilar_states} carries an important message: we can abstract any continuous state $\state_k \in \region_s$ into a single \gls{MDP} state $s \in \States$ \emph{without losing any information about the probabilistic behavior of the model}.

\paragraph{Probabilistic simulations.}
Second, we claim that our abstraction procedure creates a \emph{probabilistic feedback refinement relation} \cite<and thus a \emph{probabilistic simulation relation}; see>{hermanns2011probabilistic} from the generated \gls{MDP} to the dynamical system~\cite{DBLP:journals/tac/ReissigWR17,HSA17}.\footnote{The latter article employs alternating approximate relations, rather than simulation relations. The resulting refinement step is, however, very similar.}
Intuitively, a sufficient condition for such a relation to exist is that for any pair $(\state, s)$ of related states, i.e., for which $T(\state) = s$, and for all actions $a \in \Actions(s)$, there exists a control input $\control \in \cControlSpace$ such that the probability of transitioning to any state $s'$ in the \gls{MDP} equals the probability of transitioning to any $\state' \in T^\inv(s') = R_{s'}$ in the dynamical system.
Based on the definition from~\citeA{DBLP:journals/tac/ReissigWR17},\footnote{The definition in \citeA{DBLP:journals/tac/ReissigWR17} also contains a condition for matching observations between the two models, which we drop because we consider fully observable systems. Similarly, \citeA{HSA17} has an error metric on observations that does not apply to our setting. We also expand the original definition from non-probabilistic to probabilistic systems. Note that \cref{lemma:simulation_relation} defines sufficient conditions for a feedback refinement relation, which are stricter than those in~\citeA{DBLP:journals/tac/ReissigWR17}.} we mathematically write this condition as follows.
\begin{corollary}
    \label{lemma:simulation_relation}
    The abstraction function $T \colon \R^n \to \States$ induces a probabilistic feedback refinement relation from the \gls{MDP} to the dynamical system, because $T(\state_0) = \initState$ (i.e., the initial states match), and for any pair $(\state_k, s)$ of states related as $T(\state_k) = s$, it holds that
    \begin{equation}
        \forall a \in \Actions(s), \forall s'\in\States \, \,\colon\, 
        P(s,a)(s') = \int_{\region_{s'}} p_{\noise_k}\left(\state_{k+1} \ | \ \state_k, \control_k(\state_k, a)\right)d\state_{k+1}.
    \end{equation}
\end{corollary}
\noindent
\cref{lemma:simulation_relation} follows directly from \cref{def:transition_probabilities}.
Crucially, this probabilistic feedback refinement relation implies that \emph{any reach-avoid problem satisfiable for the \gls{MDP} is also satisfiable for the dynamical system with the same probability bound}~\cite{hermanns2011probabilistic}. 
\section{Sampling-Based Probability Intervals}
\label{sec:ScenarioApproach}

Recall that the probability density function $p_{\noise_k}\left(\state_{k+1} \ | \ \state_k, \control_k(\state_k, a)\right)$ is unknown, making a direct evaluation of \cref{eq:multivariateIntegral_mainProblem} impossible.
In this paper, we use a sampling-based method to estimate the transition probabilities described in \cref{eq:multivariateIntegral_mainProblem}, which requires a finite set of $N$ observations of the process noise, $\noise_k^{(i)} \in \Delta, \ i = 1, \ldots, N$.
Each sample has a unique index $i = 1, \ldots, N$ and is associated with a possible successor state $\state_{k+1}^{(i)} = \bm{d}_a + \noise_k^{(i)}$ under a particular action $a$. 
We assume that these samples are available from experimental data or simulations.
As such, we can generate these samples by inferring the process noise from obtained state trajectories of the dynamical system in \cref{eq:continuousSystem}, or we may sample from the noise distribution directly (e.g., if a simulator is available).
Thus, in our setting, samples of the noise can be obtained at a relatively low cost.

Recall that the process noise affecting \cref{eq:continuousSystem} is considered to be \gls{iid}. 
Due to the importance of this practically reasonable assumption to the arguments we develop in this section, we formally enshrine it in the following.
\begin{assumption}
    \label{assump:iid}
    The process noise $\noise_k$ is \gls{iid}, and we assume to have a collection of N samples from it, which we denote by the discrete set $\noise_k^{(i)} \in \Delta, \ i = 1, \ldots, N$. 
\end{assumption}
\noindent
Due to \cref{assump:iid}, the set $\noise_k^{(1)}, \ldots, \noise_k^{(N)}$ of $N$ samples is a random element from the probability space $\Delta^N$ equipped with the product probability $\amsmathbb{P}^N$ and the product $\sigma$-algebra.

\paragraph{A frequentist approach to estimation.}
As an example, we want to estimate the probability $P(s, a)(s')$ that some enabled state-action pair $(s, a)$ induces a transition to state $s' \in \States$.
A common approach to approximate the transition probability is to count the number of samples $N_{s'}^\text{in} \leq N$ leading to a transition to state $s'$ and divide it by the total of $N$ samples.
This approach is known as a \emph{frequentist approach}, and is statistically justified by the strong law of large numbers. 
Concretely, the value of $N_{s'}^\text{in}$ is obtained as follows.
\begin{definition}
    \label{def:Nj_in}
    The cardinality $N_{s'}^\text{in} \in \{0,\ldots,N\}$ 
    of the index set of the samples leading to a successor state $\state_{k+1} \in \region_{s'}$ associated with \gls{MDP} state $s' \in \States$ is defined as
    \begin{equation}
        N_{s'}^\text{in} = \Big\vert \{ i \in \{1,\ldots,N\} \, \vert \, (\bm{d}_a + \noise_k^{(i)}) \in \region_{s'} \} \Big\vert. 
        \label{eq:Nj_in}
    \end{equation}
    Similarly, we define $N_{s'}^\text{out} = N - N_{s'}^\text{in}$ as the number of samples for which $\bm{d}_a + \noise_k^{(i)} \notin \region_{s'}$.
\end{definition}
\noindent
Note that $N_{s'}^\text{in}$ and $N_{s'}^\text{out}$ depend on both the set of noise samples $\noise_k^{(1)}, \ldots, \noise_k^{(N)}$ and on the action taken.
The frequentist approach is simple, but may lead to estimates that deviate critically from their true values if the number of samples is limited (we illustrate this issue in the \gls{UAV} experiment in \cref{subsec:UAV}).
In what follows, we discuss how to render our method robust against such estimation errors.

\subsection{Sampling Techniques From the Scenario Approach}
As a key contribution, we present a method based on the scenario approach~\cite{campi2018introduction} to compute \emph{intervals of probabilities} instead of precise estimates.
Specifically, for every transition $(s,a,s')$, we compute an upper and lower bound, i.e., an interval, that contains the quantity $P(s, a)(s')$ with a user-specified (high) confidence probability.
We formalize the resulting abstraction as an \gls{iMDP}, where these intervals enter the uncertain transition function $\transfuncImdp \colon \States \times \Actions \times \States \rightharpoonup \Interval$.
As the intervals are \gls{PAC}, this \gls{iMDP} is a robust abstraction of the dynamical system.

\paragraph{Outline: PAC intervals for transition probabilities.}
Intuitively, we recast the estimation of transition probabilities into a convex optimization problem that includes a constraint for each element of a subset of noise samples.
To obtain \gls{PAC} intervals on transition probabilities, we leverage recent results from~\citeA{romao2022tac} on the probability of \emph{constraint violation} for such optimization problems.
Crucially, we show that the problem can be solved based on its geometry by an analytical \emph{counting argument}.
This counting argument makes our approach applicable to abstractions with hundreds of millions of transitions, as we do not have to solve any optimization problem explicitly.
Toward our main result for computing probability intervals (which is presented in \cref{theorem:Bounds}), we introduce a number of core concepts of the scenario approach.
Interestingly, due to the counting argument, \cref{theorem:Bounds} does not directly involve these concepts (only its derivation does, which we provide in \cref{appendix:Proofs}), so the reader may decide to skip directly to \cref{subsec:scenario:bounds}.

\paragraph{Risk of violation.}
First, we introduce the concept of \emph{risk} (or \emph{violation probability}), which is a measure of the probability that a successor state $\state_{k+1}$ \emph{is not} in a certain subset $\tilde{\region} \subset \R^n$ upon choosing an action $a \in \Actions(s)$ in state $s \in \States$~\cite{DBLP:journals/siamjo/CampiG08}.
\begin{definition}
    \label{def:risk}
    The risk $P_{\noise_k}\left(\state_{k+1} \notin \tilde{\region} \,\vert\, \state_k, \control_k(\state_k, a)\right)$ that $\state_{k+1}$ is not in region $\tilde{\region}$ is
    \begin{equation}
        P_{\noise_k}\left(\state_{k+1} \notin \tilde{\region} \,\vert\, \state_k, \control_k(\state_k, a)\right) = \amsmathbb{P} \{ \noise_k \in \Delta \ \colon \ \bm{d}_a + \noise_k \notin \tilde{\region} \}. 
        \label{eq:ViolationProbability}
    \end{equation}
\end{definition}
\noindent
Crucially, observe from \cref{eq:multivariateIntegral_mainProblem} that the transition probability $P(s,a)(s')$ that we aim to estimate is the complement of the violation probability over the fixed region $\region_{s'}$, i.e.,
\begin{equation}
    P(s,a)(s') = P_{\noise_k}\left(\state_{k+1} \in \region_{s'} \,\vert\, \state_k, \control_k(\state_k, a)\right) = 1 - P_{\noise_k}\left(\state_{k+1} \notin \region_{s'} \,\vert\, \state_k, \control_k(\state_k, a)\right).
    \label{eq:probability_complement}
\end{equation}

\paragraph{Scenario optimization problem.}
The scenario approach enables us to bound the risk over the feasible set of the optimal point of a so-called \emph{scenario optimization problem}.
Specifically, we consider scenario problems with \emph{discarded constraints}; see \citeA{DBLP:journals/jota/CampiG11} for details.
Roughly speaking, solving this problem amounts to finding, over a scalar decision variable $\lambda \in \R_+$, a convex set $\region_s(\lambda)$ of minimal size that contains a particular number of successor state samples (shortly, we shall define the geometry of $\region_s(\lambda)$ in relation to state $s \in \States$ and $\lambda$).
Concretely, the optimization problem is given as
\begin{align}
    \mathfrak{L}_{Q} \, \colon \, \minimize_{\lambda \in \R_+} \,\, & \lambda
    \label{eq:SCP}
	\\
	\nonumber
	\text{subject to} \enskip & \bm{d}_a + \noise_k^{(i)} \in \region_s (\lambda) \ \ \forall i \in \{ 1,\ldots,N \} \backslash Q,
\end{align}
where we explicitly write the dependency on the set $Q \subset \{1, \ldots, N\}$, which is a subset of samples whose constraints have been discarded.
The optimal solution $\lambda_{\vert Q \vert}^\star$ to problem $\mathfrak{L}_{Q}$ parameterizes a feasible set $\region_s(\lambda_{\vert Q \vert}^\star)$ over which we can bound the risk using the scenario approach theory.
However, to compute a transition probability $P(s,a)(s')$, we must shape the feasible set such that $\region_s(\lambda_{\vert Q \vert}^\star)$ is closely related to the fixed region $\region_{s'}$.
As explained below, we achieve this by 1) defining $\region_s(\lambda)$ as a scaled version of $R_s$, and 2) appropriately choosing the subset of discarded samples $Q$.

\paragraph{Scaled polytopes.}
We define $R_s( \lambda )$ as a version of polytope $R_s \in \Partition$ (recall from \cref{sec:Abstraction} that $R_s$ is defined by matrix $M_s$ and vector $\bm{b}_s$) which is \emph{scaled} by a factor $\lambda \geq 0$ relative to a so-called Chebyshev center $\tilde{\bm{h}}_s \in \R^n$~\cite{DBLP:books/cu/BV2014}.
As such, we obtain
\begin{equation}
    R_s( \lambda ) = \{ \state \in \R^n \ \vert \ M_s\state \leq \lambda (\bm{b}_s + \tilde{\bm{h}}_s ) - \tilde{\bm{h}}_s \}.
    \label{eq:PolytopeShifted}
\end{equation}
Note that $R_s(1) = R_s$, and that shifting by $\tilde{\bm{h}}_s$ ensures that we are scaling around a point that is by construction contained in $R_s$, such that $R_s(\lambda_1) \subset R_s(\lambda_2)$ for every $0 \leq \lambda_1 < \lambda_2$.
A visualization of the scaling for an arbitrary region $R_{s'} \in \Partition$ associated with a discrete successor state $s' \in \States$ is shown in \cref{fig:polytopeScaling}.
Note that, in this example, the Chebyshev center is not unique since the circle can be shifted while remaining within $\region_{s'}$.
\begin{figure}[t!]
\centering
\begin{minipage}[b]{.44\textwidth}
    \centering
	\usetikzlibrary{calc}
\usetikzlibrary{arrows.meta}
\usetikzlibrary{positioning}
\usetikzlibrary{fit}
\usetikzlibrary{shapes}

\tikzstyle{flowNodeRed} = [rectangle, rounded corners, minimum width=4.4cm, text width=4.3cm, minimum height=1.1cm, text centered, draw=black, fill=red!15]

\begin{tikzpicture}[scale=1.6]
	
	\newcommand\orX{0}
    \newcommand\orY{0.3}
    \newcommand\boxW{1}

    \newcommand\uppX{5}
    \newcommand\uppY{3}
    
    \newcommand\originX{1}
    \newcommand\originY{1}
    \newcommand\targetX{4}
    \newcommand\targetY{2}
	
	\newcommand\dX{\targetX+0.5*\boxW}
    \newcommand\dY{\targetY+0.5*\boxW}

	\draw[red, fill=red!05] (-5/9, 7/9) -- node[anchor=north, pos=0.3] {$R_{s'}$} (1/3, 1/3) -- (5/9, -7/9) -- (-1/3, -1/3) -- (-5/9, 7/9);
	
	\draw[red, fill=none, dashed] (-5/9*1.2, 7/9*1.2) -- node[anchor=south, pos=0.7, xshift=0.3cm] {$R_{s'}(1.2)$} (1/3*1.2, 1/3*1.2) -- (5/9*1.2, -7/9*1.2) -- (-1/3*1.2, -1/3*1.2) -- (-5/9*1.2, 7/9*1.2);
	
	\filldraw[fill=none, blue] (0,0) circle (0.3922) node[anchor=west]{};
	\filldraw[blue] (0,0) circle (0.02) node[black, anchor=west]{$\tilde{\bm{h}}_{s'}$};
	
    \draw[-{Latex[length=2mm,width=2mm]}] (-.9, -1) -- (.9, -1) node[pos=0.5, below] {$\bm{x}^{(1)}$};
    \draw[-{Latex[length=2mm,width=2mm]}] (-.9, -1) -- (-.9, 1) node[pos=0.5, left] {$\bm{x}^{(2)}$};
	
\end{tikzpicture}
	\captionof{figure}{
	    Polytope $\region_{s'}$ of state $s' \in \States$ has a Chebyshev center $\tilde{\bm{h}}_{s'}$ (note that it is not unique, as the circle can be shifted while remaining within $\region_{s'}$).
	    Polytope $R_{s'}(1.2)$ is scaled by a factor $\lambda = 1.2$ and is computed using \cref{eq:PolytopeShifted}.
	}
	\label{fig:polytopeScaling}
\end{minipage}\hfill
\begin{minipage}[b]{.52\textwidth} 
    \centering
	\begin{tikzpicture}[xscale=1.8,yscale=1.20]

\fill[red!20] (-0.73,0pt) rectangle (0.73,5pt);
\fill[blue!20] (-1.15,-5pt) rectangle (1.15,0pt);

\draw[latex-latex] (-2.2,0) -- (2.2,0) ; 
\foreach \x in  {-2,-1,0,1,2} 
\draw[shift={(\x,0)},color=black] (0pt,3pt) -- (0pt,-3pt);
\foreach \x in {-2,-1,0,1,2} 
\draw[shift={(\x,0)},color=black] (0pt,0pt) -- (0pt,-3pt) node[below] 
{$\x$};

\draw (0.73,0)  node[circle,fill,inner sep=1.5pt,red] {};
\draw (-1.39,0) node[circle,fill,inner sep=1.5pt] {};
\draw (1.15,0)  node[circle,fill,inner sep=1.5pt,blue] {};
\draw (-0.41,0) node[circle,fill,inner sep=1.5pt,red] {};
\draw (0.2,0)  node[circle,fill,inner sep=1.5pt,red] {};
\draw (1.75,0)  node[circle,fill,inner sep=1.5pt] {};
\draw (-0.06,0)  node[circle,fill,inner sep=1.5pt,red] {};
\draw (-0.65,0)  node[circle,fill,inner sep=1.5pt,red] {};
\draw (-1.61,0)  node[circle,fill,inner sep=1.5pt] {};
\draw (1.33,0)  node[circle,fill,inner sep=1.5pt] {};

\draw [blue, decorate,decoration={brace,amplitude=5pt,mirror,raise=4ex}]
  (-1.15,0.1) -- (1.15,0.1) node[blue, midway,yshift=-3em]{$\region_{s'}(\lambda^\star_{N_{s'}^\text{out}-1})$};
  
\draw [red,decorate,decoration={brace,amplitude=5pt,mirror,raise=2ex}]
  (0.73,-0.05) -- (-0.73,-0.05) node[red, midway,yshift=2em]{$\region_{s'}(\lambda^\star_{N_{s'}^\text{out}})$};

\draw[thick, shift={(-0.73,0)},color=red] (0pt,5pt) -- (0pt,-0pt);
\draw[thick, shift={(0.73,0)},color=red] (0pt,5pt) -- (0pt,-0pt);

\draw[thick, shift={(-1.15,0)},color=blue] (0pt,0pt) -- (0pt,-5pt);
\draw[thick, shift={(1.15,0)},color=blue] (0pt,0pt) -- (0pt,-5pt);

\draw[shift={(-1,0)}, dashed] (0pt,0pt) -- (0pt,30pt);
\draw[shift={(1,0)}, dashed] (0pt,0pt) -- (0pt,30pt);
\draw [decorate,decoration={brace,amplitude=5pt,mirror,raise=2ex}]
  (1,17pt) -- (-1,17pt) node[midway,yshift=2em]{$\region_{s'}$};

\end{tikzpicture}
	\captionof{figure}{
	    Bounding region $\region_{s'} = [-1, 1]$ using $N=10$ successor state samples ($N_{s'}^\text{out} = 5$).
	    Discarding $N_{s'}^\text{out} = 5$ samples defines the red region $\region_{s'}(\lambda^\star_{N_{s'}^\text{out}}) \subseteq \region_{s'}$; discarding one less sample defines the blue region $\region_{s'}(\lambda^\star_{N_{s'}^\text{out}-1}) \supset \region_{s'}$.
	}
	\label{fig:Scenario1D}
\end{minipage}
\end{figure}

\paragraph{Set of discarded samples.}
Let us now determine the subset $Q$ of samples whose constraints have been discarded from problem $\mathfrak{L}_Q$.
We obtain this set by iteratively removing individual samples based on the following rule:
\begin{proposition}
    The sample removal set $Q \subset \{ 1, \ldots, N \}$ is obtained by iteratively removing the active constraints from \cref{eq:SCP}.
    Thus, given $N$ samples and any two removal sets with cardinalities $\vert Q_1 \vert < \vert Q_2 \vert$, it holds that $Q_1 \subset Q_2$.
    Moreover, any discarded sample $i \in Q$ violates the solution $\lambda_Q^\star$ to \cref{eq:SCP}, i.e., $\bm{d}_a + \noise_k^{(i)} \notin \region_{s'}(\lambda^\star_{\vert Q \vert})$, with probability one.
    \label{lemma:Q}
\end{proposition}
\noindent
Intuitively, the successor state of a sample associated with an active constraint is on the boundary of the optimal solution $\region_{s'}(\lambda^\star_{\vert Q \vert})$ of \cref{eq:SCP}.
Under the following \emph{non-accumulation} assumption, an active constraint exists and is unique, as long as $\vert Q \vert < N$ (i.e., not all samples have been discarded).
\begin{assumption}
    \label{assumption:nonAccumulation}
    Given a noise sample $\noise_k \in \Delta$, the probability that the successor state $\state_{k+1}$ is on the boundary of any polytope $\region_{s'}(\lambda)$ is zero, for any $s' \in \States$ and $\lambda \in \R_+$.
\end{assumption}
\noindent
Note that \cref{assumption:nonAccumulation} holds for most of the smooth probability distributions commonly encountered in practice, e.g., Gaussian distributions, and is thus easily satisfied.
\cref{assumption:nonAccumulation} implies that the solution to \cref{eq:SCP} is unique with probability one and that the number of active constraints is equal to one (given $N>0$ and $\vert Q \vert < N$), as samples accumulate on the boundary of the polytope with probability zero.

\paragraph{Under/over-approximating regions.}
Recall from \cref{def:Nj_in} that $N_{s'}^\text{out}$ is the number of samples leading to a successor state outside of region $\region_{s'}$ for discrete state $s' \in \States$.
The following lemma uses $N_{s'}^\text{out}$ to define an under- or over-approximation of region $\region_{s'}$.

\begin{lemma}
    \label{lemma:Q_cardinality}
    When discarding $\vert Q \vert = N^\text{out}_{s'}$ noise samples as per \cref{lemma:Q}, it holds that $\region_{s'}(\lambda^\star_{N_{s'}^\text{out}}) \subseteq \region_{s'}$.
    When discarding $\vert Q \vert = N^\text{out}_{s'} - 1$ samples, it holds that $\region_{s'}(\lambda^\star_{N_{s'}^\text{out}-1}) \supset \region_{s'}$
\end{lemma}

\begin{proof}
    \cref{lemma:Q} states that at every step, we may only discard the sample of the active constraint.
    By construction, after discarding $\vert Q \vert = N_{s'}^\text{out}$ samples, all remaining successor state samples lie within $\region_{s'}$, so $\lambda^\star_{N_{s'}^\text{out}} \leq 1$, so $\region_{s'}(\lambda^\star_{N_{s'}^\text{out}}) \subseteq \region_{s'}$.
    When we discard one less sample, i.e., $\vert Q \vert = N_{s'}^\text{out} - 1$, we must have one sample outside of $\region_{s'}$, so $\lambda^\star_{N_{s'}^\text{out}-1} > 1$, meaning that $\region_{s'}(\lambda^\star_{N_{s'}^\text{out}-1}) \supset \region_{s'}$.
    This concludes the proof.
\end{proof}
\noindent
Intuitively, $\region_{s'}(\lambda^\star_{N_{s'}^\text{out}})$ contains \emph{exactly} those samples in $\region_{s'}$, while $\region_{s'}(\lambda^\star_{N_{s'}^\text{out}-1})$ additionally contains \emph{the sample closest outside of} $\region_{s'}$, as visualized in \cref{fig:Scenario1D} for a 1-dimensional example.
By using the scenario approach theory to bound the risk over these regions, we  can thus also obtain bounds on the transition probability $P(s,a)(s')$, as per \cref{eq:probability_complement}.

\subsection{Bounds for the Transition Probabilities}
\label{subsec:scenario:bounds}

Based on the techniques from the scenario approach introduced above, we state the main contribution of this section, as a non-trivial variant of~\citeA[Theorem~5]{romao2022tac}, adapted to our new context.
Specifically, for a given transition $(s,a,s')$ and the resulting number of samples $N_{s'}^\text{out}$ outside of region $\region_{s'}$ (as per \cref{def:Nj_in}), \cref{theorem:Bounds} returns an interval $[\munderbar{p},\bar{p}]$ that contains $P(s, a)(s')$ with at least a pre-defined confidence probability $(1-\beta)$.
\begin{theorem}[\gls{PAC} probability intervals]
    \label{theorem:Bounds}
    For $N \in \N$ samples of the noise, fix a confidence parameter $\beta \in (0,1)$.
    Given $N_{s'}^\text{out}$, the transition probability $P(s, a)(s')$ is bounded by
    \begin{equation}
        \amsmathbb{P}^N \Big\{ \munderbar{p} \leq P(s, a)(s') \leq \bar{p} \Big\} \geq 1 - \beta,
        \label{eq:TheoremBounds}
    \end{equation}
    where $\munderbar{p} = 0$ if $N_{s'}^\text{out} = N$, and otherwise $\munderbar{p}$ is the solution of
    \begin{equation}
        \frac{\beta}{2 N} = \sum_{i=0}^{N_{s'}^\text{out}} \binom Ni (1-\munderbar{p})^i \munderbar{p}^{N-i},
        \label{eq:TheoremLowerBound}
    \end{equation}
    and $\bar{p} = 1$ if $N_{s'}^\text{out} = 0$, and otherwise $\bar{p}$ is the solution of 
    \begin{equation}
        \frac{\beta}{2 N} = 1 - \sum_{i=0}^{N_{s'}^\text{out} - 1} \binom Ni (1-\bar{p})^i \bar{p}^{N-i}.
        \label{eq:TheoremUpperBound}
    \end{equation}
\end{theorem}
\noindent
We present the proof in \cref{appendix:Proofs}.
\cref{theorem:Bounds} states that, with a probability of at least $1-\beta$, the probability $P(s, a)(s')$ is bounded by the obtained interval $[\munderbar{p},\bar{p}]$. 
Importantly, this claim holds for \emph{any} $\Delta$ and $\amsmathbb{P}$ (given the previous assumptions), so we can bound the probability in \cref{eq:multivariateIntegral_mainProblem}, even when the probability distribution of the noise is unknown.

\begin{remark}[Beta distribution]
    \label{remark:beta_distribution}
    Note that \cref{eq:TheoremLowerBound,eq:TheoremUpperBound} are cumulative distribution functions of a beta distribution with parameters $N_{s'}^\text{out} + 1$ (or $N_{s'}^\text{out}$) and $N - N_{s'}^\text{out}$ (or $N - N_{s'}^\text{out} - 1$), respectively \cite{campi2018introduction}, which can directly be solved numerically for $\munderbar{p}$ or $\bar{p}$ up to arbitrary precision.
    To speed up the computations at run-time, we apply a tabular approach to compute the intervals for all relevant values of $N$, $\beta$, and $N_{s'}^\text{out}$ upfront.
\end{remark}

\paragraph{Counting argument.}
As explained in \cref{appendix:Proofs}, the proof of \cref{theorem:Bounds} is based on the solutions to a set of $N$ optimization problems.
Interestingly, as also shown in the proof, we can solve these optimization problems analytically based on their geometry.
As a result, \cref{theorem:Bounds} only requires the sample count $N_{s'}^\text{out}$, the total number of samples $N$, and the confidence parameter $\beta$, which are all quantities that are \emph{independent of the solutions to these optimization problems}.
Remarkably, this implies that we can compute \gls{PAC} probability intervals without solving any optimization program explicitly using a solver.
Since we only need the values of $N_{s'}^\text{out}$, $N$, and $\beta$, our method is in practice almost as simple as the frequentist approach but has the notable advantage that we obtain robust intervals of probabilities.
As shown in our experiments in \cref{sec:Studies}, this means we can use our abstraction procedure to generate \glspl{iMDP} with hundreds of millions of transitions.

\begin{figure}[t!]
\centering
\begin{minipage}[b]{.48\textwidth}
    \centering
	\footnotesize\begin{tikzpicture}
  \begin{axis}[
      /pgf/number format/.cd,
      use comma,
      1000 sep={\,},
      width=\linewidth,
      height=5.4cm,
      ymajorgrids,
      grid style={dashed,gray!30},
      xlabel=Number of samples ($N$),
      ylabel=Probability,
      xmode=log,
      log ticks with fixed point,
      xmin=22,
      xmax=13000,
      ymin=0,
      ymax=1.0,
      xtick={25,250,2500,25000},
      xtick pos=left,
      ytick={0,0.2,0.4,0.6,0.8,1.0},
      every axis plot/.append style={line width=0.6pt},
      legend cell align={left},
      legend columns=1,
      legend style={nodes={scale=0.78, transform shape},
                    anchor=north east, 
                    column sep=0ex,}
    ]
    
    \addplot+[
          black,
          dotted,
          mark=none
        ]
        coordinates {
        (22, 0.25)
        (12800, 0.25)
        };

    \addplot+[
      mark=x,
      color=red,
      dashed,
      error bars/.cd,
        y dir=both,
        y explicit,
    ] table [
        x=N,
        y=0.001_sc_low,
        col sep=comma,
        y error minus=0.001_sc_low_e,
        y error plus=0.001_sc_low_e,
    ] {Figures/Tikz/data/nr_samples_vs_intervals.csv};
    
    \addplot+[
      mark=o,
      color=red,
      solid,
      error bars/.cd,
        y dir=both,
        y explicit,
    ] table [
        x=N,
        y=1e-09_sc_low,
        col sep=comma,
        y error minus=1e-09_sc_low_e,
        y error plus=1e-09_sc_low_e,
    ] {Figures/Tikz/data/nr_samples_vs_intervals.csv};
    
    \addplot+[
      mark=x,
      color=MidnightBlue,
      dashed,
      error bars/.cd,
        y dir=both,
        y explicit,
    ] table [
        x=N,
        y=0.001_hf_low,
        col sep=comma,
        y error minus=0.001_hf_low_e,
        y error plus=0.001_hf_low_e,
    ] {Figures/Tikz/data/nr_samples_vs_intervals.csv};
    
    \addplot+[
      mark=o,
      color=MidnightBlue,
      solid,
      error bars/.cd,
        y dir=both,
        y explicit,
    ] table [
        x=N,
        y=1e-09_hf_low,
        col sep=comma,
        y error minus=1e-09_hf_low_e,
        y error plus=1e-09_hf_low_e,
    ] {Figures/Tikz/data/nr_samples_vs_intervals.csv};
    
    
    \addplot+[
      mark=x,
      color=red,
      dashed,
      error bars/.cd,
        y dir=both,
        y explicit,
    ] table [
        x=N,
        y=0.001_sc_upp,
        col sep=comma,
        y error minus=0.001_sc_upp_e,
        y error plus=0.001_sc_upp_e,
    ] {Figures/Tikz/data/nr_samples_vs_intervals.csv};
    
    \addplot+[
      mark=x,
      color=red,
      solid,
      error bars/.cd,
        y dir=both,
        y explicit,
    ] table [
        x=N,
        y=1e-09_sc_upp,
        col sep=comma,
        y error minus=1e-09_sc_upp_e,
        y error plus=1e-09_sc_upp_e,
    ] {Figures/Tikz/data/nr_samples_vs_intervals.csv};
    
    \addplot+[
      mark=x,
      color=MidnightBlue,
      dashed,
      error bars/.cd,
        y dir=both,
        y explicit,
    ] table [
        x=N,
        y=0.001_hf_upp,
        col sep=comma,
        y error minus=0.001_hf_upp_e,
        y error plus=0.001_hf_upp_e,
    ] {Figures/Tikz/data/nr_samples_vs_intervals.csv};
    
    \addplot+[
      mark=o,
      color=MidnightBlue,
      solid,
      error bars/.cd,
        y dir=both,
        y explicit,
    ] table [
        x=N,
        y=1e-09_hf_upp,
        col sep=comma,
        y error minus=1e-09_hf_upp_e,
        y error plus=1e-09_hf_upp_e,
    ] {Figures/Tikz/data/nr_samples_vs_intervals.csv};

    \legend{{True probability},{$\beta = 10^{-3}$ (\cref{theorem:Bounds})},{$\beta = 10^{-9}$ (\cref{theorem:Bounds})},{$\beta = 10^{-3}$ (Hoeffding)},{$\beta = 10^{-9}$ (Hoeffding)}}

 \end{axis}
\end{tikzpicture}
	\captionof{figure}{Probability intervals (with standard deviation) obtained from \cref{theorem:Bounds} versus Hoeffding's inequality, with $\beta=10^{-3}$ or $10^{-9}$ on a transition with a true probability of $0.25$ (note the log scale).}
	\label{fig:intervals_samples}
\end{minipage}\hfill
\begin{minipage}[b]{.48\textwidth} 
    \centering
	\footnotesize\begin{tikzpicture}
  \begin{axis}[
      width=\linewidth,
      height=5.4cm,
      ymajorgrids,
      grid style={dashed,gray!30},
      xlabel=Samples outside of region ($N_{s'}^\text{out}$),
      ylabel=Probability,
      xmin=0,
      xmax=800,
      ymin=-0.02,
      ymax=1.02,
      xtick={0,200,400,600,800},
      xtick pos=left,
      ytick={0,0.2,0.4,0.6,0.8,1.0},
      every axis plot/.append style={line width=0.6pt},
      legend cell align={left},
      legend columns=1,
      legend image post style={xscale=0.50},
      legend style={nodes={scale=0.78, transform shape},
                    anchor=north east, 
                    column sep=0ex,}
    ]
    
    \addplot+[
          black,
          dotted,
          mark=none
        ]
        coordinates {
        (0,1)
        (800,0)
        };

    \addplot+[
      color=red,
      mark=none
    ] table [
        x=k,
        y=1e-09_sc_low,
        col sep=comma,
    ] {Figures/Tikz/data/fraction_vs_intervals.csv};
    
    \addplot+[
      color=MidnightBlue,
      mark=none
    ] table [
        x=k,
        y=1e-09_hf_low,
        col sep=comma,
    ] {Figures/Tikz/data/fraction_vs_intervals.csv};
    
    \addplot+[
      color=red,
      mark=none
    ] table [
        x=k,
        y=1e-09_sc_upp,
        col sep=comma,
    ] {Figures/Tikz/data/fraction_vs_intervals.csv};
    
    \addplot+[
      color=MidnightBlue,
      mark=none
    ] table [
        x=k,
        y=1e-09_hf_upp,
        col sep=comma,
    ] {Figures/Tikz/data/fraction_vs_intervals.csv};
    
    \legend{{Fraction out ($N_{s'}^\text{out} / N$)},{$\beta = 10^{-9}$ (\cref{theorem:Bounds})},{$\beta = 10^{-9}$ (Hoeffding)}}

 \end{axis}
\end{tikzpicture}
	\captionof{figure}{Probability intervals
	obtained from \cref{theorem:Bounds}, with $\beta=10^{-9}$ and different values for $N_{s'}^\text{out} \in [0, N]$, versus probability intervals obtained from Hoeffding's inequality for the same value of $\beta$.}
	\label{fig:intervals_fraction}
\end{minipage}
\end{figure}

\subsection{Tightness of Probability Intervals}
Let us now explore the tightness of the obtained probability intervals.
We can interpret a transition probability $P(s,a)(s')$ as the probability of a Bernoulli random variable.
In this context, we may use Hoeffding's inequality, a well-known concentration inequality, to infer \gls{PAC} bounds on this probability~\cite{DBLP:books/daglib/0035704}.
In particular, given $N$ successor state samples of which $N_{s'}^\text{in}$ are contained in region $R_{s'}$, Hoeffding's inequality states that, for a pre-defined $\beta \in (0,1)$, it holds that
\begin{equation}
    \amsmathbb{P}^N \left\{ \frac{N_{s'}^\text{in}}{N} - \varepsilon \leq P(s,a)(s') \leq \frac{N_{s'}^\text{in}}{N} + \varepsilon \right\} \geq 1 - \beta,
    \label{eq:Hoeffding}
\end{equation}
where $\varepsilon = \sqrt{\frac{1}{2N} \log(\frac{2}{\beta})}$.
In what follows, we evaluate the tightness of our intervals obtained from \cref{theorem:Bounds}, versus those obtained from Hoeffding's inequality.

\paragraph{Number of noise samples $N$.}
To illustrate how the choice for the number of samples $N$ affects the tightness of the intervals, consider a system with a 1-dimensional state $\state_k \in \R$, where the probability density function $p_{\noise_k}(\state_{k+1} \ | \ \state_k, \control_k(\state_k, a))$ for a specific action $a \in \Actions$ with target point $\bm{d}_a$ is given by a uniform distribution over the domain $[-4, 4]$.
For a given region $R_{s'} = [-1,1]$ (also shown in \cref{fig:Scenario1D}), we want to evaluate the probability that $\state_{k+1} \in R_{s'}$, which is $0.25$. 
To this end, we apply \cref{theorem:Bounds} for different numbers of samples $N \in [25, 12\,800]$ and a confidence level of $\beta=10^{-3}$ or $10^{-9}$.
The obtained probability bounds are random variables through their dependence on the samples, so we repeat each experiment $100\,000$ times, resulting in the probability intervals shown in \cref{fig:intervals_samples}.
We observe that the uncertainty in the transition probability is reduced by increasing the number of samples.
Moreover, the lower bounds obtained from \cref{theorem:Bounds} are better than those obtained from Hoeffding's inequality, while the converse holds for the upper bounds.

\paragraph{Sample count $N_{s'}^\text{out}$.}
To explain why \cref{theorem:Bounds} yields tighter lower bounds, while Hoeffding's inequality yields tighter upper bounds, we plot in \cref{fig:intervals_fraction} the resulting probability intervals for $N=800$ samples and different values of $N_{s'}^\text{out} \in [0,N]$ (recall that $N = N_{s'}^\text{out} + N_{s'}^\text{in}$).
We observe our scenario-based approach results in \emph{significantly tighter} intervals for values of $N_{s'}^\text{out}$ close to $0$ and $N$.
On the other hand, Hoeffding's inequality leads to \emph{slightly better} intervals for moderate values of $N_{s'}^\text{out}$ around $\frac{N}{2}$.
As observed from \cref{eq:Hoeffding}, Hoeffding's inequality yields bounds given by the \emph{sample mean} $N_{s'}^\text{in}$, plus or minus a \emph{fixed} value of $\varepsilon$ which is \emph{independent of the sample mean}.
This property explains why Hoeffding's inequality leads to poor probability intervals if the probability $P(s,a)(s')$ is close to zero or one.

\subsection{iMDP Abstractions With PAC Guarantees}
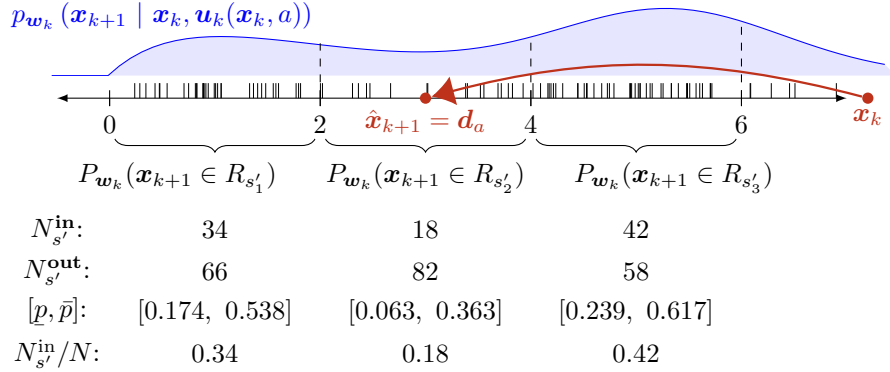
\begin{figure}[t!]
	\centering

\begin{tikzpicture}[xscale=1.4, font=\small]

\newcommand\normMean{4}
\newcommand\normScale{1}
\newcommand\gammaA{2}

\begin{axis}[
        at={(-3.55cm,0cm)},
        domain = -0.5:7,
        axis line style={draw=none},
        tick style={draw=none},
        ylabel = {},
        ylabel style={align=center,rotate=-90},
        xticklabels={,,},
        yticklabels={,,},
        yshift=0.3cm,
        xmin=-0.5, xmax=7,
        ymin=0,  ymax=0.26,
        width=9.6cm,
        height=2.5cm,
    ]
    \addplot [name path=A, color=blue] table [x=x, y=y, col sep=comma, mark=none] {Figures/Tikz/distributionPlot_sparse3.csv};
    
    \path[name path=B] (axis cs:-0.5,0) -- (axis cs:7,0);
    
    \addplot [
        thick,
        color=blue,
        fill=blue, 
        fill opacity=0.1
    ] fill between[of=A and B, soft clip={domain=-1:7},];
    
\end{axis}

\foreach \x in {1.9904310603088948,-0.5278880533336503,-1.4461112880916378,-1.5943995469103731,2.6294506104614133,-1.1867788845432967,2.037181532238858,-1.3959068973420097,2.5688452626954277,3.901003799214978,-1.5563206532574538,1.091929144540357,-1.2292227993597358,-2.063392996771781,2.255592948953187,0.6881758404254068,-2.5687775314312766,-2.2498452065637062,2.0199693617488226,1.956825926204485,0.7860453702328734,-2.52772142325479,-2.1858132538399264,2.334055970212992,2.4005500055531614,0.5210108316624362,-2.655417619739051,2.706090565629543,0.8253038068962786,-0.3318863501018159,1.7266469820105135,-1.9415769023599694,-1.5177303444790482,-2.522985295885876,-1.9401966127636552,2.479133323201429,1.6728477125759245,-0.978864651407358,0.5541711976103079,1.0204948216450314,3.5081135776747,1.4445949955555575,-2.4539701417083117,2.132481012567254,1.1618219654722655,1.5341968606117327,1.5991518762340222,-0.6935627554834376,1.3025740111999102,2.037788529815593,1.7458974425571778,2.703316071729569,-2.7131784058516173,-2.1143043485120243,2.6037162284961353,-1.974566829520481,-0.5605911771474217,2.4874738006008466,3.2832169888506773,-1.6707645617572062,1.2142909742497245,1.1912412115649103,0.013967985175968423,2.096534415839007,2.719099849235044,2.1520972561880516,-2.1091728382097736,0.9264630029369343,3.4566694045208743,1.5286715982648778,1.946569855780397,3.087362028180986,-2.1699550902726727,1.2367856267370954,2.3343339772806964,-2.2934955905575745,-1.9932353533238478,1.9395806905152062,-1.633890382078726,-2.761491108014665,1.7890688465551463,0.3807806785316634,1.447596198644714,-2.0064272349669405,-1.2056352536204722,2.236418708772468,-2.1021263761556046,2.5086611768937415,-0.637435936712774,3.0825141682213797,-2.176049547871288,-2.110686716589787,-2.0858006381367344,0.719254321697882,-1.4211937737964444,0.021350708152708897,1.5067085167756558,-0.9983214262961626,0.3976297303859848,1.173806561264576}
{
    \draw[line width=0.01mm] (\x,0) -- ++(0,0.2cm);
}

\draw[dashed] (-1cm,0cm) -- ++(0cm,0.73cm);
\draw[dashed] (1cm,0cm) -- ++(0cm,0.80cm);
\draw[dashed] (3cm,0cm)  -- ++(0cm,1.00cm);

\draw[latex-latex] (-3.5,0) -- (4,0) ; 

\foreach \x in  {-3,-1,1,3} 
    \draw[shift={(\x,0)},color=black] (0pt,3pt) -- (0pt,-3pt);

\foreach \x/\xtext in {-3/0,-1/2,1/4,3/6} 
    \draw[shift={(\x,0)},color=black] (0pt,0pt) -- (0pt,-3pt) node[below] 
    {$\xtext$};
    
\draw (-2.5cm,1.1cm) node[align=center, blue] {$p_{\noise_k}\left(\state_{k+1} \ | \ \state_k, \control_k(\state_k, a)\right)$};

\node[red, label={[red, below, label distance=-0.1cm]$\state_k$}] (predecessor) at (4.2,0) [circle,fill,inner sep=1.5pt] {};
\node[red, label={[red, below, label distance=-0.1cm]$\hat{\state}_{k+1} = \bm{d}_a$}] (target) at (0,0) [circle,fill,inner sep=1.5pt] {};

\draw[-{Latex[length=3mm,width=4mm]}, line width=0.3mm, red] (predecessor)  to [out=160,in=20] (target);

\draw [decorate,decoration={brace,amplitude=5pt,mirror,raise=2ex}]
  (-2.95,-6pt) -- (-1.05,-6pt) node[midway, yshift=-2.2em, xshift=-.5cm]{$P_{\bm{w}_k}(\state_{k+1} \in R_{s'_1})$};
  
\draw [decorate,decoration={brace,amplitude=5pt,mirror,raise=2ex}]
  (-0.95,-6pt) -- (0.95,-6pt) node[midway,yshift=-2.2em]{$P_{\bm{w}_k}(\state_{k+1} \in R_{s'_2})$};
  
\draw [decorate,decoration={brace,amplitude=5pt,mirror,raise=2ex}]
  (1.05,-6pt) -- (2.95,-6pt) node[midway,yshift=-2.2em, xshift=.5cm]{$P_{\bm{w}_k}(\state_{k+1} \in R_{s'_3})$};

\newcommand\yshift{-1.75}
\newcommand\yshiftB{-2.30}
\newcommand\yshiftC{-2.85}
\newcommand\yshiftD{-3.40}
  
\draw (-3.5, \yshift) node[align=left] {$N_{s'}^\textbf{in}$:};
\draw (-2, \yshift) node[] {34};
\draw (0,  \yshift) node[] {18};
\draw (2,  \yshift) node[] {42};

\draw (-3.5, \yshiftB) node[align=left] {$N_{s'}^\textbf{out}$:};
\draw (-2, \yshiftB) node[] {66};
\draw (0,  \yshiftB) node[] {82};
\draw (2,  \yshiftB) node[] {58};

\draw (-3.5, \yshiftC) node[align=left] {$[\munderbar{p}, \bar{p}]$:};
\draw (-2, \yshiftC) node[] {$[0.174, \ 0.538]$};
\draw (0, \yshiftC) node[]  {$[0.063, \ 0.363]$};
\draw (2, \yshiftC) node[]  {$[0.239, \ 0.617]$};

\draw (-3.5, \yshiftD) node[align=left] {$N_{s'}^\text{in}/N$:};
\draw (-2, \yshiftD) node[] {0.34};
\draw (0, \yshiftD) node[]  {0.18};
\draw (2, \yshiftD) node[]  {0.42};

\end{tikzpicture}

	
	\caption{
	    Bounds $[\underline{p},\bar{p}]$ on the probabilities $P(s,a)(s')$ for 3 regions using $N=100$ samples (black ticks) and $\beta=0.01$. 
	    The probability density function over successor states is $p_{\noise_k}\left(\state_{k+1} \ | \ \state_k, \control_k(\state_k, a)\right)$.
	    Point estimate probabilities are computed as $N_{s'}^\text{in}/N$. 
	}
	\label{fig:Scenario1D_b}
\end{figure}

We describe how we apply \cref{theorem:Bounds} to solve the overall problem statement. 
Since the process noise is independent of the state and time, we require \emph{only $N$ samples in total}.\footnote{If the noise distribution is time-varying, the transition probabilities are time-dependent.
In this case, we have at most $\vert\Actions\vert \cdot \vert S \vert \cdot \horizon$ unique probabilities and must use $N$ noise samples for each step $k = 0,\ldots,\horizon-1$. For finite-horizon properties, adapting our abstraction method for this case is straightforward.}
We can use the same samples to compute multiple intervals by shifting these samples by the appropriate target point $\bm{d}_a$ of each action $a \in \Actions$.
Recall from \cref{remark:number_probabilities} that the abstract \gls{MDP} has at most $\vert\Actions\vert \cdot \vert S \vert$ unique transition probabilities.
For every unique probability, we determine the successor state samples $\state_{k+1}^{(1)}, \ldots, \state_{k+1}^{(N)}$ under the $N$ samples of the noise.
For every possible successor state $s' \in \States$, we then determine $N_{s'}^\text{out}$ and invoke \cref{theorem:Bounds} to compute the \gls{PAC} bounds on $P(s,a)(s')$ that hold for every $s \in \States$.
\cref{fig:Scenario1D_b} shows this process, where every tick is a successor state $\state_{k+1}$ under a noise sample.
This figure also shows point estimates of the probabilities derived using the frequentist approach.
If no samples are observed in a region, we assume that $P(s, a)(s') = 0$.

\begin{theorem}
    \label{theorem:correctness}
    Given a dynamical system in \cref{eq:continuousSystem}, generate an \gls{iMDP} abstraction $\imdp$, and compute the robust reach-avoid probability $\reachProbDiscrMinStar$ under the optimal policy $\munderbar{\policy}^\star$.
    Under the controller $\phi$ that applies control inputs according to \cref{eq:controlLaw} for each $k < \horizon$, it~holds~that
    \begin{equation}
        \label{eq:correctness}
        \reachProbDiscrMinStar \geq \eta 
        \implies
        \amsmathbb{P}^N \left\{ \reachProbCont \geq \eta \right\} \geq 1 - \alpha,
    \end{equation}
    where $\alpha = \beta \cdot \vert\Actions\vert \cdot \vert\States\vert$ upper bounds the number of unique probability intervals.
\end{theorem}

\begin{proof}
    Denote by $\transfuncImdp \colon \States \times \Actions \times \States \rightharpoonup \Interval \cup \{ [0,0] \}$ the uncertain transition function of the generated \gls{iMDP}, and denote by $\transfunc \colon \States \times \Actions \rightharpoonup \distr{\States}$ the unknown transition function, defined by \cref{eq:multivariateIntegral_mainProblem}, of the underlying \gls{MDP}.
    For each $a \in \Actions$ and $s' \in \States$, it holds that
    \begin{equation}
        \label{eq:correctness_proof1}
        \amsmathbb{P}^N \left\{ P(s,a)(s') \in \transfuncImdp(s,a,s'), \,\, \forall s \in \States \right\} \geq 1-\beta.
    \end{equation}
    Recall from \cref{remark:number_probabilities} that the \gls{iMDP} has at most $|\Actions| \cdot |\States|$ unique probability intervals.
    Using Boole's inequality,\footnote{Boole's inequality (or the union bound) says that the probability that at least one of a finite set of events happens, is upper bounded by the sum of the probabilities of these events~\cite{casella2021statistical}.} we thus obtain the following correctness guarantee for the \gls{iMDP}:
    \begin{equation}
        \label{eq:correctness_proof2}
        \amsmathbb{P}^N \left\{ P(s,a)(s') \in \transfuncImdp(s,a,s), \,\, \forall s,s' \in \States, a \in \Actions \right\} 
        =
        \amsmathbb{P}^N \left\{ P \in \transfuncImdp \right\}
        \geq 
        1-\alpha,
    \end{equation}
    where $\alpha = \beta \cdot |\Actions| \cdot |\States|$.
    Next, observe that for any \gls{MDP} instantiation $P \in \transfuncImdp$, it holds by construction of \cref{eq:optimalPolicy_lb} that $\reachProbDiscrMinStar \leq \reachProbDiscrStar$.
    Moreover, \cref{lemma:simulation_relation} asserts that $\reachProbDiscrStar = \reachProbCont$ (the reach-avoid probability on the abstract \gls{MDP} equals the reach-avoid probability on the dynamical system), so we obtain
    \begin{equation}
        \amsmathbb{P}^N \left\{ \reachProbDiscrMinStar \leq \reachProbCont \right\} \geq 1-\alpha.
    \end{equation}
    Letting $\reachProbDiscrMinStar \geq \eta$, we arrive at the implication in \cref{eq:correctness}, so we conclude the proof.
\end{proof}

We remark that the tightness of \cref{eq:correctness} in \cref{theorem:correctness} depends on the value of $\alpha$, which in turn depends on the size of the generated \gls{iMDP} (through $\vert\States\vert$ and $\vert\Actions\vert$), and the confidence level $\beta$.
For large \glspl{iMDP}, one must choose a stronger confidence level (i.e., $\beta$ closer to zero), to obtain an informative bound $1-\alpha$, which is close to one.
The following corollary provides a tighter bound if the \gls{iMDP} is generated in a very specific manner.

\begin{corollary}
    \label{corollary:tighter_confidence}
    If the \gls{iMDP} in \cref{theorem:correctness} is generated under a uniform rectangular partition $\Partition$ (i.e., a grid) into $r_i$ regions in each dimension $i=1,\ldots,n$ of the state space $\R^n$, and with one action $a \in \Actions$ per region $\region \in \Partition$ whose target point $\bm{d}_a$ is the center of $\region$, then the confidence parameter becomes $\alpha = \beta \big( \prod_{i=1}^n (2 r_i - 1) + \vert\Actions\vert \big)$.
\end{corollary}

The proof of \cref{corollary:tighter_confidence} is provided in \cref{proof:cor:tight} and is analogous to the proof of \cref{theorem:correctness}, with the only difference that the number of unique probability intervals is reduced.
The key observation is that transition probabilities $P(s,a)(s')$ of the abstraction are defined by the relative position $\region_{s'} - \bm{d}_a$ between region $\region_{s'}$ and target point $\bm{d}_a$.
In particular, the corollary exploits the symmetry between the partition and the target points, which reduces the number of unique transition probabilities significantly and leads to stronger confidence levels compared to \cref{theorem:correctness}.
\section{Overall Robust Control Algorithm}

\label{sec:Algorithm}

In \cref{sec:Abstraction,sec:ScenarioApproach}, we have introduced tools for generating an \gls{iMDP} abstraction of the dynamical system in \cref{eq:continuousSystem}, which is correct with a user-specified confidence probability.
We now discuss in more detail how we use these tools to solve the overall problem statement posed in \cref{sec:prelim:problem}.
Recall that our approach, shown in \cref{fig:Approach}, consists of an offline and an online phase.
In what follows, we present an algorithm for both phases.

\subsection{Offline Planning Phase}
The offline planning phase is presented in \cref{alg:algorithm_planning}.
We first generate an \gls{MDP} abstraction by defining its states $\States$, actions $\Actions$, and initial state $\initState \in \States$ (line 1).
For each state $s \in \States$, we compute the set $\Actions(s)$ of enabled actions (line 2) and define the initial values for the sample size $N$ and for the reachability probabilities (line 3).
Recall that computing the transition probabilities of this \gls{MDP} is not possible, so we instead compute intervals of probabilities and formalize the abstract model as an \gls{iMDP} instead.

\begin{algorithm}[t!]
\caption{Offline planning (generate \gls{iMDP} and compute optimal policy)}
\label{alg:algorithm_planning}
\textbf{Input}: Linear dynamical model; property $\propertyMDP$ with probability threshold $\eta$\\
\textbf{Params}: Partition $\Partition$; confidence lvl. $\beta$; increment factor $\gamma$\\
\textbf{Output}: Optimal policy $\munderbar{\policy}^*$
\begin{algorithmic}[1] 
    \STATE Define states $\States$, actions $\Actions$, and initial state $\initState = T(\bm{x}_k)$
    \STATE Define enabled actions $\Actions(s) \subseteq \Actions$ for all $s \in \States$
    \STATE Set initial values: $N = N_0$, $\reachProbDiscrMinStar = 0$, $\reachProbDiscrMaxStar = 1$
    \WHILE{$\reachProbDiscrMinStar < \eta \leq \reachProbDiscrMaxStar$}
        \STATE Obtain $N$ samples $\noise_k^{(1)}, \noise_k^{(N)} \in \Delta$
        \FORALL{actions $a$ in $\Actions$}
            \FORALL{successor states $s'$ in $\States$}
                \STATE Compute $[\munderbar{p}, \bar{p}]$ by applying \cref{theorem:Bounds} for $N$, $N_{s'}^{out}$, and $\beta$
                \STATE Let $\transfuncImdp(s, a, s') = [\munderbar{p}, \bar{p}] \ \forall s \in \States$ such that $a \in \Actions(s)$
            \ENDFOR
        \ENDFOR
        \STATE Store \gls{iMDP} $\imdp=\iMDP$
        \STATE Compute $\munderbar{\policy}^*$, $\reachProbDiscrMinStar$, and $\reachProbDiscrMaxStar$ on $\imdp$ using \prism
        \STATE Let $N = \gamma N$
    \ENDWHILE
    \IF{$\eta > \reachProbDiscrMaxStar$}
        \STATE \textbf{return} $\mathsf{Unsatisfiable}$
    \ELSE
        \STATE \textbf{return} optimal policy $\munderbar{\policy}^*$
    \ENDIF
\end{algorithmic}
\end{algorithm}

We then enter the iterative part of our approach.
We first obtain $N$ samples of the noise (line 5).
Recall from \cref{sec:ScenarioApproach} that we can obtain these samples by inferring the process noise from previously generated state trajectories of the dynamical system or by sampling the noise distribution directly using a simulator.
For every action $a \in \Actions$ and successor state $s' \in \States$, we compute a \gls{PAC} transition probability interval $[\munderbar{p}, \bar{p}]$ using \cref{theorem:Bounds} (line 8).
These intervals are used to define the transition function $\transfuncImdp$ of the abstract \gls{iMDP} (line 9).

Recall from \cref{sec:IterativeScheme} that for the resulting \gls{iMDP} (stored in line 12), we leverage PRISM to obtain the optimal policies that maximize lower and upper bounds on the reach-avoid probability using \cref{eq:optimalPolicy_lb,eq:optimalPolicy_ub} (line 13).
If the maximum lower bound reach-avoid probability $\reachProbDiscrMinStar$ under this policy is above the required threshold $\eta$, then the algorithm returns the optimal policy $\munderbar{\policy}^\star$ and proceeds to the online control phase. 
Otherwise, if $\reachProbDiscrMinStar < \eta$ but the problem is still satisfiable (i.e., $\reachProbDiscrMaxStar \geq \eta$) we obtain additional samples by increasing $N$ by a fixed factor $\gamma > 1$ (line 14) and repeat the while loop.
If instead, it holds that $\reachProbDiscrMaxStar < \eta$, the reach-avoid problem is not satisfiable for any \gls{MDP} instantiated by $\transfunc \in \transfuncImdp$.
In this case, the algorithm terminates without returning a policy.

The main computational complexity of \cref{alg:algorithm_planning} lies within the double for loop.
In particular, generating one \gls{iMDP} abstraction\footnote{Verifying \glspl{iMDP} can be done in polynomial time, and we refer to~\citeA{DBLP:conf/cav/PuggelliLSS13} for details.} (lines 5-12 of \cref{alg:algorithm_planning}) has the following complexity with respect to the numbers of states $|\States|$ and actions $|\Actions|$ (which are directly controlled through the partitioning of the state space), and the number of noise samples~$N$.

\begin{theorem}[Complexity of generating abstractions]
    \label{thm:complexity}
    The worst-case complexity of generating one \gls{iMDP} abstraction using \cref{alg:algorithm_planning} is 
    $\mathcal{O}\big( N \cdot |\Actions| + |\States|^2 \cdot |\Actions| \big)$.
\end{theorem}

\begin{proof}
For every action $a \in \Actions$, we must check for each state $s \in \States$ if $a$ is enabled in $s$, leading to $\mathcal{O}\big( |\States| \cdot |\Actions| \big)$ operations.
Next, for each action $a \in \Actions$, we must determine for each of the $N$ samples to which state it belongs, leading to $\mathcal{O}\big( N \cdot |\Actions| \big)$, and subsequently, we compute for each action $a \in \Actions$ and successor state $s' \in \States$ the probability interval $[\munderbar{p}, \bar{p}]$, leading to $\mathcal{O}\big( |\Actions| \cdot |\States| \big)$.
Finally, we store each transition of the \gls{iMDP}, which has at most $|\States|^2 \cdot |\Actions|$ transitions.
By summing these contributions and only keeping the highest order terms, we obtain the order of complexity in \cref{thm:complexity} and conclude the proof.
\end{proof}

We remark that the number of states $|\States|$ and $|\Actions|$ depend on the partition and target points used to generate the abstraction; see \cref{sec:Abstraction} for details.
For example, under the rectangular partitioning described by \cref{corollary:tighter_confidence}, the number of states is $|\States| = \prod_{i=1}^n + 1$ and the number of actions is $|\Actions| = \prod_{i=1}^n$.
This result shows that our abstraction procedure (like any other discretization-based technique) suffers from the well-known \emph{curse of dimensionality}~\cite{LSAZ21}, i.e., the complexity is exponential in the dimension $n$ of the continuous state space $\R^n$.

\begin{algorithm}[t!]
\caption{Online control (on-the-fly controller synthesis)}
\label{alg:algorithm_control}
\textbf{Input}: Optimal policy $\munderbar{\policy}^*$; initial state $\state_0$; property $\property$ \\
\textbf{Output}: $\mathsf{SAT}$ (Boolean)
\begin{algorithmic}[1] 
\STATE Let $k = 0$
\WHILE{$k \leq \horizon$}
    \IF{$\state_k \in \goalRegion$}
        \STATE \textbf{return} $\mathsf{SAT} = \mathsf{True}$
    \ELSIF{$\state_k \in \criticalRegion$}
        \STATE \textbf{return} $\mathsf{SAT} = \mathsf{False}$
    \ELSE
        \STATE Let optimal action $a^* = \munderbar{\policy}^*(s, k)$ for state $s = T(\state_k)$
        \STATE Compute control input $u_k(a^*)$ using \cref{eq:controlLaw}
        \STATE Sample state $\state_{k+1} \sim p_{\noise_k}\left(\state_{k+1} \ | \ \state_k, \control_k(a)\right)$
    \ENDIF
    \STATE $k = k + 1$
\ENDWHILE
\STATE \textbf{return} $\mathsf{SAT} = \mathsf{False}$
\end{algorithmic}
\end{algorithm}

\subsection{Online Control Phase}
In the online control phase, we synthesize a controller for the dynamical system on the fly, based on the policy $\munderbar{\policy}^\star$ returned by \cref{alg:algorithm_planning}.
Recall that this policy is a time-varying map from \gls{iMDP} states to actions.
Intuitively, we translate the policy to a time-varying feedback controller that is piece-wise linear (namely, linear in the state $\state_k$ within each region of the partition). 
Concretely, at each time step, we use \cref{def:relation} to determine the current \gls{iMDP} state $s = T(\state_k)$ to which the continuous state $\state_k$ belongs, and then retrieve the optimal action $a^\star = \munderbar{\policy}^\star(s,k)$ from the policy (line 8).
We compute the associated control input using \cref{eq:controlLaw}, which is valid by construction, followed by sampling the successor state $\state_{k+1}$ (lines 9-10).
The algorithm returns $\mathsf{SAT} = \mathsf{True}$ if the goal region is reached within $\horizon$ steps (i.e., the reach-avoid problem is satisfied), while it returns $\mathsf{SAT} = \mathsf{False}$ if either the critical region is reached or the system fails to reach the goal region within $\horizon$ steps.

\begin{remark}[Backup controller]
    The controller obtained via \cref{alg:algorithm_planning} only yields control inputs for states in the bounded portion of the state space $\cStateSpace = \R^n \setminus \region_\ast$, which is covered by the partition $\Partition$.
    To alleviate this conservatism, one may additionally design a \emph{backup controller}, which is activated upon leaving $\cStateSpace$ and aims to return the state $\state_k$ to the bounded portion $\cStateSpace$ for which the controller generated control inputs.
    In the abstract \gls{iMDP}, visiting the absorbing state $s_\ast$ (which corresponds to the absorbing region $\region_\ast = \R^n$) means that the reach-avoid property is violated by definition.
    As such, deploying the backup controller can only increase the probability $\reachProbCont$ of satisfying the reach-avoid property, and thus, \cref{theorem:correctness} still holds regardless of the designed backup controller.
\end{remark}

\subsection{Reducing the Complexity of the iMDP Policy Synthesis}
\label{subsec:algorithmic_optimization}

The \gls{iMDP} abstractions generated using our approach can potentially have hundreds of millions of transitions, especially if the state dimension is high (see \cref{tab:spacecraft,tab:DetailedResults}).
The main bottleneck that leads to such large models, is that the number of transitions in the \gls{iMDP} is (worst-case) quadratic in the number of states, and linear in the number of actions.
Indeed, note that \cref{alg:algorithm_planning} consists of a double for-loop, enumerating over both the set of actions and the set of states, which can thus be expensive. 

To alleviate this limitation, we develop an improved policy synthesis scheme that reduces the complexity of the Bellman iterations required to compute the optimal policy over the iMDP.\footnote{While the proposed scheme reduces the complexity of computing optimal policies on the \gls{iMDP}, it does not alleviate the complexity stated in \cref{thm:complexity} for generating abstractions.}
This scheme is \emph{sound}, in the sense that the maximum lower bound reach-avoid probability $\reachProbDiscrMinStar$ that we obtain under the proposed scheme can be more conservative, but the correctness guarantee in \cref{theorem:correctness} still holds. 
In practice, instead of working with one large abstraction across the whole horizon of the reach-avoid property, we work with a smaller \gls{iMDP} at each time step, by merging states that are associated to similar reach-avoid probabilities computed via the Bellman iterations. 
In what follows, we first explain how we generate and verify this reduced model for a single time step, by merging (aggregating) states with similar reach-avoid probabilities.
Thereafter, we describe how we iterate backward over the whole time horizon, as needed to synthesize the policy.

\begin{remark}[Restriction to finite horizons]
    \label{remark:improved_scheme_restriction}
    As the improved policy synthesis scheme unfolds the \gls{iMDP} over each time step, the scheme is only applicable to finite-horizon properties.
\end{remark}

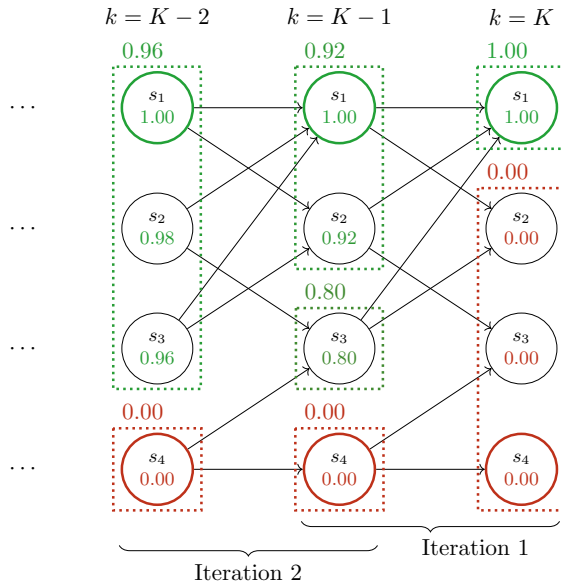
\begin{figure}[t!]
	\centering
	\resizebox{.5\linewidth}{!}{%
\begin{tikzpicture}[node distance=0.8cm]
      
      \tikzset{every state/.style={inner sep=4pt, minimum size=0pt, font=\footnotesize}};
    
      \node[state, align=center, color=darkgreen, very thick, text=black] (s1K) {$s_1$ \\ \color{darkgreen!100!red}{1.00}};
      \node[state,below=of s1K, align=center] (s2K) {$s_2$ \\ \color{darkgreen!000!red}{0.00}};
      \node[state,below=of s2K, align=center] (s3K) {$s_3$ \\ \color{darkgreen!000!red}{0.00}};
      \node[state,below=of s3K, align=center, color=red, very thick, text=black] (s4K) {$s_4$ \\ \color{darkgreen!000!red}{0.00}};
      
      \node[state,left=of s1K, xshift=-1cm, align=center, color=darkgreen, very thick, text=black] (s1K2) {$s_1$ \\ \color{darkgreen!100!red}{1.00}};
      \node[state,below=of s1K2, align=center] (s2K2) {$s_2$ \\ \color{darkgreen!92!red}{0.92}};
      \node[state,below=of s2K2, align=center] (s3K2) {$s_3$ \\ \color{darkgreen!80!red}{0.80}};
      \node[state,below=of s3K2, align=center, color=red, very thick, text=black] (s4K2) {$s_4$ \\ \color{darkgreen!00!red}{0.00}};
      
      \node[state,left=of s1K2, xshift=-1cm, align=center, color=darkgreen, very thick, text=black] (s1K3) {$s_1$ \\ \color{darkgreen!100!red}{1.00}};
      \node[state,below=of s1K3, align=center] (s2K3) {$s_2$ \\ \color{darkgreen!98!red}{0.98}};
      \node[state,below=of s2K3, align=center] (s3K3) {$s_3$ \\ \color{darkgreen!96!red}{0.96}};
      \node[state,below=of s3K3, align=center, color=red, very thick, text=black] (s4K3) {$s_4$ \\ \color{darkgreen!0!red}{0.00}};
      
      \node[above of=s1K, yshift=0.7cm] {$k = \horizon$};
      \node[above of=s1K2, yshift=0.7cm] {$k = \horizon-1$};
      \node[above of=s1K3, yshift=0.7cm] {$k = \horizon-2$};
      
      \node[left of=s1K3, xshift=-1.4cm] {$\cdots$};
      \node[left of=s2K3, xshift=-1.4cm] {$\cdots$};
      \node[left of=s3K3, xshift=-1.4cm] {$\cdots$};
      \node[left of=s4K3, xshift=-1.4cm] {$\cdots$};
      
      \draw[->] (s1K2) -- (s1K);
      \draw[->] (s1K2) -- (s2K);
      \draw[->] (s2K2) -- (s1K);
      \draw[->] (s2K2) -- (s3K);
      \draw[->] (s3K2) -- (s1K);
      \draw[->] (s3K2) -- (s2K);
      \draw[->] (s4K2) -- (s3K);
      \draw[->] (s4K2) -- (s4K);
      
      \draw[->] (s1K3) -- (s1K2);
      \draw[->] (s1K3) -- (s2K2);
      \draw[->] (s2K3) -- (s1K2);
      \draw[->] (s2K3) -- (s3K2);
      \draw[->] (s3K3) -- (s1K2);
      \draw[->] (s3K3) -- (s2K2);
      \draw[->] (s4K3) -- (s3K2);
      \draw[->] (s4K3) -- (s4K2);
      
      \draw [darkgreen!100!red,very thick,dotted] ($(s1K.north west)+(-0.3,0.25)$) node[above right] {1.00} rectangle ($(s1K.south east)+(0.3,-0.25)$);
      \draw[darkgreen!0!red,very thick,dotted] ($(s2K.north west)+(-0.3,0.25)$) node[above right] {0.00} rectangle ($(s4K.south east)+(0.3,-0.25)$);
      
      \draw[darkgreen!92!red,very thick,dotted] ($(s1K2.north west)+(-0.3,0.25)$) node[above right] {0.92} rectangle ($(s2K2.south east)+(0.3,-0.25)$);
      \draw[darkgreen!80!red,very thick,dotted] ($(s3K2.north west)+(-0.3,0.25)$) node[above right] {0.80} rectangle ($(s3K2.south east)+(0.3,-0.25)$);
      \draw[darkgreen!00!red,very thick,dotted] ($(s4K2.north west)+(-0.3,0.25)$) node[above right] {0.00} rectangle ($(s4K2.south east)+(0.3,-0.25)$);
      
      \draw[darkgreen!96!red,very thick,dotted] ($(s1K3.north west)+(-0.3,0.25)$) node[above right] {0.96} rectangle ($(s3K3.south east)+(0.3,-0.25)$);
      \draw[darkgreen!0!red,very thick,dotted] ($(s4K3.north west)+(-0.3,0.25)$) node[above right] {0.00} rectangle ($(s4K3.south east)+(0.3,-0.25)$);
      
      \draw [decorate,decoration={brace,amplitude=5pt,mirror,raise=2ex}]
      ($(s4K2.south west)+(-0.2,-0.1)$)  -- ($(s4K.south east)+(0.2,-0.1)$) node[midway, below, yshift=-.5cm, xshift=0.75cm]{Iteration 1};
      
      \draw [decorate,decoration={brace,amplitude=5pt,mirror,raise=2ex}]
      ($(s4K3.south west)+(-0.2,-0.5)$)  -- ($(s4K2.south east)+(0.2,-0.5)$) node[midway, below, yshift=-.5cm]{Iteration 2};

\end{tikzpicture}
}
	\caption{
	    The improved policy synthesis scheme (for clarity, we have omitted actions), in which we merge states based on their lower bound reach-avoid probabilities, computed by \cref{eq:optimalPolicy_lb} (shown for $\rho=10$ here). 
	    These probabilities are shown within the nodes of each state, while merged states are shown as dotted regions with their lower bound reach-avoid probability written above.
	    If a state has multiple outgoing transitions to the same merged state, e.g., $s_4$ at time $k = \horizon-1$, we replace these transitions with a single transition whose probability interval is computed as the sum over the original lower/upper bounds.
	}
	\label{fig:optimization_imdp}
\end{figure}

\paragraph{State aggregation for a single time step.}
As an illustrative example, consider the  \gls{iMDP} in \cref{fig:optimization_imdp}, which consists of four states: $s_1$ is a goal state, $s_4$ is a critical state, and $s_2$ and $s_3$ are neither.
Recall that $\horizon$ denotes the time horizon of the reach-avoid property.
If we are at time step $k = \horizon - 1$ (i.e., only one action to go), the only way to satisfy the reach-avoid problem is to reach state $s_1$. 
In other words, the reward (being the probability of satisfying the reach-avoid property) of reaching state $s_1$ is one, while the reward of the other states is zero.
Based on this intuition, we can merge states $\{s_2, s_3, s_4\}$ as a single successor state in the \gls{iMDP} (at the final time step, it does not matter if we end up in a critical state or in any other non-goal state: the property is not satisfied in either case).
More generally, we \emph{partition} the \gls{iMDP} states into $\rho \in \N$ discrete \emph{bins}, based on their lower bound reach-avoid probability.
For example, if we choose $\rho = 100$ (bins of size $1\%$), then we have at most $100$ successor states, which is often significantly less than the term $| \States |$ in the original scheme.
The reach-avoid probability of a merged state is the minimum of the lower bound reach-avoid probabilities of the original states it is comprised of.
The probability interval $[\munderbar{p}, \bar{p}]$ of transitioning under state-action pair $(s,a)$ to merged state comprised of a subset of states $M \subset \States$ is the union of the intervals over all states in $M$:
\begin{equation}
    \munderbar{p} = \sum_{s' \in M} \transfuncImdp(s,a,s')_{\text{low}}, \ \
    \bar{p} = \sum_{s' \in M} \transfuncImdp(s,a,s')_{\text{up}},
\end{equation}
where $\transfuncImdp(s,a,s')_{\text{low}}$ and $\transfuncImdp(s,a,s')_{\text{up}}$ denote the lower and upper bounds of these intervals.

\paragraph{Iterating backward over multiple time steps.}
Given the state aggregation described above for a certain time step, e.g., $k=\horizon$, we can compute the maximum lower bound reach-avoid probabilities $\reachProbDiscrMinStar$ and the corresponding optimal policy $\munderbar{\policy}^\star$ at time $\horizon-1$.
For each state $s \in \States$, we store the optimal action of the policy as $\munderbar{\policy}^\star(s, \horizon-1)$.
We then go one step backward in time (according to Bellman iterations) and create another state aggregation over the time steps $k = \horizon-2$ to $\horizon-1$.
We again partition the range of reach-avoid probabilities (this time those at time $k = \horizon-1$) into $\rho$ discrete bins and merge the successor states accordingly.
As shown in \cref{fig:optimization_imdp}, we thus \emph{iterate backward in time} to compute the optimal policy and lower bound reachability probabilities, until we have reached the initial time step of $k=0$, and thus have computed the whole policy $\munderbar{\policy}^\star$.

\paragraph{Soundness of the improved scheme.}
Due to the optimal policy being a Markovian mapping from the state and time step, we can decompose the policy synthesis into individual time steps by iterating backward in time.
\cref{theorem:correctness} is preserved under the improved policy synthesis scheme due to two reasons: 1) the partitioning of states based on their lower bound reach-avoid probabilities leads to an \emph{under-approximation} of the actual reach-avoid probabilities, and 2) summing up the probability intervals over merged states \emph{preserves the original \gls{PAC} guarantees} on the original abstract model.
However, under the proposed scheme, the number of unique intervals increases linearly with the time horizon $\horizon$.
In particular, the \glspl{iMDP} for all $\horizon$ steps combined have at most $|\Actions| \rho \horizon$ unique probability intervals (rather than $|\Actions| \cdot |\States|$, cf. \cref{remark:number_probabilities}). 
Hence, under the proposed scheme, we change the confidence parameter in \cref{theorem:correctness} to $\alpha = \beta \rho \horizon \cdot |\Actions|$.
Under this modification of the confidence parameter, the proposed scheme is sound. 

\paragraph{Using $\rho$ as a tuning parameter.}
Under the proposed scheme, $\rho$ is a tuning parameter that provides a trade-off between the size of the state space obtained from the \glspl{iMDP}, versus the level of conservatism in the obtained reach-avoid guarantees introduced by aggregating states.
A typical choice for the tuning parameter $\rho$ is $\rho = 100$ (i.e., partitioning states based on their lower bound reach-avoid probabilities with a precision of $1\%$ in reach-avoid probability).
If $|\States| > \rho \horizon$, then the worst-case number of transitions under the proposed improved policy synthesis scheme is lower than the worst-case number of transitions under the original scheme. 
However, even if this condition is not satisfied, the improved scheme may be beneficial, because the memory requirements for solving the original \gls{iMDP} can be excessively large, as demonstrated in the satellite rendezvous benchmark in \cref{subsec:spacecraft}.
\section{Numerical Examples}

\label{sec:Studies}

We implement our iterative abstraction method in Python, and tailor the model checker \prism~\cite{DBLP:conf/cav/KwiatkowskaNP11} for \glspl{iMDP} to compute robust optimal policies.
At every iteration, the obtained \gls{iMDP} is fed to \prism, which computes the optimal policy associated with the maximum reach-avoid probability, as per \cref{eq:optimalPolicy_lb}.
Our code is available via \color{Sepia}\url{https://github.com/LAVA-LAB/DynAbs}\color{black}, and all experiments are run on a computer with 32 3.7GHz cores and 64 GB of RAM.
We report the performance of our method on: (1) a \gls{UAV} motion control, (2) a building temperature regulation, and (3) a new satellite rendezvous benchmark, which was not in the earlier paper by \citeA{Badings2022AAAI}.
In addition, we benchmark our method against \sreachtools and \stochy, which are two other tools for controller synthesis based on formal abstractions.
Finally, we demonstrate the applicability of our techniques to infinite-horizon properties.

\begin{figure}[t!]
\centering
\begin{minipage}[b]{.48\textwidth}
    \centering
    \includegraphics[width=\linewidth]
    {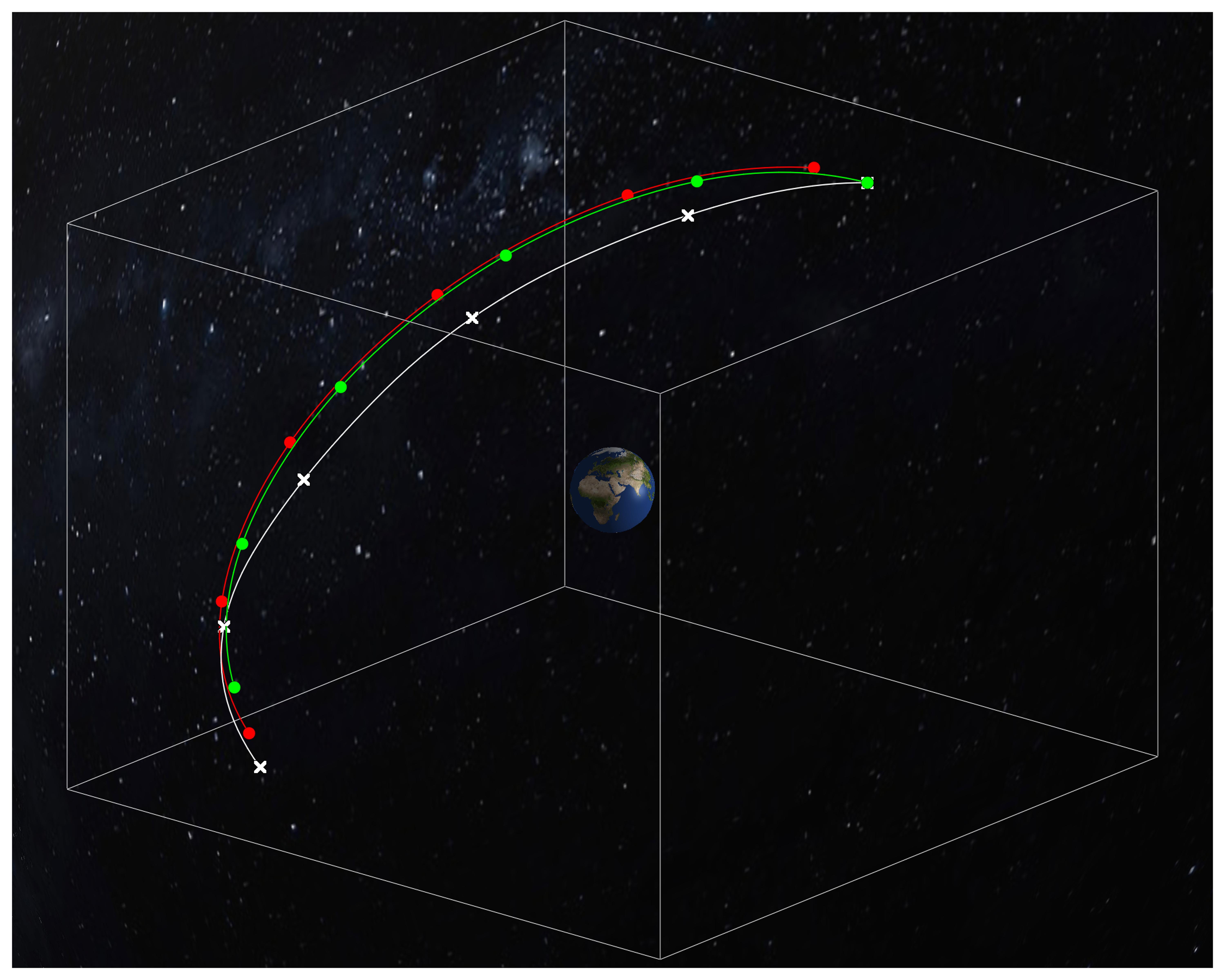}
    \captionof{figure}{State trajectory for the satellite benchmark ($N=3\,200$ with improved policy synthesis scheme). The chaser satellite (white) must navigate to the target (green) while not colliding with the one in red.}
    \label{fig:spacecraft_orbit}
\end{minipage}\hfill
\begin{minipage}[b]{.48\textwidth} 
    \centering
    \includegraphics[width=\linewidth]
    {Figures/Results/UAV/UAV_v3.png}
    \captionof{figure}{UAV reach-avoid problem (goal in green; obstacles in red), plus trajectories under the optimal \gls{iMDP}-based controller from initial state $\bm{x}_0 = [-14,0,6,0,-6,0]^\top$, under high and low turbulence.}
    \label{fig:UAV_layout}
\end{minipage}
\end{figure}

\subsection{Satellite Rendezvous Problem}
\label{subsec:spacecraft}

We consider the satellite benchmark problem from~\citeA{DBLP:conf/cdc/JewisonE16}.
More specifically, we consider phase 1 of their satellite rendezvous problem, in which a \emph{chaser} satellite needs to dock with a \emph{target} satellite while being in orbit.
The relative motion of the chaser satellite with respect to the target is modeled by the so-called Clohessy-Wiltshire-Hill (CWH) equations~\cite{clohessy1960terminal}.
We present the full 6-dimensional linear dynamical system in \cref{subappendix:spacecraft_model}.
Notably, we partition the 6-dimensional state space into $11 \times 23 \times 5 \times 5 \times 5 \times 5 = 158\,125$ discrete regions and define the same number of actions.
We define a time horizon of $\horizon=16$ steps, which becomes $8$ steps after grouping every two discrete time steps as described in \cref{sec:Abstraction}; see \cref{subappendix:spacecraft_model} for details.
The original problem from~\citeA{DBLP:conf/cdc/JewisonE16} is a reachability problem, which we extend to a reach-avoid problem by adding a third satellite that must be avoided (as shown in \cref{fig:spacecraft_orbit}).
We assume that this third satellite is located between the chaser and target satellite and has a fixed position in the CWH frame, yielding a stationary critical region.\footnote{Note that, in principle, we can also model moving obstacles (or goal regions) by modeling time explicitly in the iMDP abstraction and changing the set of critical (goal) regions at each time step.}

\begin{table*}[t!]

\caption{Model sizes and run times for the spacecraft benchmark, with $N=3\,200$ and $20\,000$ noise samples, and either the default abstraction scheme (which generates a single \gls{iMDP}) or the improved policy synthesis scheme proposed in \cref{subsec:algorithmic_optimization} (which generates an \gls{iMDP} for every time step $k=0,\ldots,\horizon$ of the property horizon, of which we show every second step in the table). The default scheme with $N=20\,000$ leads to a memory overflow.}
\label{tab:spacecraft}

\small{
\begin{tabularx}{\textwidth}{@{} lll|r|rrr @{} } \toprule
    & &
    & \multicolumn{1}{c|}{\textbf{iMDP size}} 
    & \multicolumn{3}{c}{\textbf{Run times [s]}}
    \\
    $N$ & \multicolumn{2}{l|}{Policy synthesis scheme} & Transitions & Comp. intervals & Write iMDP & Compute $\munderbar{\policy}^*$ \\ \midrule
    3.2k & \multicolumn{2}{l|}{Default (Single \gls{iMDP})} & 338\,847\,144 & 301.23 & 1103.87 & 1040.42 \\
    \midrule
    3.2k & Improved synthesis & $k=8$ & 813\,724 & 249.46 & 82.79 & 136.41 \\
    & scheme 
    & $k=6$ & 26\,892\,509 & 29.39 & 160.64 & 163.73 \\
    & & $k=4$ & 43\,868\,194 & 31.80 & 212.21 & 180.77 \\
    & & $k=2$ & 51\,816\,689 & 34.85 & 251.02 & 195.52 \\
    \midrule
    20k & \multicolumn{2}{l|}{Default (Single \gls{iMDP})} & 560\,426\,004 & -- & -- & -- \\
    \midrule
    20k &  Improved synthesis & $k=8$ & 1\,156\,654 & 1654.64 & 86.61 & 142.70 \\
    & scheme 
    & $k=6$ & 31\,914\,917 & 142.52 & 188.10 & 177.27 \\
    & & $k=4$ & 52\,517\,316 & 144.88 & 253.31 & 193.56 \\
    & & $k=2$ & 62\,934\,064 & 148.10 & 286.33 & 203.09 \\
    \bottomrule
    
\end{tabularx}
}

\end{table*}

\paragraph{Correct-by-construction control with PAC guarantees.}
First, let us show how to use \cref{theorem:correctness} in practice to determine the confidence parameter $\beta$ needed on individual transition probabilities to solve the overall problem statement with a desired confidence probability.
We choose an overall confidence probability of $1-\alpha = 0.95$.
Since we use a uniform rectangular partition of the state space, we can use \cref{corollary:tighter_confidence} to compute the corresponding confidence parameter $\beta$ on individual transitions.
For this particular experiment, we find that a confidence parameter of $\beta = \frac{\alpha}{(22-1)(46-1)(10-1)^4 + 158,125} = \num{7.86e-9}$ is sufficient to obtain an overall confidence of $1-\alpha=0.95$.

\paragraph{Improved policy synthesis scheme solves larger problems.}
We apply our method with $N=3\,200$ and $20\,000$ samples, either with or without the improved policy synthesis scheme proposed in \cref{subsec:algorithmic_optimization}.
One simulated state trajectory of the dynamical system ($N=3\,200$ and with the improved synthesis scheme enabled) is shown in \cref{fig:spacecraft_orbit}.
The figure shows that the chaser satellite (in white) is indeed able to navigate to the target (in green) without colliding with the third satellite shown in red.
For this particular case, the reach-avoid guarantee on the \gls{iMDP} is $\reachProbDiscrMinStar = 0.56$, while the empirical satisfaction of the reach-avoid property (obtained via Monte Carlo simulations) is $0.74$.\footnote{Note that we have deliberately chosen a high process noise strength in this benchmark, leading to relatively lower reach-avoid probabilities.}

The number of transitions $(s,a,s')$ of the \glspl{iMDP} and the run times for all cases are presented in \cref{tab:spacecraft}.
The time to compute the states and (enabled) actions is around $\SI{30}{\minute}$ and is equal for all cases.
Under the default synthesis scheme, we generate a single \gls{iMDP} over the whole horizon of the reach-avoid property, resulting in very large \glspl{iMDP} of about $338$ and $560$ million transitions for $N=3\,200$ and $20\,000$, respectively.
In the latter instance, the default policy synthesis scheme failed due to a memory overflow (note that we used $64$ GB of RAM).
By contrast, the improved policy synthesis scheme, which generates a significantly smaller \gls{iMDP} for each time step $k=0,\ldots,\horizon$ of the horizon, is able to solve both cases, even though the total run time (over all iterations $k=0,\ldots,\horizon$) for $N=3\,200$ is around $20\%$ higher than with the default scheme.
As shown in \cref{tab:spacecraft}, the size of the \glspl{iMDP} under the improved synthesis scheme increase per iteration, since there is more variety among the reach-avoid probabilities, and thus we can aggregate fewer states.
Nevertheless, the results show that with the improved policy synthesis scheme, we can solve reach-avoid problems leading to \glspl{iMDP} that would otherwise be infeasibly large.

\subsection{UAV Motion Planning}
\label{subsec:UAV}

We consider the reach-avoid problem for a \gls{UAV} operating under turbulence, which was introduced in \cref{sec:Introduction}.
The goal is to compute a controller that guarantees (with high confidence) that the probability to reach a goal area while also avoiding unsafe regions, is above a performance threshold of $\eta = 0.75$.
We consider a horizon of $64$ time steps, and the problem layout is displayed in \cref{fig:UAV_layout}, with the goal and unsafe regions shown in green and red, respectively.
We model the \gls{UAV} as a system of 3 double integrators
(see \cref{appendix:UAV} for details).
The state $\bm{x}_k \in \R^6$ encodes the position and velocity components, and control inputs $\bm{u}_k \in \R^3$ model actuators that change the velocity.
The effect of turbulence on the state causes (non-Gaussian) process noise, which we model using a Dryden gust model~\cite{bohn2019deep,dryden1943review}.
We compare two cases: 1) a low turbulence case, and 2) a high turbulence case.
We partition the state space into $25\,515$ regions, using \cref{theorem:Bounds} with $\beta=0.01$, and apply the iterative scheme with $\gamma = 2$, starting at $N = 25$, with an upper bound of $12\,800$ samples.

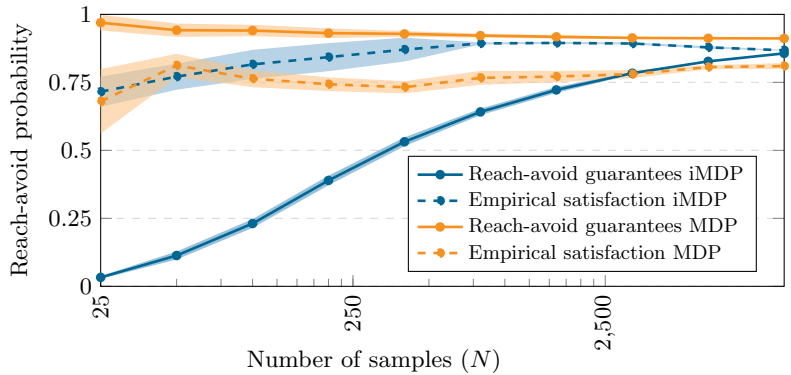
\begin{figure}[t!]
    \centering
        \footnotesize\begin{tikzpicture}
      \begin{axis}[
          width=.7\linewidth,
          height=5.2cm,
          ymajorgrids,
          grid style={dashed,gray!30},
          xlabel=Number of samples ($N$),
          ylabel=Reach-avoid probability,
          x label style={at={(axis description cs:0.4,-0.2)}},
          x tick label style={rotate=90,anchor=east},
          xmode=log,
          log ticks with fixed point,
          xmin=25,
          xmax=12800,
          ymin=0,
          ymax=1,
          xtick={25,250,2500},
          xtick pos=left,
          ytick={0,0.25,...,1.0},
          every axis plot/.append style={line width=1pt},
          legend cell align={left},
          legend columns=1,
          legend style={at={(0.96,0.48)},
                        nodes={scale=0.85, transform shape},
                        anchor=north east, 
                        column sep=0ex,}
        ]
        
        \addplot[mark=*, mark size=1.3pt, color=MidnightBlue] table[x=x, y=imdp, col sep=comma] {Figures/Results/UAV/MDP_vs_iMDP_v3.csv};
        
        \addplot [name path=upper_imdp, draw=none] table[x=x, y=imdp_error_plus, col sep=comma, forget plot] {Figures/Results/UAV/MDP_vs_iMDP_v3.csv};
        \addplot [name path=lower_imdp, draw=none] table[x=x, y=imdp_error_min, col sep=comma, forget plot] {Figures/Results/UAV/MDP_vs_iMDP_v3.csv};
        \addplot [fill=MidnightBlue!30, forget plot] fill between[of=upper_imdp and lower_imdp];
        
        \addplot[mark=*, mark size=1.3pt, color=MidnightBlue, dashed] table[x=x, y=imdp_sim, col sep=comma] {Figures/Results/UAV/MDP_vs_iMDP_v3.csv};
        
        \addplot [name path=upper_imdp_sim, draw=none] table[x=x, y=imdp_sim_error_plus, col sep=comma, forget plot] {Figures/Results/UAV/MDP_vs_iMDP_v3.csv};
        \addplot [name path=lower_imdp_sim, draw=none] table[x=x, y=imdp_sim_error_min, col sep=comma, forget plot] {Figures/Results/UAV/MDP_vs_iMDP_v3.csv};
        \addplot [fill=MidnightBlue!20, forget plot, fill opacity=1.0] fill between[of=upper_imdp_sim and lower_imdp_sim];
        
        
         \addplot[mark=*, mark size=1.3pt, color=BurntOrange] table[x=x, y=mdp, col sep=comma] {Figures/Results/UAV/MDP_vs_iMDP_v3.csv};
        
        \addplot [name path=upper_mdp, draw=none] table[x=x, y=mdp_error_plus, col sep=comma, forget plot] {Figures/Results/UAV/MDP_vs_iMDP_v3.csv};
        \addplot [name path=lower_mdp, draw=none] table[x=x, y=mdp_error_min, col sep=comma, forget plot] {Figures/Results/UAV/MDP_vs_iMDP_v3.csv};
        \addplot [fill=BurntOrange!30, forget plot] fill between[of=upper_mdp and lower_mdp];
        
        \addplot[mark=*, mark size=1.3pt, color=BurntOrange, dashed] table[x=x, y=mdp_sim, col sep=comma] {Figures/Results/UAV/MDP_vs_iMDP_v3.csv};
        
        \addplot [name path=upper_mdp_sim, draw=none] table[x=x, y=mdp_sim_error_plus, col sep=comma, forget plot] {Figures/Results/UAV/MDP_vs_iMDP_v3.csv};
        \addplot [name path=lower_mdp_sim, draw=none] table[x=x, y=mdp_sim_error_min, col sep=comma, forget plot] {Figures/Results/UAV/MDP_vs_iMDP_v3.csv};
        \addplot [fill=BurntOrange!50, forget plot, fill opacity=0.6] fill between[of=upper_mdp_sim and lower_mdp_sim];

        \legend{{Reach-avoid guarantees iMDP},{Empirical satisfaction iMDP}, {Reach-avoid guarantees MDP}, {Empirical satisfaction MDP }}
      \end{axis}
    \end{tikzpicture}
    \caption{Reach-avoid guarantees on the \glspl{iMDP} (blue) and \glspl{MDP} (orange) for their respective policies, versus the resulting empirical (simulated) performance (dashed lines) on the dynamical system. 
    Shaded areas show the standard deviation across 10 iterations.
    The empirical performance obtained from the \glspl{MDP} violates the guarantees, whereas that from the \glspl{iMDP} does not.}    
    \label{fig:MDP_vs_iMDP}
\end{figure}

\paragraph{Scalability.}
We report the model sizes and run times in \cref{subappendix:UAV_results}.
The number of \gls{iMDP} states equals the size of the partition.
Depending on the number of samples $N$, the \gls{iMDP} has $9-24$ million transitions.
The mean time to compute the set of \gls{iMDP} actions (which is only done in the first iteration) is around one minute.\footnote{We exploit the block-diagonal structure of matrices $A$ and $B$ of the dynamical system to speed up the computation of the enabled state-action pairs. In particular, we can do all necessary reachability computations separately in the spatial $(x,y,z)$ dimensions and compose the full \gls{iMDP} afterward.}
Computing the probabilities plus the verification in \prism takes $1-\SI{8}{\minute}$, depending on the number of samples $N$.

\paragraph{Accounting for noise matters for probabilistic safety.}
In \cref{fig:UAV_layout}, we show state trajectories under the optimal controller derived from the optimal \gls{iMDP} policy, under high and low turbulence (noise).
Under low noise, the controller prefers the short but narrow path; however, with the high noise level, the longer but safer path is preferred since the risk of colliding with an obstacle is too high. 
Thus, accounting for process noise is important to obtain controllers that are safe.  

\paragraph{iMDPs yield safer guarantees than MDPs.} 
To show the importance of using robust abstractions, we compare, under high turbulence, our robust \gls{iMDP} approach against a na\"ive \gls{MDP} abstraction.
This \gls{MDP} has the same states and actions as the \gls{iMDP}, but uses precise probabilities, which are computed using the frequentist approach introduced in \cref{sec:ScenarioApproach}.
The maximum reach-avoid probabilities (guarantees) for both methods are shown in \cref{fig:MDP_vs_iMDP}.
For every value of $N$, we apply the resulting controllers to the dynamical system in Monte Carlo simulations with $10\,000$ iterations, to determine the empirical reach-avoid probability as the fraction of trajectories that satisfies the reach-avoid property.
\cref{fig:MDP_vs_iMDP} shows that the non-robust \glspl{MDP} yield \emph{poor and unsafe performance guarantees}: the actual reach-avoid probability of the controller on the dynamical system is much lower than the reach-avoid guarantees obtained from \prism.
By contrast, our robust \gls{iMDP}-based approach consistently yields safe lower bound guarantees on the actual performance of controllers.
The performance threshold of $\reachProbDiscrMinStar \geq 0.75$ is guaranteed for $N = 3\,200$ and higher.

\subsection{Building Temperature Regulation}
\label{subsec:building_temperature}
Inspired by~\citeA{DBLP:journals/corr/abs-1803-06315}, we consider a temperature control problem for a building with two rooms, both having their own radiator and air supply.
The reach-avoid goal is to maximize the probability to reach a temperature within $19.8-20.2$\SI{}{\celsius} in both zones within 32 steps of 15 minutes each, while avoiding temperatures below $17.8$ or above $\SI{22.2}{\celsius}$.
The state $\bm{x}_k \in \R^4$ of the system (see \cref{appendix:BAS} for details) reflects the temperatures of both zones and radiators, and control inputs $\bm{u}_k \in \R^4$ change the air supply and boiler temperatures in both zones.
The deterministic heat gain through zone walls is modeled by the disturbance $\bm{q}_k \in \R^4$.
The noise $\bm{w}_k \in \R^4$ has a Gaussian distribution (but this assumption is not required for our approach).
We partition the state space into $35\,721$ regions ($21$ values for zone temperatures and $9$ for radiator temperatures), and we use the same values for $\beta$ and $N$ as in the \gls{UAV} benchmark in \cref{subsec:UAV}.

\begin{figure}[t!]
    \centering
    \includegraphics[width=\linewidth]{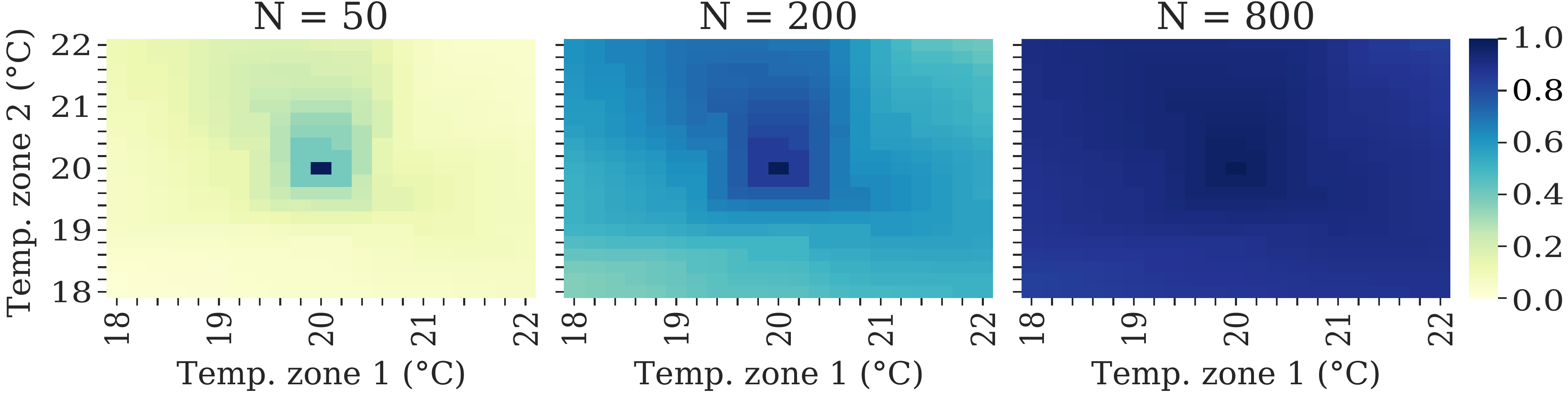}
    \caption{Cross section (for radiator temp. of \SI{38}{\celsius}) of the maximum lower bound probabilities to reach the goal of \SI{20}{\celsius} from any initial state, for either $50$, $200$ or $800$ samples.}
    \label{fig:BAS_results}
\end{figure}

\paragraph{More samples means less uncertainty.}
In \cref{fig:BAS_results}, we show (for fixed radiator temperatures) the maximum lower bound probabilities obtained from \prism, to reach the goal from any initial state within the safe set.
The results clearly show that better reach-avoid guarantees are obtained when more samples are used to compute the \gls{iMDP} probability intervals.
The higher the value of $N$, the lower the uncertainty in the intervals, leading to better reach-avoid guarantees.
Notably, as reported in \cref{subappendix:BAS_results}, the largest \gls{iMDP} has around $200$ million transitions, showing that our approach can effectively generate and verify large abstract models.

\subsection{Benchmarks Against Other Control Synthesis Tools}

\begin{figure}[t!]
    \centering
    \includegraphics[width=.65\linewidth]{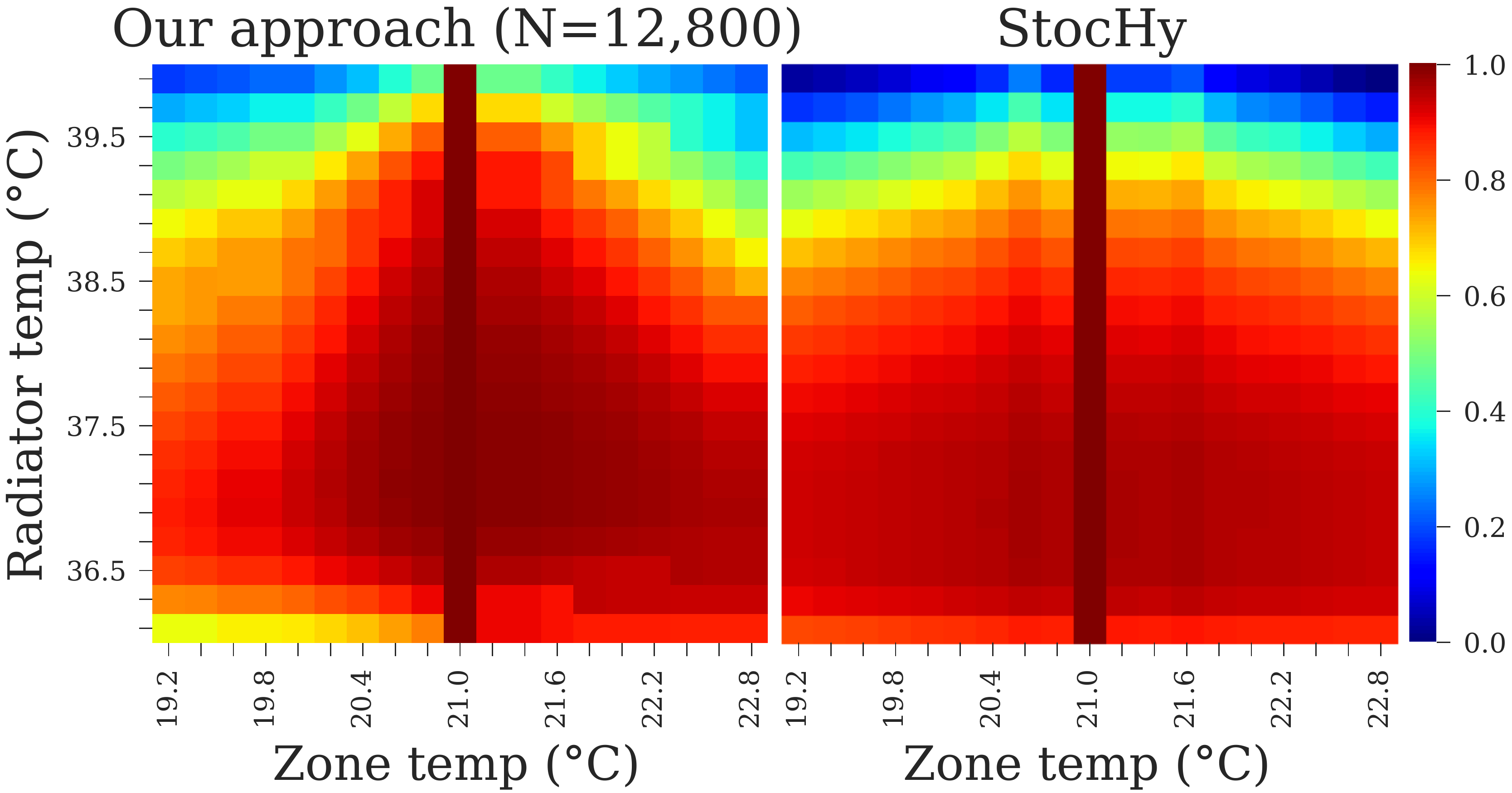}
    \caption{Maximum lower bound probabilities to reach the goal zone temperature of $\SI{21}{\celsius}$ from any initial state within $64$ steps, for our approach ($N=12\,800$) and \stochy.}
    \label{fig:BAS_stochy}
\end{figure}

\paragraph{StocHy.} 
We benchmark our method on a building temperature regulation problem against \stochy~\cite{DBLP:conf/tacas/CauchiA19}, a verification and synthesis tool based on formal abstractions (see \cref{appendix:StocHy} for details on the setup and results).
Similar to our approach, \stochy also derives robust \gls{iMDP} abstractions.
However, \stochy requires precise knowledge of the noise distribution, and it discretizes the control input space of the dynamical system, to obtain a finite action space.
The maximum probabilities to reach the goal zone temperature from any initial state obtained for both methods are presented in \cref{fig:BAS_stochy}.
The obtained results are qualitatively similar and, close to the goal zone temperature, our lower bound reach-avoid guarantees are \emph{slightly higher} than those obtained from \stochy.
However, when starting at temperatures close to the boundary (e.g., at both low radiator and zone temperature), the guarantees obtained from our approach are \emph{slightly more conservative}.
This is due to the fact that our approach relies on \gls{PAC} guarantees on the transition probabilities, while \stochy gives straight probabilistic outcomes, thanks to the assumed precise knowledge of the noise distribution.
While both methods yield results that are qualitatively similar, our approach is an order of magnitude faster ($\SI{45}{\minute}$ for \stochy, vs. $3-\SI{9}{\second}$ for our approach; see \cref{appendix:StocHy,tab:DetailedResults} for details). 

\begin{figure}[t!]
    \centering
    \includegraphics[width=.7\linewidth]{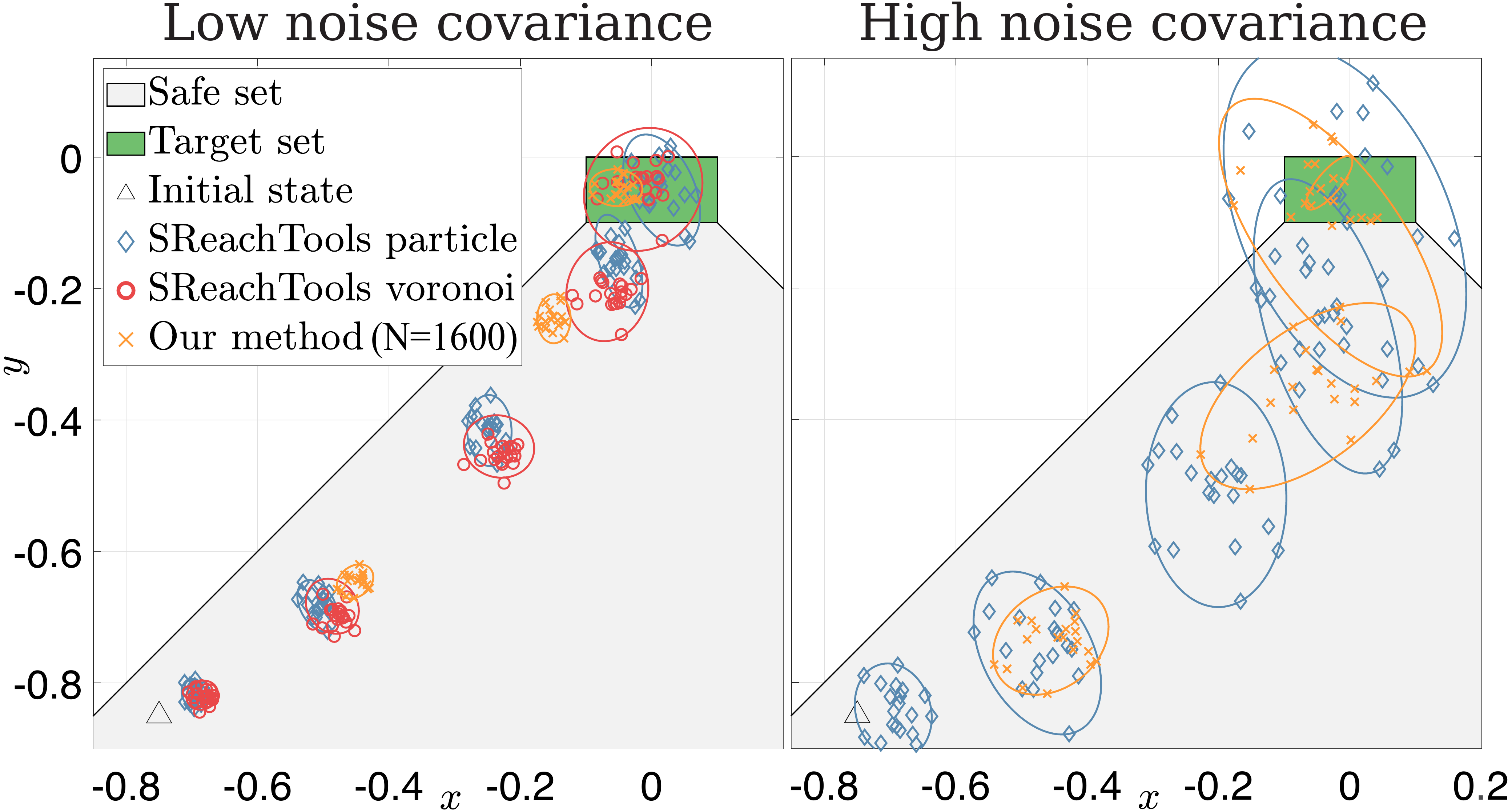}
    \caption{Simulated state trajectories for the spacecraft docking problem, under low and high noise covariance.
    Our feedback controllers are more robust, as shown by the smaller error in the state trajectories over time (the Voronoi method under high covariance failed to generate a solution).}
    \label{fig:SReachTools_results}
\end{figure}

\paragraph{\sreachtools.}
We apply our method to the spacecraft docking benchmark\footnote{Note that the dynamical system used in this spacecraft docking benchmark differs significantly from the system used in the satellite rendezvous problem in \cref{subsec:spacecraft}; see \cref{appendix:spacecraft,appendix:SReachTools} for details.} (see \cref{fig:SReachTools_results}) of \sreachtools~\cite{DBLP:conf/hybrid/VinodGO19}, an optimization-based toolbox for probabilistic reach-avoid problems (see \cref{appendix:SReachTools} for details).
While we use samples to generate a model abstraction, \sreachtools employs sample-based methods over the properties directly.
Distinctively, \sreachtools does not create abstractions (as in our case) and is thus generally faster than our method.
However, its complexity is exponential in the number of samples (versus the linear complexity for our method).
Importantly, we derive \emph{feedback} controllers, while the sampling-based methods of \sreachtools compute \emph{open-loop} controllers.
Feedback controllers respond to state observations over time and are, therefore, more robust against strong disturbances from noise, as also shown in \cref{fig:SReachTools_results}.

\begin{figure}[t!]
    \centering
    \includegraphics[width=\linewidth]{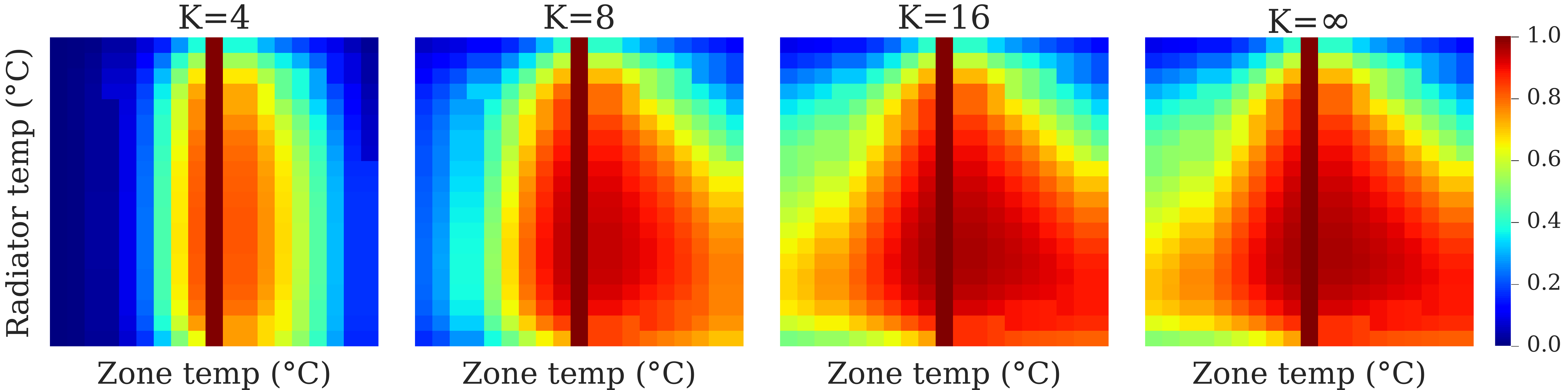}
    \caption{Obtained lower bound reach-avoid guarantees for the 1-room temperature control problem with time horizons between $\horizon=4$ steps (left) and infinity (right).}
    \label{fig:BAS_increasing_horizon}.
\end{figure}

\subsection{Infinite-horizon properties}
To demonstrate that our techniques are also applicable to infinite-horizon properties, we consider again the 1-room version of the temperature control problem (on which we benchmarked against \stochy).
We consider reach-avoid properties with increasing time bounds of $\horizon \in \{ 4, 8, 16, \infty \}$ and show the corresponding maximum lower bounds on the reach-avoid probability in \cref{fig:BAS_increasing_horizon}.
As expected, these figures show that the reach-avoid probability increases with the time horizon (even though the difference between $\horizon=16$ and $\horizon=\infty$ is marginal).
The time to verify the \gls{iMDP} (which has $383$ states and $64\,029$ transitions) in \prism is about $\SI{1.7}{\second}$ for the property with a horizon of $K=4$ steps, versus $\SI{3.4}{\second}$ for the infinite-horizon property.
\section{Related Work}
\label{sec:Related}

We summarize the literature most related to this paper.
We start with a discussion of robust \glspl{MDP}, (probabilistic) bisimulation, and controller synthesis with formal guarantees.
Then, we introduce the scenario approach, as well as other sampling techniques and distributionally robust optimization.
Finally, we discuss the concept of \gls{PAC} and briefly touch upon safe and robust learning methods.

\paragraph{Robust MDPs.}
In robust (also called uncertain) \glspl{MDP}, the transition function (in particular, each distribution $P(s,a)$ over successor states, cf. \cref{def:MDP}) is only known to be an element of an uncertainty set~\cite{DBLP:journals/ior/WiesemannKS14,DBLP:conf/nips/XuM10}.
This set is often assumed to be convex for computational reasons, but generalizations to nonconvex sets exist as well~\cite{DBLP:journals/ior/NilimG05}.
An \gls{iMDP} is a special case of robust \gls{MDP}, where the uncertainty set is described by interval constraints on each transition probability.
Specifically, \glspl{iMDP} have uncertainty sets as \emph{probability simplexes with interval constraints}, which are convex polytopes by construction~\cite{DBLP:books/degruyter/Ben-TalGN09}.

\paragraph{Probabilistic bisimulation.}
Bisimulation is a key concept used to establish equivalence in the behavior between different systems with respect to a particular logic property~\cite{DBLP:books/daglib/BaierKatoen2008}.
Probabilistic (or stochastic) notions of bisimulation have been studied for \glspl{MDP}~\cite{DBLP:conf/birthday/FernsPK14,DBLP:journals/ai/GivanDG03} and other Markov models~\cite{DBLP:journals/iandc/LarsenS91,DBLP:journals/iandc/DesharnaisEP02}.
Since exact probabilistic bisimulation requires transition probabilities to agree \emph{exactly}, various metrics and notions of approximate bisimulation have been developed as well~\cite{DBLP:conf/aaai/FernsPP04,DBLP:journals/entcs/Abate13}.
More recently, probabilistic bisimulation has been studied for interval \glspl{MDP}~\cite{DBLP:conf/lata/HashemiH0STW16,DBLP:conf/qest/HahnHHT16}, and notions of approximate probabilistic bisimulation have been leveraged for robust model checking of probabilistic computation tree logic (PCTL) for interval Markov models~\cite{DBLP:conf/hybrid/DInnocenzoAK12,DBLP:journals/corr/HashemiHK14,DBLP:conf/adhs/LunWDA18}.

In \cref{lemma:simulation_relation}, we establish a probabilistic feedback refinement relation from the abstract \gls{MDP} to the dynamical system.
This relation is exact and yields a \emph{simulation} rather than a \emph{bisimulation} relation: due to the abstraction, not every move by the dynamical system can be matched by the abstract \gls{MDP}.
Estimating the abstract \gls{MDP} by an \gls{iMDP} is, on the other hand, approximate since (as we have shown in \cref{theorem:correctness}) there is a probability of at most $\beta$ that any of the probability intervals is incorrect.
As such, our approach creates a probabilistic closeness guarantee on the satisfaction of a reach-avoid property between the dynamical system and the \gls{iMDP}, similar to the relations presented in~\cite{LSAZ21}

\paragraph{Formal controller synthesis.}
Formal verification and controller synthesis for reachability and reach-avoid problems in stochastic continuous-state systems is an active field of research in safety-critical engineering~\cite{Abate2008probabilisticSystems,LSAZ21}.
Most approaches are either based on formal abstractions of these continuous systems~\cite{Alur2000,DBLP:journals/tac/LahijanianAB15,DBLP:journals/siamads/SoudjaniA13} or work in the continuous domain directly, e.g., using Hamilton-Jacobi reachability analysis~\cite{DBLP:conf/cdc/BansalCHT17,DBLP:conf/cdc/HerbertCHBFT17} or optimization~\cite{Rosolia2020unified}.
Several controller synthesis tools have been developed in this research area, such as \stochy~\cite{DBLP:conf/tacas/CauchiA19}, \probreach~\cite{DBLP:conf/hybrid/ShmarovZ15} and \sreachtools~\cite{DBLP:conf/hybrid/VinodGO19}.
Despite intense activity, however, the majority of these papers and tools rely on \emph{full knowledge} of the probabilistic models.

\paragraph{The scenario approach.}
Building upon this motivation, we break away from this literature by developing a method that can be used to generate formal abstractions \emph{without requiring any knowledge of the noise distribution}.
The main theoretical tool that we leverage in this paper is the scenario approach~\cite{DBLP:journals/mp/CalafioreC05,campi2018introduction}.
The scenario approach has been used for the verification of \glspl{MDP}~\cite{Badings2022STTT} and continuous-time Markov chains~\cite{DBLP:conf/cav/BadingsJJSV22} with uncertain parameters, albeit only for finite-state systems.
However, the non-trivial connection that we make in this paper between the scenario approach theory and techniques for formal abstractions had not been established.

\paragraph{Other sampling techniques.}
The aforementioned tool \sreachtools also exhibits a sampling-based method but relies on Hoeffding's inequality to obtain confidence guarantees~\cite{DBLP:conf/amcc/SartipizadehVAO19}, so the noise is assumed to be sub-Gaussian~\cite{DBLP:books/daglib/0035704}.
By contrast, the scenario approach is \emph{completely distribution-free}~\cite{campi2018introduction}.
Moreover, as we have shown in \cref{fig:intervals_samples,fig:intervals_fraction}, the scenario approach may lead to abstract models with significantly better probability intervals compared to Hoeffding's inequality.
In addition, \sreachtools is limited to problems with convex safe sets (a restrictive assumption in many problems) and its sampling-based methods can only synthesize open-loop controllers, which may undermine the robustness of the overall approach.

Another body of relevant literature entails sampling-based feedback motion planning algorithms, e.g., LQR-Trees~\cite{DBLP:journals/ijrr/TedrakeMTR10}.
However, sampling in LQR-Trees relates to random exploration of the state space and not to stochastic noise affecting the dynamics as in our setting~\cite{DBLP:journals/ijrr/ReistPT16}.

Monte Carlo methods (e.g., particle methods) have also been used to solve stochastic reach-avoid problems~\cite{DBLP:journals/trob/BlackmoreOBW10,DBLP:conf/cdc/LesserOE13}.
These methods simulate the system via many samples of the uncertain variable~\cite{Smith2013sequentialMonteCarlo}.
Monte Carlo methods \emph{approximate} stochastic problems but do not provide rigorous \emph{bounds} with a desired \emph{confidence level} on the obtained results as our approach does.

\paragraph{Distributionally robust optimization.}
In \gls{DRO}, decisions are robust with respect to \emph{ambiguity sets} of distributions~\cite{DBLP:journals/mp/EsfahaniK18,goh2010distributionally,DBLP:journals/ior/WiesemannKS14}.
While the scenario approach uses samples of the uncertain variable, \gls{DRO} works on the domain of uncertainty directly, thus involving potentially complex ambiguity sets~\cite{garatti2019risk}.
Designing robust policies for \glspl{iMDP} with known uncertainty sets was studied by~\citeA{DBLP:conf/cav/PuggelliLSS13}, and~\citeA{DBLP:conf/cdc/WolffTM12}.
Finally, hybrid methods between the scenario approach and robust optimization also exist~\cite{Margellos2014OnProblems}.

\paragraph{PAC literature.}
The term \gls{PAC} refers to obtaining, with high probability, a hypothesis that is a good approximation of some unknown phenomenon~\cite{DBLP:conf/aaai/Haussler90}.
\gls{PAC} learning methods for discrete-state \glspl{MDP} are developed in \citeA{DBLP:journals/jmlr/BrafmanT02},~\citeA{DBLP:conf/rss/FuT14}, and~\citeA{DBLP:journals/ml/KearnsS02}, and \gls{PAC} statistical model checking for \glspl{MDP} in~\citeA{DBLP:conf/cav/AshokKW19}.

\paragraph{Safe and robust learning methods.}
We briefly discuss the emerging field of safe learning~\cite{Brunke2021safelearning,DBLP:journals/jmlr/GarciaF15}.
Recent works use Gaussian processes for learning-based model predictive control~\cite{DBLP:journals/tcst/HewingKZ20,DBLP:conf/cdc/KollerBT018} or reinforcement learning with safe exploration~\cite{DBLP:conf/nips/BerkenkampTS017}, and control barrier functions to reduce model uncertainty~\cite{DBLP:conf/l4dc/TaylorSYA20}.
Safe learning control concerns learning unknown, \emph{deterministic} system dynamics, while imposing strong assumptions on stochasticity~\cite{DBLP:journals/tac/FisacAZKGT19}.
By contrast, our problem setting is fundamentally different: we reason about \emph{stochastic} noise of a completely unknown distribution.

Finally, various methods have been proposed to render reinforcement learning more robust against differences in the dynamics between the training and testing environments~\cite{DBLP:journals/neco/MorimotoD05,DBLP:journals/make/MoosHASCP22}.
For example,~\citeA{DBLP:conf/icra/PengAZA18} train neural network controllers for robotic control tasks on simulation models with randomized dynamics.
The authors show empirically that this randomization leads to controllers that are significantly more robust against discrepancies in the dynamics of the real robot on which they are deployed.
Similarly,~\citeA{DBLP:conf/icml/PintoDSG17} and~\citeA{Vinitsky2020} propose adversarial methods to improve the robustness of reinforcement learning by introducing an adversary which applies maximally destabilizing disturbances to the system.
While these methods have been shown empirically to improve the robustness of reinforcement learning, we remark that providing formal guarantees about safety or robustness (as we do in this paper) is rarely possible.
\section{Concluding Remarks and Future Work}
\label{sec:Conclusion}

We have presented a novel approach for robust control of continuous-state dynamical systems with (stochastic) process noise of unknown distribution.
Our approach is based on a finite abstraction of the continuous-state system in the form of an \gls{iMDP}.
As a key contribution, we have developed a rigorous method to use the scenario approach theory for computing \gls{PAC} probability intervals for such an abstract \gls{iMDP}.
The \gls{PAC}-correctness of these intervals, and thus of the whole abstract \gls{iMDP}, is valid irrespective of the noise distribution.
We have shown how to use the \gls{iMDP} to compute feedback controllers with \gls{PAC} guarantees on the performance on the continuous system.
Our experiments have confirmed this claim, showing that our method effectively solves realistic problems and provides safe lower-bound guarantees on the performance of controllers.

\paragraph{State space discretization.}
The discretization of the state space influences the quality of the reach-avoid guarantees.
Partitions into regions that are smaller are more easily contained in the backward reachable sets of actions, thus enabling more actions in the abstraction.
Moreover, defining more target points leads to an abstract model that captures more actions in general.
Hence, a more fine-grained partition with more target points yields an abstraction that is a more accurate representation of the dynamical system but also increases the computational complexity.
Whilst the time-dependent improved policy synthesis scheme based on state aggregation, as discussed in \cref{subsec:algorithmic_optimization}, balances this trade-off to some extent, we plan to employ more general and sophisticated adaptive discretization schemes in the future, such as those proposed in~\citeA{DBLP:journals/siamads/SoudjaniA13}.

\paragraph{Extensions to general PCTL properties.}
As described in \cref{sec:prelim:problem}, we have considered control objectives in the form of (in)finite-horizon reach-avoid properties.
Formally, these reach-avoid properties contain formulae expressed in probabilistic computation tree logic, or PCTL~\cite{DBLP:conf/voss/CiesinskiG04,DBLP:journals/fac/HanssonJ94}.
Our abstraction approach is valid for properties over both finite and infinite horizons, except for the improved policy synthesis scheme proposed in \cref{subsec:algorithmic_optimization}, which only applies to finite horizons (as this scheme relies on modeling each time step separately).
By contrast, techniques that result in abstraction errors accumulating with time are generally not applicable to infinite-horizon properties~\cite{LSAZ21}.

Moreover, while we have left out rewards for brevity, our approach is also amenable to expected reward properties~\cite{DBLP:books/daglib/BaierKatoen2008} with state-based\footnote{We cannot consider action-based reward functions because there is no direct correspondence between control inputs $\control_k \in \cControlSpace$ for the dynamical system and actions $a \in \Actions$ of the abstract \gls{iMDP}.} reward functions (mapping states to rewards) as long as the reward function is aligned with the partition, similar to \cref{assumption:RegionAlignment}.
More precisely, all states $\state_k \in \region_s$ belonging to the region of abstract state $s \in \States$ must be assigned the same reward.
Under this assumption, our approaches are equally applicable to general PCTL properties, although we leave a formal derivation of these results as an avenue for future work.

\paragraph{Non-stationary noise distributions.}
For our controller synthesis approach to be correct, the samples used for computing the probability intervals of the \gls{iMDP} must be \gls{iid} and drawn from the same distribution as the noise $\noise_k$ affecting the linear dynamical system in \cref{eq:continuousSystem}.
Thus, our approach requires that the probability distribution of the noise is fixed, which may be a questionable assumption in practice~\cite{Brunke2021safelearning}.
As such, one open research question is to what extent the correctness of our approach can be guaranteed if the noise distribution is not stationary but instead perturbed by some (bounded) distance.

\paragraph{Extensions to systems with uncertain dynamics.}
In this paper, we have considered linear dynamical systems whose dynamics (apart from the distribution of the noise) are precisely known.
Therefore, for physics-based systems, such as a \gls{UAV}, system parameters, including its mass or friction coefficient, must be known exactly.
Our ongoing research, recently presented in~\citeA{Badings2023AAAI}, lifts this restrictive assumption by considering systems with both stochastic noise and uncertain dynamics, e.g., due to imprecisely known parameters.
In particular, such uncertainty in the dynamics causes \emph{epistemic uncertainty}, which is fundamentally different from the probability distributions over outcomes due to process noise, leading to \emph{aleatoric uncertainty}~\cite{sullivan2015introduction}.

In a setting with both epistemic and aleatoric uncertainty, we must simultaneously learn about the unknown deterministic dynamics and the stochastic noise.
Learning deterministic dynamics is common in safe learning control~\cite{Brunke2021safelearning}, but enabling safe exploration requires strong assumptions on stochastic uncertainty.
As such, we see this direction as a challenging avenue for future work.

Extensions to dynamical systems with both aleatoric and epistemic uncertainty also open the door to capturing other types of uncertainty beyond process noise.
Besides the uncertainty in system parameters described above, we may, for example, deal with state/control-dependent process noise or systems with nonlinear dynamics (see below).
Moreover, such an extension may also enable us to lift \cref{assump:BackwardReachRank}, which is needed because our current method requires the system to be fully actuated.

\paragraph{Nonlinear systems.}
While we have focused on linear systems, we wish to develop extensions to nonlinear systems, as discussed in \cref{remark:LinearSystems}.
Such extensions are non-trivial and may require more involved reachability computations~\cite{DBLP:conf/cdc/BansalCHT17,DBLP:conf/cav/ChenAS13}.
Specifically, the challenge is to compute the enabled \gls{iMDP} actions via the backward reachable set defined in \cref{eq:BackwardReach}, which may become non-convex under nonlinear dynamics.
Note that computing the \gls{PAC} probability intervals remains unchanged, as the scenario approach relies on the convexity of the target set only, and not on that of the backward reachable set.
Alternatively, we may apply our method on a linearized version of the nonlinear system, in which case we must account for any linearization error to preserve guarantees.
Similar to the extension with uncertain parameters, this linearization error may potentially be captured by epistemic uncertainty in the system dynamics.

\section*{Acknowledgments}
This work was partially funded by 
the NWO grant NWA.1160.18.238 (\emph{PrimaVera}),
the 2022 JPMorgan Chase Faculty Research Award ``Learning and Reasoning in Repeated Games with Partial Information'',
the European Union’s Horizon 2020 research and innovation programme under the Marie Skłodowska-Curie grant agreement No. 101008233,
the ERC Consolidator Grant 864075 (\emph{CAESAR}), 
the EPSRC IAA Award EP/X525777/1,
and the ERC Advanced Grant 834115 (\emph{FUN2MODEL}).

\newpage

\appendix
\newcommand{\PxNotinR}{\ensuremath{P \big( \bm{x} \notin R_j}}
\newcommand{\PxInR}{\ensuremath{P \big( \bm{x} \in R_j}}

\section{Proof of \cref{theorem:Bounds}}
\label{appendix:Proofs}

The proof of \cref{theorem:Bounds} is adapted from~\citeA[Theorem~5]{romao2022tac}, which is based on three key assumptions: (1) the considered scenario problem belongs to the class of so-called \emph{fully-supported problems} (see~\citeA{DBLP:journals/siamjo/CampiG08} for a definition), (2) its solution is unique, and (3) discarded samples violate all interim optimal solutions with probability one.
In our case, requirement (2) is implied by \cref{assumption:nonAccumulation}, (3) is satisfied by \cref{lemma:Q}, and the problem is fully-supported because the number of decision variables is one.
Under these requirements,~\citeA[Theorem~5]{romao2022tac} states that the risk associated with an optimal solution $\lambda^\star_{\vert Q \vert}$ to \cref{eq:SCP} for $\vert Q \vert$ discarded samples satisfies the following expression:
\begin{equation}
\begin{split}
    \amsmathbb{P}^N \Big\{ \PxNotinR (\lambda^\star_{\vert Q \vert}) \big) \leq \epsilon \Big\}
    = 1 - \sum_{i=0}^{\vert Q \vert} \binom Ni \epsilon^i (1-\epsilon)^{N-i},
\end{split}
\label{eq:Proof1}
\end{equation}
where we omit the subscripts in $P_{\bm{w}_k}(\cdot)$ and $\bm{x}_{k+1}$ for brevity.
\cref{eq:Proof1} is the cumulative distribution function (CDF) of a beta distribution with parameters $\vert Q \vert + 1$ and $N-\vert Q \vert$.
We denote this CDF by $F_{\vert Q \vert}(\epsilon)$, where we explicitly write the dependency on $\vert Q \vert$.
Hence, we obtain
\begin{equation}
    F_{\vert Q \vert}(\epsilon) = \amsmathbb{P}^N \Big\{ \PxNotinR (\lambda^\star_{\vert Q \vert}) \big) \leq \epsilon \Big\} = \tilde{\beta}.
\label{eq:Proof1b}
\end{equation}
Thus, if we discard $\vert Q \vert$ samples, \cref{eq:Proof1b} returns the confidence probability $\tilde{\beta} \in (0,1)$ by which the probability of $\bm{x}_{k+1} \notin R_j(\lambda^\star_{\vert Q \vert})$ is upper bounded by $\epsilon$.
Conversely, we can compute the value of $\epsilon$ needed to obtain a confidence probability of $\tilde{\beta}$, using the percent point function (PPF) $G_{\vert Q \vert}(\tilde{\beta})$:
\begin{equation}
    \amsmathbb{P}^N \Big\{ \PxNotinR (\lambda^\star_{\vert Q \vert}) \big) \leq G_{\vert Q \vert}(\tilde{\beta}) \Big\} = \tilde{\beta}.
    \label{eq:Proof2}
\end{equation}
The PPF is the inverse of the CDF, so by definition, we have
\begin{equation}
    \epsilon = G_{\vert Q \vert}(\tilde{\beta}) = G_{\vert Q \vert}\big( F_{\vert Q \vert} (\epsilon) \big).
    \label{eq:Proof2b}
\end{equation}
Note that $P(\bm{x} \in R) + P(\bm{x} \notin R) = 1$, so \cref{eq:Proof1b} equals
\begin{equation}
    F_{\vert Q \vert}(\epsilon) = \amsmathbb{P}^N \Big\{ \PxInR (\lambda^\star_{\vert Q \vert}) \big) \geq 1-\epsilon \Big\} = \tilde{\beta}.
\label{eq:Proof3}
\end{equation}
By defining $p = 1 - \epsilon$, \cref{eq:Proof3,eq:Proof2b} are combined as
\begin{equation}
\begin{split}
    \amsmathbb{P}^N \Big\{ 1 - G_{\vert Q \vert}(\tilde{\beta}) \leq \PxInR (\lambda^\star_{\vert Q \vert}) \big) \Big\}
    = 1 - \sum_{i=0}^{\vert Q \vert} \binom Ni (1-p)^i p^{N-i} = \tilde{\beta}.
    \label{eq:Proof4}
\end{split}
\end{equation}
In what follows, we separately use \cref{eq:Proof3} to prove the lower and upper bound of the probability interval in \cref{theorem:Bounds}.

\paragraph{Lower bound.}
There are $N$ possible values for $\vert Q \vert$, ranging from $0$ to $N-1$.
The case $\vert Q \vert = N$ (i.e. all samples are discarded) is treated as a special case in \cref{theorem:Bounds}.
We fix $\tilde{\beta} = 1 - \frac{\beta}{2 N}$ in \cref{eq:Proof4}, yielding the series of equations
\begin{alignat}{3}
    & \amsmathbb{P}^N \Big\{ 1 - G_{0} \big( 1 - \frac{\beta}{2 N} \big) \leq \PxInR (\lambda^\star_{0}) \big) \Big\} && = && 1 - \frac{\beta}{2 N}
    \nonumber
    \\
    & \qquad\qquad\qquad\qquad \vdots && && \vdots
    \label{eq:Proof_sysEquations_low}
    \\
    & \amsmathbb{P}^N \Big\{ 1 - G_{N-1} \big( 1 - \frac{\beta}{2 N} \big) \leq \PxInR (\lambda^\star_{N-1}) \big) \Big\} && = && 1 - \frac{\beta}{2 N}.
    \nonumber
\end{alignat}
Denote the event that $1 - G_{n} \big( 1 - \frac{\beta}{2 N} \big) \leq \PxInR (\lambda^\star_{n}) \big)$ for $n = 0,\ldots,N-1$ by $\mathcal{A}_n$.
Regardless of $n$, this event has a probability of $\amsmathbb{P}^N \{ \mathcal{A}_n \} = 1 - \frac{\beta}{2N}$, and its complement $\mathcal{A}_n'$ of $\amsmathbb{P}^N \{ \mathcal{A}_n' \} = \frac{\beta}{2N}$.
Via Boole's inequality, we know that 
\begin{equation}
    \amsmathbb{P}^N \Big\{ \bigcup_{i = 0}^{N-1} \mathcal{A}_n' \Big\} \leq 
    \sum_{i=0}^{N-1} \amsmathbb{P}^N \big\{ \mathcal{A}_n' \big\}
    =
    \frac{\beta}{2N} N
    =
    \frac{\beta}{2}.
    \label{eq:UnionBound}
\end{equation}
Thus, for the intersection of all events in \cref{eq:Proof_sysEquations_low} we have
\begin{equation}
    \amsmathbb{P}^N \Big\{ \bigcap_{i = 0}^{N-1} \mathcal{A}_n \Big\} 
    = 
    1 - \amsmathbb{P}^N \Big\{ \bigcup_{i = 0}^{N-1} \mathcal{A}_n' \Big\}
    \geq 1 - \frac{\beta}{2}.
    \label{eq:UnionBound2}
\end{equation}
After observing the samples at hand, we replace $\vert Q \vert$ by the actual value of $N_j^\text{out}$ (as per \cref{def:Nj_in}), giving one of the expressions in \cref{eq:Proof_sysEquations_low}.
The probability that this expression holds cannot be smaller than that of the intersection of all events in \cref{eq:UnionBound2}. 
Thus, we obtain
\begin{equation}
    \amsmathbb{P}^N \Big\{
        \munderbar{p} 
        \leq 
        \PxInR (\lambda^\star_{N_j^\text{out}}) \big)
    \Big\} \geq 1 - \frac{\beta}{2},
    \label{eq:Proof6_lowerBound}
\end{equation}
where $\munderbar{p} = 0$ if $N_j^\text{out} = N$ (which trivially holds with probability one), and otherwise $\munderbar{p} = 1 - G_{N_j^\text{out}} (1 - \frac{\beta}{2 N})$ is the solution for $p$ to \cref{eq:Proof4}, with $\vert Q \vert = N_j^\text{out}$ and $\tilde{\beta} = 1-\frac{\beta}{2 N}$:
\begin{equation}
\begin{split}
    1 - \frac{\beta}{2 N} &= 1 - \sum_{i=0}^{N_j^\text{out}} \binom Ni (1-p)^i p^{N-i},
    \label{eq:Proof7}
\end{split}
\end{equation}
which is equivalent to \cref{eq:TheoremLowerBound}.

\paragraph{Upper bound.}
\cref{eq:Proof4} is rewritten as an upper bound as
\begin{equation}
    \amsmathbb{P}^N \Big\{ \PxInR (\lambda^\star_{\vert Q \vert}) \big) < 1 - G_{\vert Q \vert}(\tilde{\beta}) \Big\} = 1 - \tilde{\beta},
    \label{eq:Proof8}
\end{equation}
where again, $\vert Q \vert$ can range from $0$ to $N-1$.
However, to obtain high-confidence guarantees on the upper bound, we now fix $\tilde{\beta} = \frac{\beta}{2N}$, which yields the series of equations
\begin{alignat}{3}
    & \amsmathbb{P}^N \Big\{ \PxInR (\lambda^\star_{0}) \big) < 1 - G_{0} \big( \frac{\beta}{2 N} \big) \Big\} & = & 1 - \frac{\beta}{2 N}
    \nonumber
    \\
    & \qquad\qquad\qquad\qquad \vdots & & \vdots
    \label{eq:Proof_sysEquations_upp}
    \\
    & \amsmathbb{P}^N \Big\{ \PxInR (\lambda^\star_{N-1}) \big) < 1 - G_{N-1} \big( \frac{\beta}{2 N} \big) \Big\} & = & 1 - \frac{\beta}{2 N}.
    \nonumber
\end{alignat}
Analogous to the lower bound case, Boole's inequality implies that the intersection of all expressions in \cref{eq:Proof_sysEquations_upp} has a probability of at least $1 - \frac{\beta}{2}$. 
After observing the samples at hand, we replace $\vert Q \vert$ by $N_j^\text{out} - 1$, yielding one of the expressions in \cref{eq:Proof_sysEquations_upp}.
For this expression, it holds that
\begin{equation}
    \amsmathbb{P}^N \Big\{
        \PxInR (\lambda^\star_{N_j^\text{out} - 1}) \big) \leq \bar{p}
    \Big\} \geq 1 - \frac{\beta}{2},
    \label{eq:Proof10_upperBound}
\end{equation}
where $\bar{p} = 1$ if $N_j^\text{out} = 0$ (which trivially holds with probability one), and otherwise $\bar{p} = 1 - G_{N_j^\text{out} - 1}(\frac{\beta}{2 N})$ is the solution for $p$ to \cref{eq:Proof4}, with $\vert Q \vert = N_j^\text{out} - 1$ and $\tilde{\beta} = \frac{\beta}{2 N}$:
\begin{equation}
    \frac{\beta}{2 N} = 1 - \sum_{i=0}^{N_j^\text{out} -1 } \binom Ni (1-p)^i p^{N-i},
    \label{eq:Proof11}
\end{equation}
which is equivalent to \cref{eq:TheoremUpperBound}.

\paragraph{Probability interval.}
We invoke \cref{lemma:Q_cardinality}, which states that $R_j(\lambda^\star_{N_j^\text{out}}) \subseteq R_j \subset R_j(\lambda^\star_{N_j^\text{out} - 1})$, so we have
\begin{equation}
\begin{split}
    \PxInR (\lambda^\star_{N_j^\text{out}}) \big)
    & \leq 
    P (\bm{x} \in R_j)
    =
    P(s_i, a_l)(s_j)
    \label{eq:Proof12}
    \\
    & < \PxInR (\lambda^\star_{N_j^\text{out} - 1}) \big).
\end{split}
\end{equation}
We use \cref{eq:Proof12} to write \cref{eq:Proof6_lowerBound,eq:Proof10_upperBound} in terms of the transition probability $P(s_i, a_l)(s_j)$.
Finally, by applying Boole's inequality, we combine \cref{eq:Proof6_lowerBound,eq:Proof10_upperBound} as follows:
\begin{equation}
\begin{split}
    \amsmathbb{P}^N \Big\{
        \munderbar{p} \leq P(s_i, a_l)(s_j) \, \bigcap \, P(s_i, a_l)(s_j) \leq \bar{p}
    \Big\} 
    & = \amsmathbb{P}^N \Big\{
        \munderbar{p} \leq P(s_i, a_l)(s_j) \leq \bar{p}
    \Big\}
    \\
    & \geq 1 - \beta,
    \label{eq:Proof13_interval}
\end{split}
\end{equation}
which is equal to \cref{eq:TheoremBounds}, so we conclude the proof.

\section{Proof of \cref{corollary:tighter_confidence}}
\label{proof:cor:tight}

\begin{proof}
    The derivation of \cref{corollary:tighter_confidence} is analogous to the proof of \cref{theorem:correctness}, with the only difference that the number of unique probability intervals is reduced.
    The key observation is that the successor state $\bm{x}_{k+1}$ in \cref{eq:continuous_successor} is linear in the target point $\bm{d}_a$ and the noise $\noise_k$.
    Thus, by denoting $p(\noise_k)$ as the (unknown) probability density function of the noise, we rewrite \cref{eq:multivariateIntegral_mainProblem} as
    \begin{equation}
    \begin{split}
        \label{eq:cor:tight1}
        P(s,a)(s') 
        &= \int_{\region_{s'}} p_{\noise_k}\left(\state_{k+1} \ | \ \state_k, \control_k(\state_k, a)\right)d\state_{k+1} 
        \\
        &= \int_{\region_{s'} - \bm{d}_a} p\left(\noise_k\right)d\noise_k,
    \end{split}
    \end{equation}
    that is, for a fixed density function of $\noise_k$ the transition probability $P(s,a)(s')$ is defined by the relative position $\region_{s'} - \bm{d}_{a}$ between the region and the target point.
    Thus, for any two regions $\region_{s'}$, $\region_{s''}$ and two target points $\bm{d}_a$, $\bm{d}_{a'}$ whose relative positions are the same, i.e. $\region_{s'} - \bm{d}_a = \region_{s''} - \bm{d}_{a'}$, it holds that
    \begin{equation}
        P(s,a)(s') 
        = \int_{\region_{s'} - \bm{d}_a} p\left(\noise_k\right)d\noise_k
        = \int_{\region_{s''} - \bm{d}_{a'}} p\left(\noise_k\right)d\noise_k
        = P(s,a')(s'').
    \end{equation}
    Now, observe that under the proposed partitioning and target points, the number of pairs $(R_{s'}, \bm{d}_a)$ with equal relative position is maximized.
    Formally, let $\mu_{s'}$ be the center of $\region_{s'}$ for each $s' \in \States \setminus \{s_\star\}$, and let $w_i$ be the width of the rectangular partitioning in each dimension $i=1,\ldots,n$.
    Then, the difference $\mu_{s'} - \bm{d}_a$ for any $s' \in \States$, $a \in \Actions$ can be written as
    \begin{equation}
        \mu_{s'} - \bm{d}_a = \begin{bmatrix}
            \xi_1 w_1
            \\
            \vdots
            \\
            \xi_n w_n
        \end{bmatrix}, \,\,
        \xi_i \in \{-r_i+1, \ldots, r_i-1 \} \, \forall i = 1,\ldots,n,
    \end{equation}
    except when $s'$ is the absorbing state.
    Since each integer variable $\xi_i$ can take one of $2 r_i - 1$ different values, we observe that $\mu_{s'} - \bm{d}_a$ can take one of $\prod_{i=1}^n (2r_i - 1)$ different values.
    Thus, the \gls{iMDP} has exactly $\prod_{i=1}^n (2r_i - 1)$ unique transition probabilities (and hence the same number of unique probability intervals), plus one unique probability $P(s,a)(s_\star)$ of transitioning to the absorbing state $s_\star$ for each $a \in \Actions$.
    Plugging in this total number of $\prod_{i=1}^n (2r_i - 1) + |\Actions|$ unique probability intervals, we obtain the desired expression in \cref{eq:cor:tight1} and thus conclude the proof.
\end{proof}
\newcommand\scalemath[2]{\scalebox{#1}{\mbox{\ensuremath{\displaystyle #2}}}}

\section{Details on Satellite Rendezvous Benchmark}
\label{appendix:spacecraft}

\subsection{Explicit Model Formulation}
\label{subappendix:spacecraft_model}

The dynamical system of this benchmark problem is defined on the so-called Hill's frame, which represents the relative coordinate frame describing the difference in position and velocity between a chaser and target satellite~\cite{DBLP:conf/cdc/JewisonE16}.
The dynamics of this 6-dimensional system are defined as follows:
\begin{equation*}
\begin{split}
    \state_{k+1} =& \scalemath{0.9}{\begin{bmatrix}
        4-3\cos(n \tau) & 0 & 0 & \frac{1}{n} \sin(n \tau) & \frac{2}{n} (1-\cos(n \tau)) & 0 \\
        6(\sin(n \tau) - n\tau) & 1 & 0 & \frac{-2}{n} (1-\cos(n \tau)) & \frac{4}{n} \sin(n \tau) - 3n\tau & 0 \\
        0 & 0 & \cos(n \tau) & 0 & 0 & \frac{1}{n} \sin(n \tau) \\
        3n \sin(n \tau) & 0 & 0 & \cos(n \tau) & 2 \sin(n \tau) & 0 \\
        -6n(1-\cos(n \tau)) & 0 & 0 & -2 \sin(n \tau) & 4 \cos(n \tau) - 3 & 0 \\
        0 & 0 & -n \sin(n \tau) & 0 & 0 & \cos(n \tau)
    \end{bmatrix}}
    \state_{k} \\
    &+ \scalemath{0.9}{\begin{bmatrix}
        \frac{1}{n}\sin(n \tau) & \frac{2}{n}(1 - \cos(n\tau)) & 0 \\
        \frac{-2}{n}(1-\cos(n \tau)) & \frac{1}{n}(4 \sin(n \tau) - 3n\tau) & 0 \\
        0 & 0 & \frac{1}{n}\sin(n \tau) \\
        \cos(n \tau) & 2\sin(n \tau) & 0 \\
        -2\sin(n \tau) & 4\cos(n \tau) - 3 & 0 \\
        0 & 0 & \cos(n \tau)
    \end{bmatrix}} \control_{k} + \mathcal{N} \Big( \scalemath{0.9}{\begin{bmatrix}0 \\ 0 \\ 0 \\ 0 \\ 0 \\ 0 \end{bmatrix}}, \text{diag} \big( \scalemath{0.9}{\begin{bmatrix}
        0.1 \\ 0.1 \\ 0.01 \\ 0.01 \\ 0.01 \\ 0.01
    \end{bmatrix}} \big) \Big),
    \label{eq:spacecraft_explicitSystem}
\end{split}
\end{equation*}
where $\tau$ is the discretization time, and $n$ is a constant; see \citeA{DBLP:conf/cdc/JewisonE16} for details.
Every element of the 3-dimensional control input vector $\control_k$ is constrained to the interval $[-2, 2]$.
Since the model in \cref{eq:spacecraft_explicitSystem} is not fully actuated (it has only $3$ controls, versus a state of dimension $6$), we group every two discrete time steps together (as described in \cref{sec:Abstraction}) and rewrite the model as follows:
\begin{equation}
    \state_{k+2} = \bar{A} \state_k + \bar{B} \control_{k,k+1} + \noise_{k,k+1},
    \label{eq:UAV_fullyActuated}
\end{equation}
where $\bar{A} \in \R^{6 \times 6}$ and $\bar{B} \in \R^{6 \times 6}$ are properly redefined matrices, and $\control_{k,k+1} \in \R^6$ and $\noise_{k,k+1} \in \R^6$ reflect the control inputs and process noise at both time steps $k$ and $k+1$, combined in one vector.

\section{Details on UAV Benchmark}
\label{appendix:UAV}

\subsection{Explicit Model Formulation}
\label{subappendix:UAV_model}

The 6-dimensional state vector of the \gls{UAV} model is $\state = [p_x, v_x, p_y, v_y, p_z, v_z]^\top \in \R^6$, where $p_i$ and $v_i$ are the position and velocity in the direction $i$.
Every element of the vector of control inputs $\control = [f_x, f_y, f_z]^\top \in \R^3$ is constrained to the interval $[-4, 4]$.
The discrete-time dynamics are an extension of the lower-dimensional case in~\citeA{Badings2021FilterUncertainty}, and are written in the form of \cref{eq:continuousSystem} as follows:
\begin{equation}
    \state_{k+1} =	\begin{bmatrix}
        1 & 1 & 0 & 0 & 0 & 0 \\
        0 & 1 & 0 & 0 & 0 & 0 \\
        0 & 0 & 1 & 1 & 0 & 0 \\
        0 & 0 & 0 & 1 & 0 & 0 \\
        0 & 0 & 0 & 0 & 1 & 1 \\
        0 & 0 & 0 & 0 & 0 & 1
    \end{bmatrix}
    \state_{k} + \begin{bmatrix}
        0.5 & 0 & 0 \\
        1 & 0 & 0 \\
        0 & 0.5 & 0 \\
        0 & 1 & 0 \\
        0 & 0 & 0.5 \\
        0 & 0 & 1
    \end{bmatrix} \control_{k} + \noise_k,
    \label{eq:UAV_explicitSystem}
\end{equation}
where $\noise_k$ is the effect of turbulence on the continuous state, which we model using a Dryden gust model; we use a Python implementation by~\citeA{bohn2019deep}.
Note that the drift term $\drift_k = 0$, and is thus omitted from  \cref{eq:UAV_explicitSystem}.
Similar to the dynamical system of the satellite benchmark in \cref{eq:spacecraft_explicitSystem}, we group every two discrete time steps together to render the system in \cref{eq:UAV_explicitSystem} fully actuated.

The objective is to reach the goal region, which is also depicted in \cref{fig:UAV_layout}, within $64$ steps (i.e. $32$ steps with the model in \cref{eq:UAV_fullyActuated}).
This goal region is written as the following set of continuous states (for brevity, we omit an explicit definition of the critical regions):
\begin{equation}
    \goalRegion = \{ \state \in \R^6 \ \vert \ 11 \leq p_x \leq 15, \ \ 1 \leq p_y \leq 5, \ \ -7 \leq p_z \leq -3 \}.
\end{equation}

\subsection{Run Times and Size of Model Abstraction}

\label{subappendix:UAV_results}

The average run times and \gls{iMDP} model sizes (over 10 runs) for all iterations of the \gls{UAV} benchmark are presented in \cref{tab:DetailedResults}.
Computing the states and actions of the underlying \gls{MDP} took one minute but since this step is only performed in the first iteration, we omit the value from the table.
The run times per iteration are the times for those steps within the while-loop in \cref{alg:algorithm_planning}.
The number of states (which equals the number of regions in the partition) and choices (the total number of state-action pairs) of the \gls{iMDP} are independent of the value of $N$.
By contrast, the number of transitions increases with $N$, because the additional noise samples may reveal more possible outcomes of a state-action pair.

\begin{table*}[t!]

\caption{Run times per iteration (which excludes the computation of states and actions) and model sizes for the \gls{UAV} (under high turbulence), and 2-zone and 1-zone building temperature regulation (BAS) benchmarks.}
\label{tab:DetailedResults}

\scriptsize{
\begin{tabularx}{\textwidth}{@{} lr|rrr|rrr @{} } \toprule
    \multicolumn{2}{c|}{\textbf{Instance}} & \multicolumn{3}{c|}{\textbf{Run times (per iteration, in seconds)}} & \multicolumn{3}{c}{\textbf{\gls{iMDP} model size}}
    \\
    Benchmark & Samples $N$ & {Update intervals} & Write \gls{iMDP} & Compute $\munderbar{\policy}^*$ & States & Choices & Transitions
    \\ \midrule
    UAV & 25   & 11.4  & 29.0  & 39.2  & 25\,516 & 1\,228\,749 & 9\,477\,795 \\
    UAV & 50   & 15.4  & 34.4  & 46.0  & 25\,516 & 1\,228\,749 & 11\,350\,692 \\
    UAV & 100  & 18.6  & 39.3  & 51.8  & 25\,516 & 1\,228\,749 & 13\,108\,282 \\
    UAV & 200  & 24.3  & 44.4  & 59.1  & 25\,516 & 1\,228\,749 & 14\,921\,925 \\
    UAV & 400  & 35.4  & 50.6  & 68.5  & 25\,516 & 1\,228\,749 & 17\,203\,829 \\
    UAV & 800  & 55.4  & 58.3  & 80.2  & 25\,516 & 1\,228\,749 & 19\,968\,663 \\
    UAV & 1600 & 92.2  & 64.9  & 90.6  & 25\,516 & 1\,228\,749 & 22\,419\,161 \\
    UAV & 3200 & 164.1 & 68.6  & 94.4  & 25\,516 & 1\,228\,749 & 23\,691\,107 \\
    UAV & 6400 & 312.1 & 69.8  & 95.5  & 25\,516 & 1\,228\,749 & 23\,958\,183 \\
    UAV & 12800& 611.2 & 69.8  & 95.1  & 25\,516 & 1\,228\,749 & 23\,969\,028 \\ \midrule
    BAS (2-zone) & 25   & 13.9  & 73.6  & 105.9  & 35\,722 & 1\,611\,162 & 24\,929\,617 \\
    BAS (2-zone) & 50   & 43.2  & 107.7 & 160.5  & 35\,722 & 1\,611\,162 & 37\,692\,421 \\
    BAS (2-zone) & 100  & 50.8  & 153.2 & 230.2  & 35\,722 & 1\,611\,162 & 53\,815\,543 \\
    BAS (2-zone) & 200  & 59.1  & 205.8 & 315.2  & 35\,722 & 1\,611\,162 & 72\,588\,298 \\
    BAS (2-zone) & 400  & 72.3  & 260.6 & 423.2  & 35\,722 & 1\,611\,162 & 91\,960\,970 \\
    BAS (2-zone) & 800  & 96.2  & 316.8 & 504.3  & 35\,722 & 1\,611\,162 & 112\,028\,449 \\
    BAS (2-zone) & 1600 & 132.6 & 370.0 & 614.8  & 35\,722 & 1\,611\,162 & 131\,844\,146 \\
    BAS (2-zone) & 3200 & 200.1 & 432.9 & 789.4  & 35\,722 & 1\,611\,162 & 154\,155\,445 \\
    BAS (2-zone) & 6400 & 331.8 & 513.2 & 982.2  & 35\,722 & 1\,611\,162 & 182\,175\,229 \\
    BAS (2-zone) & 12800& 545.5 & 601.8 & 1177.8 & 35\,722 & 1\,611\,162 & 218\,031\,384 \\ \midrule
    BAS (1-zone) & 25   & 1.62 & 0.06 & 1.39 & 381 & 1\,511 & 20\,494 \\
    BAS (1-zone) & 50   & 1.81 & 0.08 & 1.39 & 381 & 1\,511 & 27\,327 \\
    BAS (1-zone) & 100  & 1.87 & 0.09 & 1.40 & 381 & 1\,511 & 33\,871 \\
    BAS (1-zone) & 200  & 1.96 & 0.11 & 1.40 & 381 & 1\,511 & 40\,368 \\
    BAS (1-zone) & 400  & 2.06 & 0.13 & 1.42 & 381 & 1\,511 & 47\,333 \\
    BAS (1-zone) & 800  & 2.26 & 0.14 & 1.40 & 381 & 1\,511 & 54\,155 \\
    BAS (1-zone) & 1600 & 2.59 & 0.16 & 1.39 & 381 & 1\,511 & 59\,686 \\
    BAS (1-zone) & 3200 & 3.24 & 0.17 & 1.40 & 381 & 1\,511 & 65\,122 \\
    BAS (1-zone) & 6400 & 4.50 & 0.18 & 1.40 & 381 & 1\,511 & 70\,245 \\
    BAS (1-zone) & 12800& 7.01 & 0.20 & 1.38 & 381 & 1\,511 & 76,076 \\ \bottomrule
    
\end{tabularx}
}

\end{table*}

\section{Details on Building Temperature Regulation Benchmark}
\label{appendix:BAS}

\subsection{Explicit Model Formulation}
\label{subappendix:BAS_model}

This model is a variant of the two-zone building automation system benchmark with stochastic dynamics in~\citeA{DBLP:journals/corr/abs-1803-06315}.
The state vector of the model is $\state = [T^z_1, T^z_2, T^r_1, T^r_2]^\top \in \R^4$, reflecting both room (zone) temperatures ($T^z$) and both radiator temperatures ($T^r$).
The control inputs $\control = [T^\text{ac}_1, T^\text{ac}_2, T^\text{boil}_1, T^\text{boil}_2]^\top \in \R^4$ are the air conditioning (ac) and boiler temperatures, respectively, which are constrained within $14 \leq T^\text{ac} \leq 26$ and $65 \leq T^\text{boil} \leq 85$.
The changes in the temperature of both zones are governed by the following thermodynamics:
\begin{subequations}
\begin{align}
    \dot{T}^z_1 =& \frac{1}{C_1} \Big[ \frac{T^z_2 - T^z_1}{R_{1,2}} + \frac{T_\text{wall} - T^z_1}{R_{1,\text{wall}}} 
    + m C_{pa} (T^\text{ac}_1 - T^z_1) + P_\text{out} (T_1^r - T_1^z) \Big]
    \nonumber
    \\
    \dot{T}^z_2 =& \frac{1}{C_2} \Big[ \frac{T^z_1 - T^z_2}{R_{1,2}} + \frac{T_\text{wall} - T^z_2}{R_{2,\text{wall}}} 
    + m C_{pa} (T^\text{ac}_2 - T^z_2) + P_\text{out} (T_2^r - T_2^z) \Big],
    \nonumber
\end{align}
\label{eq:BAS_zoneTemps}
\end{subequations}
where $C_i$ is the thermal capacitance of zone $i$, and $R_{i,j}$ is the resistance between zones $i$ and $j$, $m$ is the air mass flow, $C_{pa}$ is the specific heat capacity of air, and $P_\text{out}$ is the rated output of the radiator.
Similarly, the dynamics of the radiator in room $i$ are governed by the following equation:
\begin{equation}
    \dot{T}_i^r = k_1 (T_i^z - T_i^r) + k_0 w (T_i^\text{boil} - T_i^r), 
    \label{eq:BAS_radiatorTemp}
\end{equation}
where $k_0$ and $k_1$ are constant parameters, and $w$ is the water mass flow from the boiler.
For the precise parameter values used, we refer to our codes, which are provided in the supplementary material.
By discretizing \cref{eq:BAS_zoneTemps,eq:BAS_radiatorTemp} by a forward Euler method at a time resolution of $\SI{15}{\minute}$, we obtain the following model in explicit form:
\begin{align}
    \state_{k+1} = & \begin{bmatrix}
        0.8425 & 0.0537 & -0.0084 &  0.0000 \\
        0.0515 & 0.8435 &  0.0000 & -0.0064 \\
        0.0668 & 0.0000 &  0.8971 &  0.0000 \\
        0.0000 & 0.0668 &  0.0000 &  0.8971
    \end{bmatrix} \state_k
    \nonumber
    + \begin{bmatrix}
        0.0584 & 0.0000 & 0.0000 & 0.0000 \\
        0.0000 & 0.0599 & 0.0000 & 0.0000 \\
        0.0000 & 0.0000 & 0.0362 & 0.0000 \\
        0.0000 & 0.0000 & 0.0000 & 0.0362
    \end{bmatrix} \control_k 
    \\ & 
    + \begin{bmatrix}
        1.2291 \\ 1.0749 \\ 0.0000 \\ 0.0000
    \end{bmatrix} + \Gauss[\begin{bmatrix}0 \\ 0 \\ 0 \\ 0 \end{bmatrix}]{\begin{bmatrix}
        0.01 & 0 & 0 & 0 \\
        0 & 0.01 & 0 & 0 \\
        0 & 0 & 0.01 & 0 \\
        0 & 0 & 0 & 0.01
    \end{bmatrix}},
    \nonumber
\end{align}
where $\Gauss[\bm\mean]{\cov} \in \R^n$ is a multivariate Gaussian random variable with mean $\bm\mean \in \R^n$ and covariance matrix $\cov$. 

The goal in this problem is to reach a temperature of $19.9 - \SI{20.1}{\celsius}$ in both zones within $32$ discrete time steps of $\SI{15}{\minute}$.
As such, the goal region $\goalRegion$ and critical region $\criticalRegion$ are:
\begin{equation}
    \goalRegion = \{ \state \in \R^2 \ \vert \ 19.9 \leq T^z_1 \leq 20.1, \ 19.9 \leq T^z_2 \leq 20.1 \},
    \quad
    \criticalRegion = \varnothing.
    \nonumber
\end{equation}

\subsection{Run Times and Size of Model Abstraction}

\label{subappendix:BAS_results}

The run times and \gls{iMDP} model sizes for all iterations of the 2-zone building temperature regulation benchmark are presented in \cref{tab:DetailedResults}.
Computing the states and actions took $\SI{5.5}{\minute}$, but we omit this value from the table as this step is only performed in the first iteration.
The discussion of model sizes is analogous to \cref{subappendix:UAV_results} on the \gls{UAV} benchmark.
\section{Benchmarks Against Other Tools}
\label{appendix:Benchmarks}

\subsection{Benchmark Against \stochy}
\label{appendix:StocHy}
We provide a comparison between our approach and \stochy~\cite{DBLP:conf/tacas/CauchiA19} on a reduced version of the temperature regulation problem in \cref{subappendix:BAS_model} with only one zone.
Thus, the state vector becomes $\bm{x} = [T^z, T^r]^\top$, and the control inputs are $\bm{u} = [T^{\text{ac}}, T^{\text{boil}}]^\top$, which are constrained to $14 \leq T^\text{ac} \leq 28$ and $-10 \leq T^\text{boil} \leq 10$ (note that $T^\text{boil}$ is now relative to the nominal boiler temperature of $\SI{75}{\celsius}$).
Using the same dynamics as in \cref{eq:BAS_zoneTemps,eq:BAS_radiatorTemp}, we obtain the following model in explicit form:
\begin{equation*}
    \bm{x}_{k+1} = \begin{bmatrix}
        0.8820 & 0.0058 \\
        0.0134 & 0.9625 
    \end{bmatrix} \bm{x}_k + \begin{bmatrix}
        0.0584 & 0.0000 \\
        0.0000 & 0.0241
    \end{bmatrix} \bm{u}_k 
    + \begin{bmatrix}
        0.9604 \\
        1.3269
    \end{bmatrix} + \Gauss[\begin{bmatrix} 0 \\ 0 \end{bmatrix}]{\begin{bmatrix}
        0.02 & 0 \\
        0 & 0.1
    \end{bmatrix}}. 
\end{equation*}
The objective in this problem is to reach a zone temperature of $20.9 - \SI{21.1}{\celsius}$ within $64$ discrete time steps of $\SI{15}{\minute}$.
As such, the goal region $\goalRegion$ and critical region $\criticalRegion$ are:
\begin{equation}
    \goalRegion = \{ \bm{x} \in \R^2 \ \vert \ 20.9 \leq T^z \leq 21.1 \}, \quad \criticalRegion = \varnothing.
\end{equation}
We partition the continuous state space in a rectangular grid of $19 \times 20$ regions of width $\SI{0.2}{\celsius}$, which is centered at $T^z = \SI{21}{\celsius}$ and $T^r = \SI{38}{\celsius}$.
As such, the partition covers zone temperatures between $19.1 - \SI{22.9}{\celsius}$, and radiator temperatures between $36 - \SI{40}{\celsius}$.

\paragraph{Abstraction method.}
Similar to our approach, \stochy generates robust \gls{iMDP} abstractions of the model above.
However, while our approach also works when the distribution of the noise is unknown, \stochy requires precise knowledge of the noise distribution.
Another difference is that \stochy discretizes the control inputs of the dynamical system, to obtain a finite action space.
An \gls{iMDP} action is then associated with the execution of a specific value of the control input $\bm{u}_k$.
In our approach, an \gls{iMDP} action instead reflects an attempted transition to a given target point, and the actual control input $\bm{u}_k$ (calculated using the control law \cref{eq:controlLaw}) depends on the precise continuous state $\bm{x}_k$.

\begin{table*}[t!]

\caption{Maximum lower bound reach-avoid probability obtained using our method, the average simulated reach-avoid probability, and run times for our method and \sreachtools, under the default (x1) and increased (x10) noise covariance.}
\label{tab:SReachTools}

\footnotesize{
\begin{tabularx}{\textwidth}{@{} l|lllll|ll @{} } \toprule
    \multicolumn{1}{c|}{}
    & 
    \multicolumn{5}{c|}{\textbf{Our method}} 
    & 
    \multicolumn{2}{c}{\textbf{\sreachtools}}
    \\
    & Init. abstr. & $N=25$ & $N=100$ & $N=400$ & $N=1600$ & Particle & Voronoi \\ \midrule
    \textbf{Noise covariance x1} & & & & & & & \\
    Reach-avoid probability    & ---   & 0.44  & 0.81  & 0.95   & 0.98   & 0.88  & 0.85 \\ 
    Simulated probability       & ---   & 1.00  & 1.00  & 1.00   & 1.00   & 0.83  & 0.86 \\ 
    Run time (s)    & 3.420 & 6.966 & 9.214 & 11.016 & 17.294 & 0.703 & 3.462 \\ \midrule
    \textbf{Noise covariance x10} & & & & & & & \\
    Reach-avoid probability    & ---   & 0.19   & 0.36   & 0.52   & 0.62   & 0.40  & NaN \\ 
    Simulated probability       & ---   & 0.59   & 0.63   & 0.59   & 0.65   & 0.19  & NaN \\ 
    Run time (s)    & 3.372 & 11.433 & 15.909 & 18.671 & 25.452 & 2.495 & NaN \\ \bottomrule
    
\end{tabularx}
}

\end{table*}

\paragraph{Our approach is faster.}
The run times and model sizes of our approach are reported in \cref{tab:DetailedResults} under BAS (1-zone).
For our approach, the time to compute the states and actions (needed only in the first iteration) is $\SI{0.1}{\second}$, and run times per iteration (excluding verification times) vary from $1.7-\SI{7.2}{\second}$, depending on the value of $N$.
Notably, the \stochy run times are \emph{orders of magnitude higher}: the abstraction procedure of \stochy (excluding verification time) took $\SI{45}{\minute}$, compared to $\SI{7.2}{\second}$ for our approach with the maximum of $N=12\,800$ samples (as reported in \cref{tab:DetailedResults}).
We omit a comparison of the verification times, since \stochy does not leverage \prism like our method does.

These results provide a clear message: our approach is more general and significantly faster, and (as shown earlier in \cref{fig:BAS_stochy}) generates results that are qualitatively similar to those obtained from \stochy.

\subsection{Benchmark Against \sreachtools}
\label{appendix:SReachTools}

We benchmark our approach against the spacecraft docking problem supplied with the \matlab toolbox \sreachtools~\cite{DBLP:conf/hybrid/VinodGO19}.
In this problem, one spacecraft must dock with another within $8$ time steps, while remaining in a target tube. 
The goal is to compute a controller that maximizes the probability of achieving this objective. 
The 4-dimensional state $\bm{x} = [p_x, p_x, v_x, v_y]^\top \in \R^4$ describes the position and velocity in both directions.
We adopt the same discrete-time dynamics as used in~\citeA{DBLP:conf/hybrid/VinodGO19} and assume the same Gaussian noise (see the reference for details).

This problem is a finite-horizon reach-avoid problem with a rectangular goal region (target set), while the safe set is a convex tube, as also shown in  \cref{fig:SReachTools_results}.
For our approach, we partition the state space into $3\,200$ rectangular regions.
It is fair to note that the safe set in SReachTools is smooth, while ours is not (due to our partition-based approach). 
While our current implementation is limited to regions as hyper-rectangles and parallelepipeds, our method can in theory be used with all convex polytopic regions.

\paragraph{Reach-avoid guarantees.}
We compare our method (with $N=25,\ldots,1\,600$ samples of the noise) to the results obtained via the different methods in \sreachtools. 
We only consider the particle and Voronoi methods of \sreachtools, because only these methods are sampling-based like our approach (although only the Voronoi method can give confidence guarantees on the results).
The reach-avoid guarantees and run times are presented in \cref{tab:SReachTools}, and simulated trajectories in \cref{fig:SReachTools_results}.
As expected, our reach-avoid guarantees, as well as for the Voronoi method, are a lower bound on the average simulated performance (and are thus safe), while this is not the case for the particle method of \sreachtools.

\paragraph{Complexity and run time.}
As shown in \cref{tab:SReachTools}, the grid-free methods of \sreachtools are generally faster than our abstraction-based approach. 
However, we note that our method was designed for non-convex problems, which cannot be solved by \sreachtools.
Interestingly, the complexity of the particle~\cite{DBLP:conf/cdc/LesserOE13} and Voronoi partition method~\cite{DBLP:conf/amcc/SartipizadehVAO19} increases \emph{exponentially} with the number of samples, because their optimization problems have a binary variable for every particle.
By contrast, the complexity of our method increases only \emph{linearly} with the number of samples, because it suffices to count the number of samples in every region, as discussed in \cref{subsec:scenario:bounds}.

\paragraph{Controller type.}
The particle and Voronoi methods of \sreachtools synthesize \emph{open-loop} controllers, which means that they cannot act in response to observed disturbances of the state.
Open-loop controllers do not consider any feedback, making such controllers unsafe in settings with significant noise levels~\cite{aastrom2010feedback}.
By contrast, we derive \emph{piecewise linear feedback} controllers.
This structure is obtained because our controllers are based on a state-based control policy that maps every continuous state within the same region to the same abstract action.

\paragraph{Robustness against disturbances.}
To demonstrate the importance of feedback control, we test both methods under stronger disturbances, by increasing the covariance of the noise by a factor of 10.
As shown in \cref{tab:SReachTools}, the particle method yields a low reach-avoid probability (with the simulated performance being even lower), and the Voronoi method was not able to generate a solution at all.
By contrast, our method still provides safe lower bound guarantees, which become tighter for increasing sample sizes.
Our state-based controller is robust against the stronger noise, because it is able to act in response to observed disturbances, unlike the open-loop methods of \sreachtools that results in a larger error in the simulated state trajectories, as also shown in \cref{fig:SReachTools_results}.

\bibliography{references,scenarioReferences}

\begin{thebibliography}{}

\bibitem[\protect\BCAY{Abate, Katoen, Lygeros,\ \BBA\ Prandini}{Abate
  et~al.}{2010}]{AKLP10}
Abate, A., Katoen, J., Lygeros, J., \BBA\ Prandini, M. \BBOP2010\BBCP.
\newblock \BBOQ Approximate model checking of stochastic hybrid systems\BBCQ\
\newblock {\Bem European Journal of Control}, {\Bem 16\/}(6), 624--641.

\bibitem[\protect\BCAY{Abate}{Abate}{2011}]{DBLP:journals/entcs/Abate13}
Abate, A. \BBOP2011\BBCP.
\newblock \BBOQ Approximation metrics based on probabilistic bisimulations for
  general state-space markov processes: {A} survey\BBCQ\
\newblock In {\Bem Hybrid Autonomous Systems@ETAPS}, \lowercase{\BVOL}\ 297 of
  {\Bem Electronic Notes in Theoretical Computer Science}, \BPGS\ 3--25.
  Elsevier.

\bibitem[\protect\BCAY{Abate, Prandini, Lygeros,\ \BBA\ Sastry}{Abate
  et~al.}{2008}]{Abate2008probabilisticSystems}
Abate, A., Prandini, M., Lygeros, J., \BBA\ Sastry, S. \BBOP2008\BBCP.
\newblock \BBOQ Probabilistic reachability and safety for controlled discrete
  time stochastic hybrid systems\BBCQ\
\newblock {\Bem Automatica}, {\Bem 44\/}(11), 2724 -- 2734.

\bibitem[\protect\BCAY{Alur, Henzinger, Lafferriere,\ \BBA\ Pappas}{Alur
  et~al.}{2000}]{Alur2000}
Alur, R., Henzinger, T.~A., Lafferriere, G., \BBA\ Pappas, G.~J.
  \BBOP2000\BBCP.
\newblock \BBOQ {Discrete abstractions of hybrid systems}\BBCQ\
\newblock {\Bem Proceedings of the IEEE}, {\Bem 88\/}(7), 971--984.

\bibitem[\protect\BCAY{Anderson\ \BBA\ Moore}{Anderson\ \BBA\
  Moore}{2007}]{anderson2007optimal}
Anderson, B.~D.\BBACOMMA\  \BBA\ Moore, J.~B. \BBOP2007\BBCP.
\newblock {\Bem Optimal control: linear quadratic methods}.
\newblock Courier Corporation.

\bibitem[\protect\BCAY{Ashok, Kret{\'{\i}}nsk{\'{y}},\ \BBA\ Weininger}{Ashok
  et~al.}{2019}]{DBLP:conf/cav/AshokKW19}
Ashok, P., Kret{\'{\i}}nsk{\'{y}}, J., \BBA\ Weininger, M. \BBOP2019\BBCP.
\newblock \BBOQ {PAC} statistical model checking for markov decision processes
  and stochastic games\BBCQ\
\newblock In {\Bem {CAV} {(1)}}, \lowercase{\BVOL}\ 11561 of {\Bem Lecture
  Notes in Computer Science}, \BPGS\ 497--519. Springer.

\bibitem[\protect\BCAY{{\AA}str{\"o}m\ \BBA\ Murray}{{\AA}str{\"o}m\ \BBA\
  Murray}{2010}]{aastrom2010feedback}
{\AA}str{\"o}m, K.~J.\BBACOMMA\  \BBA\ Murray, R.~M. \BBOP2010\BBCP.
\newblock {\Bem Feedback systems: an introduction for scientists and
  engineers}.
\newblock Princeton university press.

\bibitem[\protect\BCAY{Badings, Abate, Jansen, Parker, Poonawala,\ \BBA\
  Stoelinga}{Badings et~al.}{2022a}]{Badings2022AAAI}
Badings, T.~S., Abate, A., Jansen, N., Parker, D., Poonawala, H.~A., \BBA\
  Stoelinga, M. \BBOP2022a\BBCP.
\newblock \BBOQ Sampling-based robust control of autonomous systems with
  non-gaussian noise\BBCQ\
\newblock In {\Bem {AAAI}}, \BPGS\ 9669--9678. {AAAI} Press.

\bibitem[\protect\BCAY{Badings, Cubuktepe, Jansen, Junges, Katoen,\ \BBA\
  Topcu}{Badings et~al.}{2022b}]{Badings2022STTT}
Badings, T.~S., Cubuktepe, M., Jansen, N., Junges, S., Katoen, J., \BBA\ Topcu,
  U. \BBOP2022b\BBCP.
\newblock \BBOQ Scenario-based verification of uncertain parametric mdps\BBCQ\
\newblock {\Bem Int. J. Softw. Tools Technol. Transf.}, {\Bem 24\/}(5),
  803--819.

\bibitem[\protect\BCAY{Badings, Jansen, Junges, Stoelinga,\ \BBA\ Volk}{Badings
  et~al.}{2022c}]{DBLP:conf/cav/BadingsJJSV22}
Badings, T.~S., Jansen, N., Junges, S., Stoelinga, M., \BBA\ Volk, M.
  \BBOP2022c\BBCP.
\newblock \BBOQ Sampling-based verification of ctmcs with uncertain rates\BBCQ\
\newblock In {\Bem {CAV} {(2)}}, \lowercase{\BVOL}\ 13372 of {\Bem Lecture
  Notes in Computer Science}, \BPGS\ 26--47. Springer.

\bibitem[\protect\BCAY{Badings, Jansen, Poonawala,\ \BBA\ Stoelinga}{Badings
  et~al.}{2021}]{Badings2021FilterUncertainty}
Badings, T.~S., Jansen, N., Poonawala, H.~A., \BBA\ Stoelinga, M.
  \BBOP2021\BBCP.
\newblock \BBOQ Filter-based abstractions with correctness guarantees for
  planning under uncertainty\BBCQ\
\newblock {\Bem CoRR}, {\Bem abs/2103.02398}.

\bibitem[\protect\BCAY{Badings, Romao, Abate,\ \BBA\ Jansen}{Badings
  et~al.}{2022}]{Badings2023AAAI}
Badings, T.~S., Romao, L., Abate, A., \BBA\ Jansen, N. \BBOP2022\BBCP.
\newblock \BBOQ Probabilities are not enough: Formal controller synthesis for
  stochastic dynamical models with epistemic uncertainty\BBCQ\
\newblock {\Bem CoRR}, {\Bem abs/2210.05989}.

\bibitem[\protect\BCAY{Baier\ \BBA\ Katoen}{Baier\ \BBA\
  Katoen}{2008}]{DBLP:books/daglib/BaierKatoen2008}
Baier, C.\BBACOMMA\  \BBA\ Katoen, J. \BBOP2008\BBCP.
\newblock {\Bem Principles of model checking}.
\newblock {MIT} Press.

\bibitem[\protect\BCAY{Bansal, Chen, Herbert,\ \BBA\ Tomlin}{Bansal
  et~al.}{2017}]{DBLP:conf/cdc/BansalCHT17}
Bansal, S., Chen, M., Herbert, S.~L., \BBA\ Tomlin, C.~J. \BBOP2017\BBCP.
\newblock \BBOQ Hamilton-jacobi reachability: {A} brief overview and recent
  advances\BBCQ\
\newblock In {\Bem {CDC}}, \BPGS\ 2242--2253. {IEEE}.

\bibitem[\protect\BCAY{Ben{-}Tal, Ghaoui,\ \BBA\ Nemirovski}{Ben{-}Tal
  et~al.}{2009}]{DBLP:books/degruyter/Ben-TalGN09}
Ben{-}Tal, A., Ghaoui, L.~E., \BBA\ Nemirovski, A. \BBOP2009\BBCP.
\newblock {\Bem Robust Optimization}, \lowercase{\BVOL}~28 of {\Bem Princeton
  Series in Applied Mathematics}.
\newblock Princeton University Press.

\bibitem[\protect\BCAY{Berkenkamp, Turchetta, Schoellig,\ \BBA\
  Krause}{Berkenkamp et~al.}{2017}]{DBLP:conf/nips/BerkenkampTS017}
Berkenkamp, F., Turchetta, M., Schoellig, A.~P., \BBA\ Krause, A.
  \BBOP2017\BBCP.
\newblock \BBOQ Safe model-based reinforcement learning with stability
  guarantees\BBCQ\
\newblock In {\Bem {NIPS}}, \BPGS\ 908--918.

\bibitem[\protect\BCAY{Bertsimas, Brown,\ \BBA\ Caramanis}{Bertsimas
  et~al.}{2011}]{DBLP:journals/siamrev/BertsimasBC11}
Bertsimas, D., Brown, D.~B., \BBA\ Caramanis, C. \BBOP2011\BBCP.
\newblock \BBOQ Theory and applications of robust optimization\BBCQ\
\newblock {\Bem {SIAM} Rev.}, {\Bem 53\/}(3), 464--501.

\bibitem[\protect\BCAY{Blackmore, Ono, Bektassov,\ \BBA\ Williams}{Blackmore
  et~al.}{2010}]{DBLP:journals/trob/BlackmoreOBW10}
Blackmore, L., Ono, M., Bektassov, A., \BBA\ Williams, B.~C. \BBOP2010\BBCP.
\newblock \BBOQ A probabilistic particle-control approximation of
  chance-constrained stochastic predictive control\BBCQ\
\newblock {\Bem {IEEE} Trans. Robotics}, {\Bem 26\/}(3), 502--517.

\bibitem[\protect\BCAY{B{\o}hn, Coates, Moe,\ \BBA\ Johansen}{B{\o}hn
  et~al.}{2019}]{bohn2019deep}
B{\o}hn, E., Coates, E.~M., Moe, S., \BBA\ Johansen, T.~A. \BBOP2019\BBCP.
\newblock \BBOQ Deep reinforcement learning attitude control of fixed-wing uavs
  using proximal policy optimization\BBCQ\
\newblock In {\Bem ICUAS}, \BPGS\ 523--533. IEEE.

\bibitem[\protect\BCAY{Boucheron, Lugosi,\ \BBA\ Massart}{Boucheron
  et~al.}{2013}]{DBLP:books/daglib/0035704}
Boucheron, S., Lugosi, G., \BBA\ Massart, P. \BBOP2013\BBCP.
\newblock {\Bem Concentration Inequalities - {A} Nonasymptotic Theory of
  Independence}.
\newblock Oxford University Press.

\bibitem[\protect\BCAY{Boyd\ \BBA\ Vandenberghe}{Boyd\ \BBA\
  Vandenberghe}{2014}]{DBLP:books/cu/BV2014}
Boyd, S.~P.\BBACOMMA\  \BBA\ Vandenberghe, L. \BBOP2014\BBCP.
\newblock {\Bem Convex Optimization}.
\newblock Cambridge University Press.

\bibitem[\protect\BCAY{Brafman\ \BBA\ Tennenholtz}{Brafman\ \BBA\
  Tennenholtz}{2002}]{DBLP:journals/jmlr/BrafmanT02}
Brafman, R.~I.\BBACOMMA\  \BBA\ Tennenholtz, M. \BBOP2002\BBCP.
\newblock \BBOQ {R-MAX} - {A} general polynomial time algorithm for
  near-optimal reinforcement learning\BBCQ\
\newblock {\Bem J. Mach. Learn. Res.}, {\Bem 3}, 213--231.

\bibitem[\protect\BCAY{Brunke, Greeff, Hall, Yuan, Zhou, Panerati,\ \BBA\
  Schoellig}{Brunke et~al.}{2022}]{Brunke2021safelearning}
Brunke, L., Greeff, M., Hall, A.~W., Yuan, Z., Zhou, S., Panerati, J., \BBA\
  Schoellig, A.~P. \BBOP2022\BBCP.
\newblock \BBOQ Safe learning in robotics: From learning-based control to safe
  reinforcement learning\BBCQ\
\newblock {\Bem Annual Review of Control, Robotics, and Autonomous Systems},
  {\Bem 5\/}(1), 411--444.

\bibitem[\protect\BCAY{Calafiore\ \BBA\ Campi}{Calafiore\ \BBA\
  Campi}{2005}]{DBLP:journals/mp/CalafioreC05}
Calafiore, G.~C.\BBACOMMA\  \BBA\ Campi, M.~C. \BBOP2005\BBCP.
\newblock \BBOQ Uncertain convex programs: randomized solutions and confidence
  levels\BBCQ\
\newblock {\Bem Math. Program.}, {\Bem 102\/}(1), 25--46.

\bibitem[\protect\BCAY{Campi, Car{\`{e}},\ \BBA\ Garatti}{Campi
  et~al.}{2021}]{DBLP:journals/arc/CampiCG21}
Campi, M.~C., Car{\`{e}}, A., \BBA\ Garatti, S. \BBOP2021\BBCP.
\newblock \BBOQ The scenario approach: {A} tool at the service of data-driven
  decision making\BBCQ\
\newblock {\Bem Annu. Rev. Control.}, {\Bem 52}, 1--17.

\bibitem[\protect\BCAY{Campi\ \BBA\ Garatti}{Campi\ \BBA\
  Garatti}{2008}]{DBLP:journals/siamjo/CampiG08}
Campi, M.~C.\BBACOMMA\  \BBA\ Garatti, S. \BBOP2008\BBCP.
\newblock \BBOQ The exact feasibility of randomized solutions of uncertain
  convex programs\BBCQ\
\newblock {\Bem {SIAM} J. Optim.}, {\Bem 19\/}(3), 1211--1230.

\bibitem[\protect\BCAY{Campi\ \BBA\ Garatti}{Campi\ \BBA\
  Garatti}{2011}]{DBLP:journals/jota/CampiG11}
Campi, M.~C.\BBACOMMA\  \BBA\ Garatti, S. \BBOP2011\BBCP.
\newblock \BBOQ A sampling-and-discarding approach to chance-constrained
  optimization: Feasibility and optimality\BBCQ\
\newblock {\Bem J. Optim. Theory Appl.}, {\Bem 148\/}(2), 257--280.

\bibitem[\protect\BCAY{Campi\ \BBA\ Garatti}{Campi\ \BBA\
  Garatti}{2018}]{campi2018introduction}
Campi, M.~C.\BBACOMMA\  \BBA\ Garatti, S. \BBOP2018\BBCP.
\newblock {\Bem Introduction to the scenario approach}.
\newblock SIAM.

\bibitem[\protect\BCAY{Casella\ \BBA\ Berger}{Casella\ \BBA\
  Berger}{2021}]{casella2021statistical}
Casella, G.\BBACOMMA\  \BBA\ Berger, R.~L. \BBOP2021\BBCP.
\newblock {\Bem Statistical inference}.
\newblock Cengage Learning.

\bibitem[\protect\BCAY{Cauchi\ \BBA\ Abate}{Cauchi\ \BBA\
  Abate}{2018}]{DBLP:journals/corr/abs-1803-06315}
Cauchi, N.\BBACOMMA\  \BBA\ Abate, A. \BBOP2018\BBCP.
\newblock \BBOQ Benchmarks for cyber-physical systems: {A} modular model
  library for building automation systems (extended version)\BBCQ\
\newblock {\Bem CoRR}, {\Bem abs/1803.06315}.

\bibitem[\protect\BCAY{Cauchi\ \BBA\ Abate}{Cauchi\ \BBA\
  Abate}{2019}]{DBLP:conf/tacas/CauchiA19}
Cauchi, N.\BBACOMMA\  \BBA\ Abate, A. \BBOP2019\BBCP.
\newblock \BBOQ Stochy: Automated verification and synthesis of stochastic
  processes\BBCQ\
\newblock In {\Bem {TACAS} {(2)}}, \lowercase{\BVOL}\ 11428 of {\Bem Lecture
  Notes in Computer Science}, \BPGS\ 247--264. Springer.

\bibitem[\protect\BCAY{Cauchi, Laurenti, Lahijanian, Abate, Kwiatkowska,\ \BBA\
  Cardelli}{Cauchi et~al.}{2019}]{DBLP:conf/hybrid/CauchiLLAKC19}
Cauchi, N., Laurenti, L., Lahijanian, M., Abate, A., Kwiatkowska, M., \BBA\
  Cardelli, L. \BBOP2019\BBCP.
\newblock \BBOQ Efficiency through uncertainty: scalable formal synthesis for
  stochastic hybrid systems\BBCQ\
\newblock In {\Bem {HSCC}}, \BPGS\ 240--251. {ACM}.

\bibitem[\protect\BCAY{Chen, {\'{A}}brah{\'{a}}m,\ \BBA\ Sankaranarayanan}{Chen
  et~al.}{2013}]{DBLP:conf/cav/ChenAS13}
Chen, X., {\'{A}}brah{\'{a}}m, E., \BBA\ Sankaranarayanan, S. \BBOP2013\BBCP.
\newblock \BBOQ Flow*: An analyzer for non-linear hybrid systems\BBCQ\
\newblock In {\Bem {CAV}}, \lowercase{\BVOL}\ 8044 of {\Bem Lecture Notes in
  Computer Science}, \BPGS\ 258--263. Springer.

\bibitem[\protect\BCAY{Ciesinski\ \BBA\ Gr{\"{o}}{\ss}er}{Ciesinski\ \BBA\
  Gr{\"{o}}{\ss}er}{2004}]{DBLP:conf/voss/CiesinskiG04}
Ciesinski, F.\BBACOMMA\  \BBA\ Gr{\"{o}}{\ss}er, M. \BBOP2004\BBCP.
\newblock \BBOQ On probabilistic computation tree logic\BBCQ\
\newblock In {\Bem Validation of Stochastic Systems}, \lowercase{\BVOL}\ 2925
  of {\Bem Lecture Notes in Computer Science}, \BPGS\ 147--188. Springer.

\bibitem[\protect\BCAY{Clohessy\ \BBA\ Wiltshire}{Clohessy\ \BBA\
  Wiltshire}{1960}]{clohessy1960terminal}
Clohessy, W.\BBACOMMA\  \BBA\ Wiltshire, R. \BBOP1960\BBCP.
\newblock \BBOQ Terminal guidance system for satellite rendezvous\BBCQ\
\newblock {\Bem Journal of the Aerospace Sciences}, {\Bem 27\/}(9), 653--658.

\bibitem[\protect\BCAY{Desharnais, Edalat,\ \BBA\ Panangaden}{Desharnais
  et~al.}{2002}]{DBLP:journals/iandc/DesharnaisEP02}
Desharnais, J., Edalat, A., \BBA\ Panangaden, P. \BBOP2002\BBCP.
\newblock \BBOQ Bisimulation for labelled markov processes\BBCQ\
\newblock {\Bem Inf. Comput.}, {\Bem 179\/}(2), 163--193.

\bibitem[\protect\BCAY{D'Innocenzo, Abate,\ \BBA\ Katoen}{D'Innocenzo
  et~al.}{2012}]{DBLP:conf/hybrid/DInnocenzoAK12}
D'Innocenzo, A., Abate, A., \BBA\ Katoen, J. \BBOP2012\BBCP.
\newblock \BBOQ Robust {PCTL} model checking\BBCQ\
\newblock In {\Bem {HSCC}}, \BPGS\ 275--286. {ACM}.

\bibitem[\protect\BCAY{Dryden}{Dryden}{1943}]{dryden1943review}
Dryden, H.~L. \BBOP1943\BBCP.
\newblock \BBOQ A review of the statistical theory of turbulence\BBCQ\
\newblock {\Bem Quarterly of Applied Mathematics}, {\Bem 1\/}(1), 7--42.

\bibitem[\protect\BCAY{Esfahani\ \BBA\ Kuhn}{Esfahani\ \BBA\
  Kuhn}{2018}]{DBLP:journals/mp/EsfahaniK18}
Esfahani, P.~M.\BBACOMMA\  \BBA\ Kuhn, D. \BBOP2018\BBCP.
\newblock \BBOQ Data-driven distributionally robust optimization using the
  wasserstein metric: performance guarantees and tractable reformulations\BBCQ\
\newblock {\Bem Math. Program.}, {\Bem 171\/}(1-2), 115--166.

\bibitem[\protect\BCAY{Fan, Qin, Mathur, Ning, Mitra,\ \BBA\ Viswanathan}{Fan
  et~al.}{2022}]{DBLP:journals/tac/FanQMNMV22}
Fan, C., Qin, Z., Mathur, U., Ning, Q., Mitra, S., \BBA\ Viswanathan, M.
  \BBOP2022\BBCP.
\newblock \BBOQ Controller synthesis for linear system with reach-avoid
  specifications\BBCQ\
\newblock {\Bem {IEEE} Trans. Autom. Control.}, {\Bem 67\/}(4), 1713--1727.

\bibitem[\protect\BCAY{Ferns, Panangaden,\ \BBA\ Precup}{Ferns
  et~al.}{2004}]{DBLP:conf/aaai/FernsPP04}
Ferns, N., Panangaden, P., \BBA\ Precup, D. \BBOP2004\BBCP.
\newblock \BBOQ Metrics for finite markov decision processes\BBCQ\
\newblock In {\Bem {AAAI}}, \BPGS\ 950--951. {AAAI} Press / The {MIT} Press.

\bibitem[\protect\BCAY{Ferns, Precup,\ \BBA\ Knight}{Ferns
  et~al.}{2014}]{DBLP:conf/birthday/FernsPK14}
Ferns, N., Precup, D., \BBA\ Knight, S. \BBOP2014\BBCP.
\newblock \BBOQ Bisimulation for markov decision processes through families of
  functional expressions\BBCQ\
\newblock In {\Bem Horizons of the Mind}, \lowercase{\BVOL}\ 8464 of {\Bem
  Lecture Notes in Computer Science}, \BPGS\ 319--342. Springer.

\bibitem[\protect\BCAY{Fisac, Akametalu, Zeilinger, Kaynama, Gillula,\ \BBA\
  Tomlin}{Fisac et~al.}{2019}]{DBLP:journals/tac/FisacAZKGT19}
Fisac, J.~F., Akametalu, A.~K., Zeilinger, M.~N., Kaynama, S., Gillula, J.~H.,
  \BBA\ Tomlin, C.~J. \BBOP2019\BBCP.
\newblock \BBOQ A general safety framework for learning-based control in
  uncertain robotic systems\BBCQ\
\newblock {\Bem {IEEE} Trans. Autom. Control.}, {\Bem 64\/}(7), 2737--2752.

\bibitem[\protect\BCAY{Fu\ \BBA\ Topcu}{Fu\ \BBA\
  Topcu}{2014}]{DBLP:conf/rss/FuT14}
Fu, J.\BBACOMMA\  \BBA\ Topcu, U. \BBOP2014\BBCP.
\newblock \BBOQ Probably approximately correct {MDP} learning and control with
  temporal logic constraints\BBCQ\
\newblock In {\Bem Robotics: Science and Systems}.

\bibitem[\protect\BCAY{Garatti\ \BBA\ Campi}{Garatti\ \BBA\
  Campi}{2022}]{garatti2019risk}
Garatti, S.\BBACOMMA\  \BBA\ Campi, M.~C. \BBOP2022\BBCP.
\newblock \BBOQ Risk and complexity in scenario optimization\BBCQ\
\newblock {\Bem Math. Program.}, {\Bem 191\/}(1), 243--279.

\bibitem[\protect\BCAY{Garc{\'{\i}}a\ \BBA\ Fern{\'{a}}ndez}{Garc{\'{\i}}a\
  \BBA\ Fern{\'{a}}ndez}{2015}]{DBLP:journals/jmlr/GarciaF15}
Garc{\'{\i}}a, J.\BBACOMMA\  \BBA\ Fern{\'{a}}ndez, F. \BBOP2015\BBCP.
\newblock \BBOQ A comprehensive survey on safe reinforcement learning\BBCQ\
\newblock {\Bem J. Mach. Learn. Res.}, {\Bem 16}, 1437--1480.

\bibitem[\protect\BCAY{Givan, Dean,\ \BBA\ Greig}{Givan
  et~al.}{2003}]{DBLP:journals/ai/GivanDG03}
Givan, R., Dean, T.~L., \BBA\ Greig, M. \BBOP2003\BBCP.
\newblock \BBOQ Equivalence notions and model minimization in markov decision
  processes\BBCQ\
\newblock {\Bem Artif. Intell.}, {\Bem 147\/}(1-2), 163--223.

\bibitem[\protect\BCAY{Givan, Leach,\ \BBA\ Dean}{Givan
  et~al.}{2000}]{DBLP:journals/ai/GivanLD00}
Givan, R., Leach, S.~M., \BBA\ Dean, T.~L. \BBOP2000\BBCP.
\newblock \BBOQ Bounded-parameter markov decision processes\BBCQ\
\newblock {\Bem Artif. Intell.}, {\Bem 122\/}(1-2), 71--109.

\bibitem[\protect\BCAY{Goh\ \BBA\ Sim}{Goh\ \BBA\
  Sim}{2010}]{goh2010distributionally}
Goh, J.\BBACOMMA\  \BBA\ Sim, M. \BBOP2010\BBCP.
\newblock \BBOQ Distributionally robust optimization and its tractable
  approximations\BBCQ\
\newblock {\Bem Operations research}, {\Bem 58\/}(4-part-1), 902--917.

\bibitem[\protect\BCAY{Haesaert, Soudjani,\ \BBA\ Abate}{Haesaert
  et~al.}{2017}]{HSA17}
Haesaert, S., Soudjani, S., \BBA\ Abate, A. \BBOP2017\BBCP.
\newblock \BBOQ Verification of general markov decision processes by
  approximate similarity relations and policy refinement\BBCQ\
\newblock {\Bem SIAM Journal on Control and Optimisation}, {\Bem 55\/}(4),
  2333--2367.

\bibitem[\protect\BCAY{Hahn, Hashemi, Hermanns, Lahijanian,\ \BBA\
  Turrini}{Hahn et~al.}{2017}]{DBLP:conf/qest/HahnHHLT17}
Hahn, E.~M., Hashemi, V., Hermanns, H., Lahijanian, M., \BBA\ Turrini, A.
  \BBOP2017\BBCP.
\newblock \BBOQ Multi-objective robust strategy synthesis for interval markov
  decision processes\BBCQ\
\newblock In {\Bem {QEST}}, \lowercase{\BVOL}\ 10503 of {\Bem Lecture Notes in
  Computer Science}, \BPGS\ 207--223. Springer.

\bibitem[\protect\BCAY{Hahn, Hashemi, Hermanns,\ \BBA\ Turrini}{Hahn
  et~al.}{2016}]{DBLP:conf/qest/HahnHHT16}
Hahn, E.~M., Hashemi, V., Hermanns, H., \BBA\ Turrini, A. \BBOP2016\BBCP.
\newblock \BBOQ Exploiting robust optimization for interval probabilistic
  bisimulation\BBCQ\
\newblock In {\Bem {QEST}}, \lowercase{\BVOL}\ 9826 of {\Bem Lecture Notes in
  Computer Science}, \BPGS\ 55--71. Springer.

\bibitem[\protect\BCAY{Hansson\ \BBA\ Jonsson}{Hansson\ \BBA\
  Jonsson}{1994}]{DBLP:journals/fac/HanssonJ94}
Hansson, H.\BBACOMMA\  \BBA\ Jonsson, B. \BBOP1994\BBCP.
\newblock \BBOQ A logic for reasoning about time and reliability\BBCQ\
\newblock {\Bem Formal Aspects Comput.}, {\Bem 6\/}(5), 512--535.

\bibitem[\protect\BCAY{Hashemi, Hatefi,\ \BBA\ Krc{\'{a}}l}{Hashemi
  et~al.}{2014}]{DBLP:journals/corr/HashemiHK14}
Hashemi, V., Hatefi, H., \BBA\ Krc{\'{a}}l, J. \BBOP2014\BBCP.
\newblock \BBOQ Probabilistic bisimulations for {PCTL} model checking of
  interval mdps (extended version)\BBCQ\
\newblock In {\Bem SynCoP}, \lowercase{\BVOL}\ 145 of {\Bem {EPTCS}}, \BPGS\
  19--33.

\bibitem[\protect\BCAY{Hashemi, Hermanns, Song, Subramani, Turrini,\ \BBA\
  Wojciechowski}{Hashemi et~al.}{2016}]{DBLP:conf/lata/HashemiH0STW16}
Hashemi, V., Hermanns, H., Song, L., Subramani, K., Turrini, A., \BBA\
  Wojciechowski, P. \BBOP2016\BBCP.
\newblock \BBOQ Compositional bisimulation minimization for interval markov
  decision processes\BBCQ\
\newblock In {\Bem {LATA}}, \lowercase{\BVOL}\ 9618 of {\Bem Lecture Notes in
  Computer Science}, \BPGS\ 114--126. Springer.

\bibitem[\protect\BCAY{Haussler}{Haussler}{1990}]{DBLP:conf/aaai/Haussler90}
Haussler, D. \BBOP1990\BBCP.
\newblock \BBOQ Probably approximately correct learning\BBCQ\
\newblock In {\Bem {AAAI}}, \BPGS\ 1101--1108. {AAAI} Press / The {MIT} Press.

\bibitem[\protect\BCAY{Herbert, Chen, Han, Bansal, Fisac,\ \BBA\
  Tomlin}{Herbert et~al.}{2017}]{DBLP:conf/cdc/HerbertCHBFT17}
Herbert, S.~L., Chen, M., Han, S., Bansal, S., Fisac, J.~F., \BBA\ Tomlin,
  C.~J. \BBOP2017\BBCP.
\newblock \BBOQ Fastrack: {A} modular framework for fast and guaranteed safe
  motion planning\BBCQ\
\newblock In {\Bem {CDC}}, \BPGS\ 1517--1522. {IEEE}.

\bibitem[\protect\BCAY{Hermanns, Parma, Segala, Wachter,\ \BBA\ Zhang}{Hermanns
  et~al.}{2011}]{hermanns2011probabilistic}
Hermanns, H., Parma, A., Segala, R., Wachter, B., \BBA\ Zhang, L.
  \BBOP2011\BBCP.
\newblock \BBOQ Probabilistic logical characterization\BBCQ\
\newblock {\Bem Information and Computation}, {\Bem 209\/}(2), 154--172.

\bibitem[\protect\BCAY{Hewing, Kabzan,\ \BBA\ Zeilinger}{Hewing
  et~al.}{2020}]{DBLP:journals/tcst/HewingKZ20}
Hewing, L., Kabzan, J., \BBA\ Zeilinger, M.~N. \BBOP2020\BBCP.
\newblock \BBOQ Cautious model predictive control using gaussian process
  regression\BBCQ\
\newblock {\Bem {IEEE} Trans. Control. Syst. Technol.}, {\Bem 28\/}(6),
  2736--2743.

\bibitem[\protect\BCAY{Jewison\ \BBA\ Erwin}{Jewison\ \BBA\
  Erwin}{2016}]{DBLP:conf/cdc/JewisonE16}
Jewison, C.\BBACOMMA\  \BBA\ Erwin, R.~S. \BBOP2016\BBCP.
\newblock \BBOQ A spacecraft benchmark problem for hybrid control and
  estimation\BBCQ\
\newblock In {\Bem {CDC}}, \BPGS\ 3300--3305. {IEEE}.

\bibitem[\protect\BCAY{Kearns\ \BBA\ Singh}{Kearns\ \BBA\
  Singh}{2002}]{DBLP:journals/ml/KearnsS02}
Kearns, M.~J.\BBACOMMA\  \BBA\ Singh, S.~P. \BBOP2002\BBCP.
\newblock \BBOQ Near-optimal reinforcement learning in polynomial time\BBCQ\
\newblock {\Bem Mach. Learn.}, {\Bem 49\/}(2-3), 209--232.

\bibitem[\protect\BCAY{Koller, Berkenkamp, Turchetta,\ \BBA\ Krause}{Koller
  et~al.}{2018}]{DBLP:conf/cdc/KollerBT018}
Koller, T., Berkenkamp, F., Turchetta, M., \BBA\ Krause, A. \BBOP2018\BBCP.
\newblock \BBOQ Learning-based model predictive control for safe
  exploration\BBCQ\
\newblock In {\Bem {CDC}}, \BPGS\ 6059--6066. {IEEE}.

\bibitem[\protect\BCAY{Kulakowski, Gardner,\ \BBA\ Shearer}{Kulakowski
  et~al.}{2007}]{Kulakowski2007dynamic}
Kulakowski, B.~T., Gardner, J.~F., \BBA\ Shearer, J.~L. \BBOP2007\BBCP.
\newblock {\Bem Dynamic modeling and control of engineering systems}.
\newblock Cambridge University Press.

\bibitem[\protect\BCAY{Kwiatkowska, Norman,\ \BBA\ Parker}{Kwiatkowska
  et~al.}{2011}]{DBLP:conf/cav/KwiatkowskaNP11}
Kwiatkowska, M.~Z., Norman, G., \BBA\ Parker, D. \BBOP2011\BBCP.
\newblock \BBOQ {PRISM} 4.0: Verification of probabilistic real-time
  systems\BBCQ\
\newblock In {\Bem {CAV}}, \lowercase{\BVOL}\ 6806 of {\Bem Lecture Notes in
  Computer Science}, \BPGS\ 585--591. Springer.

\bibitem[\protect\BCAY{Kwiatkowska\ \BBA\ Parker}{Kwiatkowska\ \BBA\
  Parker}{2013}]{DBLP:conf/atva/KwiatkowskaP13}
Kwiatkowska, M.~Z.\BBACOMMA\  \BBA\ Parker, D. \BBOP2013\BBCP.
\newblock \BBOQ Automated verification and strategy synthesis for probabilistic
  systems\BBCQ\
\newblock In {\Bem {ATVA}}, \lowercase{\BVOL}\ 8172 of {\Bem Lecture Notes in
  Computer Science}, \BPGS\ 5--22. Springer.

\bibitem[\protect\BCAY{Lahijanian, Andersson,\ \BBA\ Belta}{Lahijanian
  et~al.}{2015}]{DBLP:journals/tac/LahijanianAB15}
Lahijanian, M., Andersson, S.~B., \BBA\ Belta, C. \BBOP2015\BBCP.
\newblock \BBOQ Formal verification and synthesis for discrete-time stochastic
  systems\BBCQ\
\newblock {\Bem {IEEE} Trans. Autom. Control.}, {\Bem 60\/}(8), 2031--2045.

\bibitem[\protect\BCAY{Larsen\ \BBA\ Skou}{Larsen\ \BBA\
  Skou}{1991}]{DBLP:journals/iandc/LarsenS91}
Larsen, K.~G.\BBACOMMA\  \BBA\ Skou, A. \BBOP1991\BBCP.
\newblock \BBOQ Bisimulation through probabilistic testing\BBCQ\
\newblock {\Bem Inf. Comput.}, {\Bem 94\/}(1), 1--28.

\bibitem[\protect\BCAY{Lavaei, Soudjani, Abate,\ \BBA\ Zamani}{Lavaei
  et~al.}{2022}]{LSAZ21}
Lavaei, A., Soudjani, S., Abate, A., \BBA\ Zamani, M. \BBOP2022\BBCP.
\newblock \BBOQ Automated verification and synthesis of stochastic hybrid
  systems: A survey\BBCQ\
\newblock {\Bem Automatica}, {\Bem 146}, 110617.

\bibitem[\protect\BCAY{Lesser, Oishi,\ \BBA\ Erwin}{Lesser
  et~al.}{2013}]{DBLP:conf/cdc/LesserOE13}
Lesser, K., Oishi, M. M.~K., \BBA\ Erwin, R.~S. \BBOP2013\BBCP.
\newblock \BBOQ Stochastic reachability for control of spacecraft relative
  motion\BBCQ\
\newblock In {\Bem {CDC}}, \BPGS\ 4705--4712. {IEEE}.

\bibitem[\protect\BCAY{Lun, Wheatley, D'Innocenzo,\ \BBA\ Abate}{Lun
  et~al.}{2018}]{DBLP:conf/adhs/LunWDA18}
Lun, Y.~Z., Wheatley, J., D'Innocenzo, A., \BBA\ Abate, A. \BBOP2018\BBCP.
\newblock \BBOQ Approximate abstractions of markov chains with interval
  decision processes\BBCQ\
\newblock In {\Bem {ADHS}}, \lowercase{\BVOL}~51 of {\Bem IFAC-PapersOnLine},
  \BPGS\ 91--96. Elsevier.

\bibitem[\protect\BCAY{Margellos, Goulart,\ \BBA\ Lygeros}{Margellos
  et~al.}{2014}]{Margellos2014OnProblems}
Margellos, K., Goulart, P., \BBA\ Lygeros, J. \BBOP2014\BBCP.
\newblock \BBOQ {On the road between robust optimization and the scenario
  approach for chance constrained optimization problems}\BBCQ\
\newblock {\Bem IEEE Transactions on Automatic Control}, {\Bem 59\/}(8),
  2258--2263.

\bibitem[\protect\BCAY{Moos, Hansel, Abdulsamad, Stark, Clever,\ \BBA\
  Peters}{Moos et~al.}{2022}]{DBLP:journals/make/MoosHASCP22}
Moos, J., Hansel, K., Abdulsamad, H., Stark, S., Clever, D., \BBA\ Peters, J.
  \BBOP2022\BBCP.
\newblock \BBOQ Robust reinforcement learning: {A} review of foundations and
  recent advances\BBCQ\
\newblock {\Bem Mach. Learn. Knowl. Extr.}, {\Bem 4\/}(1), 276--315.

\bibitem[\protect\BCAY{Morimoto\ \BBA\ Doya}{Morimoto\ \BBA\
  Doya}{2005}]{DBLP:journals/neco/MorimotoD05}
Morimoto, J.\BBACOMMA\  \BBA\ Doya, K. \BBOP2005\BBCP.
\newblock \BBOQ Robust reinforcement learning\BBCQ\
\newblock {\Bem Neural Comput.}, {\Bem 17\/}(2), 335--359.

\bibitem[\protect\BCAY{Nilim\ \BBA\ Ghaoui}{Nilim\ \BBA\
  Ghaoui}{2005}]{DBLP:journals/ior/NilimG05}
Nilim, A.\BBACOMMA\  \BBA\ Ghaoui, L.~E. \BBOP2005\BBCP.
\newblock \BBOQ Robust control of markov decision processes with uncertain
  transition matrices\BBCQ\
\newblock {\Bem Oper. Res.}, {\Bem 53\/}(5), 780--798.

\bibitem[\protect\BCAY{Ogata et~al.}{Ogata et~al.}{2010}]{ogata2010modern}
Ogata, K.\BBACOMMA\  et~al. \BBOP2010\BBCP.
\newblock {\Bem Modern control engineering}, \lowercase{\BVOL}~5.
\newblock Prentice hall Upper Saddle River, NJ.

\bibitem[\protect\BCAY{Park, Serpedin,\ \BBA\ Qaraqe}{Park
  et~al.}{2013}]{DBLP:journals/spm/ParkSQ13}
Park, S., Serpedin, E., \BBA\ Qaraqe, K.~A. \BBOP2013\BBCP.
\newblock \BBOQ Gaussian assumption: The least favorable but the most useful
  [lecture notes]\BBCQ\
\newblock {\Bem {IEEE} Signal Process. Mag.}, {\Bem 30\/}(3), 183--186.

\bibitem[\protect\BCAY{Peng, Andrychowicz, Zaremba,\ \BBA\ Abbeel}{Peng
  et~al.}{2018}]{DBLP:conf/icra/PengAZA18}
Peng, X.~B., Andrychowicz, M., Zaremba, W., \BBA\ Abbeel, P. \BBOP2018\BBCP.
\newblock \BBOQ Sim-to-real transfer of robotic control with dynamics
  randomization\BBCQ\
\newblock In {\Bem {ICRA}}, \BPGS\ 1--8. {IEEE}.

\bibitem[\protect\BCAY{Pinto, Davidson, Sukthankar,\ \BBA\ Gupta}{Pinto
  et~al.}{2017}]{DBLP:conf/icml/PintoDSG17}
Pinto, L., Davidson, J., Sukthankar, R., \BBA\ Gupta, A. \BBOP2017\BBCP.
\newblock \BBOQ Robust adversarial reinforcement learning\BBCQ\
\newblock In {\Bem {ICML}}, \lowercase{\BVOL}~70 of {\Bem Proceedings of
  Machine Learning Research}, \BPGS\ 2817--2826. {PMLR}.

\bibitem[\protect\BCAY{Puggelli, Li, Sangiovanni{-}Vincentelli,\ \BBA\
  Seshia}{Puggelli et~al.}{2013}]{DBLP:conf/cav/PuggelliLSS13}
Puggelli, A., Li, W., Sangiovanni{-}Vincentelli, A.~L., \BBA\ Seshia, S.~A.
  \BBOP2013\BBCP.
\newblock \BBOQ Polynomial-time verification of {PCTL} properties of mdps with
  convex uncertainties\BBCQ\
\newblock In {\Bem {CAV}}, \lowercase{\BVOL}\ 8044 of {\Bem Lecture Notes in
  Computer Science}, \BPGS\ 527--542. Springer.

\bibitem[\protect\BCAY{Puterman}{Puterman}{1994}]{DBLP:books/wi/Puterman94}
Puterman, M.~L. \BBOP1994\BBCP.
\newblock {\Bem {M}arkov Decision Processes: Discrete Stochastic Dynamic
  Programming}.
\newblock Wiley Series in Probability and Statistics. Wiley.

\bibitem[\protect\BCAY{Reissig, Weber,\ \BBA\ Rungger}{Reissig
  et~al.}{2017}]{DBLP:journals/tac/ReissigWR17}
Reissig, G., Weber, A., \BBA\ Rungger, M. \BBOP2017\BBCP.
\newblock \BBOQ Feedback refinement relations for the synthesis of symbolic
  controllers\BBCQ\
\newblock {\Bem {IEEE} Trans. Autom. Control.}, {\Bem 62\/}(4), 1781--1796.

\bibitem[\protect\BCAY{Reist, Preiswerk,\ \BBA\ Tedrake}{Reist
  et~al.}{2016}]{DBLP:journals/ijrr/ReistPT16}
Reist, P., Preiswerk, P., \BBA\ Tedrake, R. \BBOP2016\BBCP.
\newblock \BBOQ Feedback-motion-planning with simulation-based lqr-trees\BBCQ\
\newblock {\Bem Int. J. Robotics Res.}, {\Bem 35\/}(11), 1393--1416.

\bibitem[\protect\BCAY{Romao, Papachristodoulou,\ \BBA\ Margellos}{Romao
  et~al.}{2022}]{romao2022tac}
Romao, L., Papachristodoulou, A., \BBA\ Margellos, K. \BBOP2022\BBCP.
\newblock \BBOQ On the exact feasibility of convex scenario programs with
  discarded constraints\BBCQ\
\newblock {\Bem IEEE Transactions on Automatic Control}, {\Bem to appear}.

\bibitem[\protect\BCAY{Rosolia, Singletary,\ \BBA\ Ames}{Rosolia
  et~al.}{2022}]{Rosolia2020unified}
Rosolia, U., Singletary, A., \BBA\ Ames, A.~D. \BBOP2022\BBCP.
\newblock \BBOQ Unified multirate control: From low-level actuation to
  high-level planning\BBCQ\
\newblock {\Bem IEEE Transactions on Automatic Control}, {\Bem 67\/}(12),
  6627--6640.

\bibitem[\protect\BCAY{Sartipizadeh, Vinod, Acikmese,\ \BBA\
  Oishi}{Sartipizadeh et~al.}{2019}]{DBLP:conf/amcc/SartipizadehVAO19}
Sartipizadeh, H., Vinod, A.~P., Acikmese, B., \BBA\ Oishi, M. \BBOP2019\BBCP.
\newblock \BBOQ Voronoi partition-based scenario reduction for fast
  sampling-based stochastic reachability computation of linear systems\BBCQ\
\newblock In {\Bem {ACC}}, \BPGS\ 37--44. {IEEE}.

\bibitem[\protect\BCAY{Shmarov\ \BBA\ Zuliani}{Shmarov\ \BBA\
  Zuliani}{2015}]{DBLP:conf/hybrid/ShmarovZ15}
Shmarov, F.\BBACOMMA\  \BBA\ Zuliani, P. \BBOP2015\BBCP.
\newblock \BBOQ Probreach: verified probabilistic delta-reachability for
  stochastic hybrid systems\BBCQ\
\newblock In {\Bem {HSCC}}, \BPGS\ 134--139. {ACM}.

\bibitem[\protect\BCAY{Smith}{Smith}{2013}]{Smith2013sequentialMonteCarlo}
Smith, A. \BBOP2013\BBCP.
\newblock {\Bem Sequential Monte Carlo methods in practice}.
\newblock Springer Science \& Business Media.

\bibitem[\protect\BCAY{Soudjani\ \BBA\ Abate}{Soudjani\ \BBA\
  Abate}{2013}]{DBLP:journals/siamads/SoudjaniA13}
Soudjani, S. E.~Z.\BBACOMMA\  \BBA\ Abate, A. \BBOP2013\BBCP.
\newblock \BBOQ Adaptive and sequential gridding procedures for the abstraction
  and verification of stochastic processes\BBCQ\
\newblock {\Bem {SIAM} J. Appl. Dyn. Syst.}, {\Bem 12\/}(2), 921--956.

\bibitem[\protect\BCAY{Sullivan}{Sullivan}{2015}]{sullivan2015introduction}
Sullivan, T.~J. \BBOP2015\BBCP.
\newblock {\Bem Introduction to uncertainty quantification},
  \lowercase{\BVOL}~63.
\newblock Springer.

\bibitem[\protect\BCAY{Taylor, Singletary, Yue,\ \BBA\ Ames}{Taylor
  et~al.}{2020}]{DBLP:conf/l4dc/TaylorSYA20}
Taylor, A.~J., Singletary, A., Yue, Y., \BBA\ Ames, A.~D. \BBOP2020\BBCP.
\newblock \BBOQ Learning for safety-critical control with control barrier
  functions\BBCQ\
\newblock In {\Bem {L4DC}}, \lowercase{\BVOL}\ 120 of {\Bem Proceedings of
  Machine Learning Research}, \BPGS\ 708--717. {PMLR}.

\bibitem[\protect\BCAY{Tedrake, Manchester, Tobenkin,\ \BBA\ Roberts}{Tedrake
  et~al.}{2010}]{DBLP:journals/ijrr/TedrakeMTR10}
Tedrake, R., Manchester, I.~R., Tobenkin, M.~M., \BBA\ Roberts, J.~W.
  \BBOP2010\BBCP.
\newblock \BBOQ Lqr-trees: Feedback motion planning via sums-of-squares
  verification\BBCQ\
\newblock {\Bem Int. J. Robotics Res.}, {\Bem 29\/}(8), 1038--1052.

\bibitem[\protect\BCAY{Tkachev\ \BBA\ Abate}{Tkachev\ \BBA\ Abate}{2014}]{TA14}
Tkachev, I.\BBACOMMA\  \BBA\ Abate, A. \BBOP2014\BBCP.
\newblock \BBOQ Characterization and computation of infinite horizon
  specifications over markov processes\BBCQ\
\newblock {\Bem Theoretical Computer Science}, {\Bem 515}, 1--18.

\bibitem[\protect\BCAY{Vinitsky, Du, Parvate, Jang, Abbeel,\ \BBA\
  Bayen}{Vinitsky et~al.}{2020}]{Vinitsky2020}
Vinitsky, E., Du, Y., Parvate, K., Jang, K., Abbeel, P., \BBA\ Bayen, A.~M.
  \BBOP2020\BBCP.
\newblock \BBOQ Robust reinforcement learning using adversarial
  populations\BBCQ\
\newblock {\Bem CoRR}, {\Bem abs/2008.01825}.

\bibitem[\protect\BCAY{Vinod, Gleason,\ \BBA\ Oishi}{Vinod
  et~al.}{2019}]{DBLP:conf/hybrid/VinodGO19}
Vinod, A.~P., Gleason, J.~D., \BBA\ Oishi, M. M.~K. \BBOP2019\BBCP.
\newblock \BBOQ Sreachtools: a {MATLAB} stochastic reachability toolbox\BBCQ\
\newblock In {\Bem {HSCC}}, \BPGS\ 33--38. {ACM}.

\bibitem[\protect\BCAY{Wiesemann, Kuhn,\ \BBA\ Sim}{Wiesemann
  et~al.}{2014}]{DBLP:journals/ior/WiesemannKS14}
Wiesemann, W., Kuhn, D., \BBA\ Sim, M. \BBOP2014\BBCP.
\newblock \BBOQ Distributionally robust convex optimization\BBCQ\
\newblock {\Bem Oper. Res.}, {\Bem 62\/}(6), 1358--1376.

\bibitem[\protect\BCAY{Wolff, Topcu,\ \BBA\ Murray}{Wolff
  et~al.}{2012}]{DBLP:conf/cdc/WolffTM12}
Wolff, E.~M., Topcu, U., \BBA\ Murray, R.~M. \BBOP2012\BBCP.
\newblock \BBOQ Robust control of uncertain markov decision processes with
  temporal logic specifications\BBCQ\
\newblock In {\Bem {CDC}}, \BPGS\ 3372--3379. {IEEE}.

\bibitem[\protect\BCAY{Xu\ \BBA\ Mannor}{Xu\ \BBA\
  Mannor}{2010}]{DBLP:conf/nips/XuM10}
Xu, H.\BBACOMMA\  \BBA\ Mannor, S. \BBOP2010\BBCP.
\newblock \BBOQ Distributionally robust markov decision processes\BBCQ\
\newblock In {\Bem {NIPS}}, \BPGS\ 2505--2513. Curran Associates, Inc.

\bibitem[\protect\BCAY{Zikelic, Lechner, Henzinger,\ \BBA\ Chatterjee}{Zikelic
  et~al.}{2022}]{Zikelic2022}
Zikelic, D., Lechner, M., Henzinger, T.~A., \BBA\ Chatterjee, K.
  \BBOP2022\BBCP.
\newblock \BBOQ Learning control policies for stochastic systems with
  reach-avoid guarantees\BBCQ\
\newblock {\Bem CoRR}, {\Bem abs/2210.05308}.

\end{thebibliography}
\bibliographystyle{theapa}

\end{document}